Review

# A comprehensive survey on smart contract construction and execution: paradigms, tools, and systems

Bin Hu,[1] Zongyang Zhang,[1,*] Jianwei Liu,[1,*] Yizhong Liu,[1] Jiayuan Yin,[1] Rongxing Lu,[2] and Xiaodong Lin[3]
[1]School of Cyber Science and Technology, Beihang University, Beijing 100191, China
[2]Faculty of Computer Science, University of New Brunswick, Fredericton, NB E3B 5A3, Canada
[3]School of Computer Science, University of Guelph, Guelph, ON N1G 2W1, Canada
*Correspondence: zongyangzhang@buaa.edu.cn (Z.Z.), liujianwei@buaa.edu.cn (J.L.)

**THE BIGGER PICTURE** Smart contracts are one of the most promising and appealing notions in blockchain technology. We provide a comprehensive review of state-of-the-art smart contract construction and execution schemes. We classify three major categories: (1) design paradigms that give examples and patterns on contract construction, (2) design tools that facilitate the development of secure smart contracts, and (3) extensions and alternatives that improve privacy or efficiency. We found that frequently occurred vulnerabilities, incomplete paradigms, inefficient analysis tools, low processing rate, limited contract complexity, and the lack of privacy are the main challenges that hinder the adoption of smart contracts. We identify several future research directions, including fair off-chain networks, practical implementations, scalable and automatic analysis tools, and private contracts with practical compilers.

## SUMMARY

Smart contracts are regarded as one of the most promising and appealing notions in blockchain technology. Their self-enforcing and event-driven features make some online activities possible without a trusted third party. Nevertheless, problems such as miscellaneous attacks, privacy leakage, and low processing rates prevent them from being widely applied. Various schemes and tools have been proposed to facilitate the construction and execution of secure smart contracts. However, a comprehensive survey for these proposals is absent, hindering new researchers and developers from a quick start. This paper surveys the literature and online resources on smart contract construction and execution over the period 2008–2020. We divide the studies into three categories: (1) design paradigms that give examples and patterns on contract construction, (2) design tools that facilitate the development of secure smart contracts, and (3) extensions and alternatives that improve the privacy or efficiency of the system. We start by grouping the relevant construction schemes into the first two categories. We then review the execution mechanisms in the last category and further divide the state-of-the-art solutions into three classes: private contracts with extra tools, off-chain channels, and extensions on core functionalities. Finally, we summarize several challenges and identify future research directions toward developing secure, privacy-preserving, and efficient smart contracts.

## 1. INTRODUCTION

The advent of Bitcoin[1] in 2008 marks the birth of cryptocurrencies, which are maintained without trusted third parties (TTP). Since then, numerous cryptocurrencies have emerged. Unlike traditional fiat currencies issued by governments, cryptocurrencies circulate in specific computer programs through peer-to-peer (P2P) technology, whereby no leader or dominant node is responsible for message transmission. Cryptocurrencies are equipped with numerous cryptographic and game-theoretic schemes that ensure their safe circulation on the Internet.

The core technology that enables the cryptocurrencies is blockchain, which ensures data consistency among distributed nodes in a P2P network without mutual trust. Later in 2014, Ethereum[2,3] extended Bitcoin and introduced the smart contract[4] into the blockchain, which greatly enriches the application scenarios of blockchain. Ethereum thus becomes one of the most promoting motivations of blockchain technology.

After the birth of Ethereum, applications of smart contracts have gradually become prevalent, and many other platforms have been derived. In 2016, Corda,[5] a distributed ledger platform for the financial-service industry, was proposed to improve the transaction-processing rate. For the privacy concerns on smart contracts, Quorum[6] has introduced private state trie and other technical methods into Ethereum to support the execution of private contracts. Besides, the Hyperledger Fabric[7] system led by IBM facilitates the application of smart contracts. It allows companies or consortia to run it collaboratively to improve the transaction-processing rate while keeping the data consistent and non-malleable. Since 2015, smart contracts in Hyperledger Fabric have been widely used in the supply chain, education, business, and other domains[8] to support reliable data communication and value exchanges by maintaining data consistency and verifiability among various departments and organizations.

From the academic perspective, researchers are mostly dedicated to the improvements on public blockchains where everyone could join, especially Bitcoin and Ethereum, since public blockchains have received the most attention by users and researchers, and many systems are derived from Bitcoin and Ethereum (e.g., Corda[5] and Quorum[6]). Moreover, problems that occur on these platforms are usually universal to other derived systems. Developers might migrate the research results to other platforms with slight modifications.

One of the most significant issues of blockchain and smart contract technology is that all transaction details are public. This might cause the leakage of privacy despite the pseudonym mechanism by identifying and clustering users with pseudonyms.[9,10] Moreover, since transactions are executed and validated by all participating nodes in a duplicated way, the transaction rate (or throughput) is quite limited in the blockchain system. Both the privacy and efficiency problems hinder many applications from being implemented as smart contracts and make smart contracts only applicable in a small number of fields.

There are also concerns that smart contracts are vulnerable to hacker attacks. One infamous example is the attack against the crowdfunding project Decentralized Autonomous Organization (DAO) in 2016, which will be introduced in more detail in Section 5.2.1. DAO relies on a smart contract on Ethereum, through which developers collect crowdfunding for their blockchain-based applications, and investors are rewarded in return. However, there is a vulnerability in the contract code,[11,12] which led to an economic loss worth about 60 million dollars at that time. This so-called re-entrancy attack changes investors' attitudes toward smart contracts, which hinders the development of smart contracts and blockchain.

The DAO event is just one of the most typical attacks against smart contracts. Since smart contracts usually involve financial transactions, any attack may cause a severe economic loss. Consequently, compared with traditional programming, the design of smart contracts has higher security requirements, which makes it more difficult for ordinary users to write secure smart contracts by themselves, further inhibiting the popularity of smart contracts in other industries.

To sum up, security,[12,13] privacy,[13] and efficiency[14] are the main obstacles to smart contracts' universal adoption. There have been various schemes to overcome such barriers and promote the development of smart contracts. However, the quick evolution of the blockchain and smart contracts leaves a gap between the research and implementation, which may confuse new incomers and prevent them from getting involved. In this paper, we conduct a systematic review of the schemes on contract construction and execution, aiming at providing a comprehensive review of the smart contract technology. In the following, we specify our research questions that will be addressed in this paper.

RQ 1. What are the state-of-the-art design paradigms, design tools, and alternative systems to develop or execute smart contracts securely?
RQ 2. What are the current challenges for the efficient development of secure smart contracts, and for these contracts to be adopted in various application scenarios?
RQ 3. What are the potential research directions that may overcome the challenges and limitations mentioned in RQ 2?

### 1.1. Methodology

We follow the systematic review methodology proposed by Kitchenham.[15] With the above research questions in mind, to provide a systematic survey on smart contract construction and execution, we first collect the most relevant literature through searches in Google Scholar, ACM, IEEE, Web of Science, Springer, IACR ePrint, and arXiv. These databases are the most prevalent repositories for papers related to blockchain and smart contracts.

As our research focuses on the construction and execution schemes of smart contracts, we only include the papers that discuss the specific design paradigms, design tools, or execution schemes from a technical point of view. We do not consider the problems related to blockchain but not smart contracts, such as sharding, side-chains, and child-chains. The high-level descriptions of combining smart contracts with other technologies, such as artificial intelligence, cloud computing, or the Internet of Things, are also not included.

According to the criteria described above, we use the search keywords "smart contract," "contract construction," "contract design," and "contract execution" on our selected databases. The search was first conducted in February 2020. We collect the first 20 pages of the search results. We then manually check their relevance to our research questions through the title and abstract: whether they are focused on the design paradigms, design tools, or alternative smart contract execution systems. Such a search is conducted every three weeks to keep track of the latest studies. We also perform a manual selection by searching the proceedings of famous conferences and workshops, including ACM CCS, USENIX Security, NDSS, IEEE S&P, CRYPTO, ASIACRYPT, EUROCRYPT, and FC, as these conferences usually reserve sections especially for studies on blockchain and smart contracts. We further refer to the references of collected papers and the citation API provided by websites such as Google Scholar, IEEE Xplore, and ACM Digital Library. Until August 2020, we have collected 159 papers (or online resources) directly relevant to our research and 12 related surveys, as will be discussed later.

### 1.2. Our contributions

We conduct a comprehensive survey on smart contract construction and execution from the perspectives of paradigms, tools, and systems over the period 2008–2020. Although the first published paper relevant to smart contracts dates back to 2013, there are still several online resources such as wiki pages and forums discussing smart contracts during the period 2008–2013. For example, the famous forum Bitcointalk[16] was established in 2009, and the earliest Bitcoin wiki page[17] was created in 2010. We believe that our work will provide insights to researchers and developers who are new to smart contracts and offer a holistic technical perspective upon contract construction and execution schemes. Our main contributions are as follows.

(1) We provide the essential background knowledge of blockchain and smart contracts, especially the contract execution mechanisms, to provide new incomers an overall impression of the related concepts and help experienced readers have better comprehension. Besides, we give several necessary definitions to form a systematization of knowledge.
(2) We provide a taxonomy of the contract construction and execution schemes, according to the 159 papers and online resources we have collected. We divide existing blockchain systems that support smart contracts into script-based and Turing-complete blockchains, with Bitcoin and Ethereum as representatives, respectively. We then discuss the design paradigms and tools on these two types of platforms and make a further categorization on each topic.
(3) We investigate and categorize the extensions and alternative systems for contract execution, aiming to mitigate the problems and limitations on the existing mainstream contract execution mechanisms.
(4) We discuss the strengths and weaknesses of the state-of-the-art solutions that address the privacy and efficiency issues in both contract construction and execution aspects, and point out promising future research directions during our specification of each topic and at the end of this paper.

### 1.3. Organization

The remainder of this paper is as follows. Section 2 introduces the background, preliminaries, and related work. Section 3 provides the systematization methodology of this paper. Section 4 and Section 5 describe smart contract construction schemes in script-based and Turing-complete blockchains, taking Bitcoin and Ethereum as representatives, respectively. Section 6 discusses various solutions and extensions to improve the privacy and efficiency of contract execution mechanisms. Section 7 outlines our observations on the research questions, and summarizes the challenges and future research directions on the construction and execution of smart contracts. Finally, Section 8 provides a conclusion.

## 2. BACKGROUND, NOTATIONS, AND RELATED WORK

In this section, we first provide the essential background in Section 2.1, specify the notations in Section 2.2, and finally present the related work in Section 2.3. We also provide several essential definitions in the Definitions section.

### 2.1. Background

The background information for this study is given in the following: we first give a brief impression about blockchain technology in Section 2.1.1 and then explain the concept of smart contracts in mainstream blockchains in Section 2.1.2.

#### *2.1.1. Blockchain*

Informally, a blockchain is a sequence of blocks linked with hash values. Transactions that deliver messages among users and interact with the blockchain are stored in the block body, and digest information and other identifiers are recorded in the block header. A blockchain is maintained by the nodes participating in the network, and the data consistency among the nodes is ensured according to some predetermined rules called consensus.

We take the Bitcoin blockchain as an example and illustrate its structure in Figure 1. It is formed by linking multiple blocks in sequence with their hash values. Each block consists of a block header and a block body. Specifically, a block header includes a hash value of the previous block $H_{\text{Prev}}$, a version number $v$ of the current consensus, a current mining difficulty parameter $d$, a timestamp $t$, a Merkle root of transactions $H_{\text{root}}$, and a random nonce $n_r$. A block body includes transactions $Tx_{j\in\mathbb{N}^*}$ that are used to calculate $H_{\text{root}}$. Figure 1 shows that every two adjacent hash values are combined to calculate the hash in the upper layer. If there is a single node left at the end, it will be duplicated and combined with itself, as shown in the path of $H_5 \rightarrow H_{55} \rightarrow H_{5555}$. Note that the contents in the dotted box in Figure 1 are only used to illustrate the calculation of $H_{\text{root}}$ and are not included in the block.

In the Bitcoin blockchain, the verification of new blocks is simplified due to the separation of block headers and bodies. Cryptographic schemes (such as hash function and Merkle tree) are adopted to guarantee the tamper resistance and data consistency. Each node can individually calculate the final state by executing all transactions in order from the genesis block (the initial block of the blockchain). In this way, a central trusted third party is eliminated from the system, and any individual party cannot interrupt the operations in the blockchain. For more technical details about Bitcoin blockchain, readers may refer to Tschorsch and Scheuermann.[18]

We remark that the blockchain structure discussed above is widely adopted by other derived systems, e.g., Ethereum, Corda, and Quorum, and the introduction of these blockchains is omitted here. The blockchain serves as an infrastructure for data communication and value exchanges in their particular application scenarios.

#### *2.1.2. Smart contract*

The concept of smart contracts was first proposed by Szabo [4] in 1997, referred to as a multi-party protocol that could be automatically enforced without a trusted third party. It did not receive enough attention since it was impractical at that time. Several years later, with the birth and development of blockchain, smart contracts were brought back into practice.

Smart contracts are usually defined as event-driven computer programs executed and enforced by all participants in a P2P network in the blockchain context. In each smart contract, there are public interfaces that handle the relevant events. These interfaces are invoked by the transactions with proper payload data, and all valid transactions are recorded on the blockchain. Formal definitions of smart contracts and other related concepts are provided in the Definitions section, which serves as a glossary for readers new to this area.

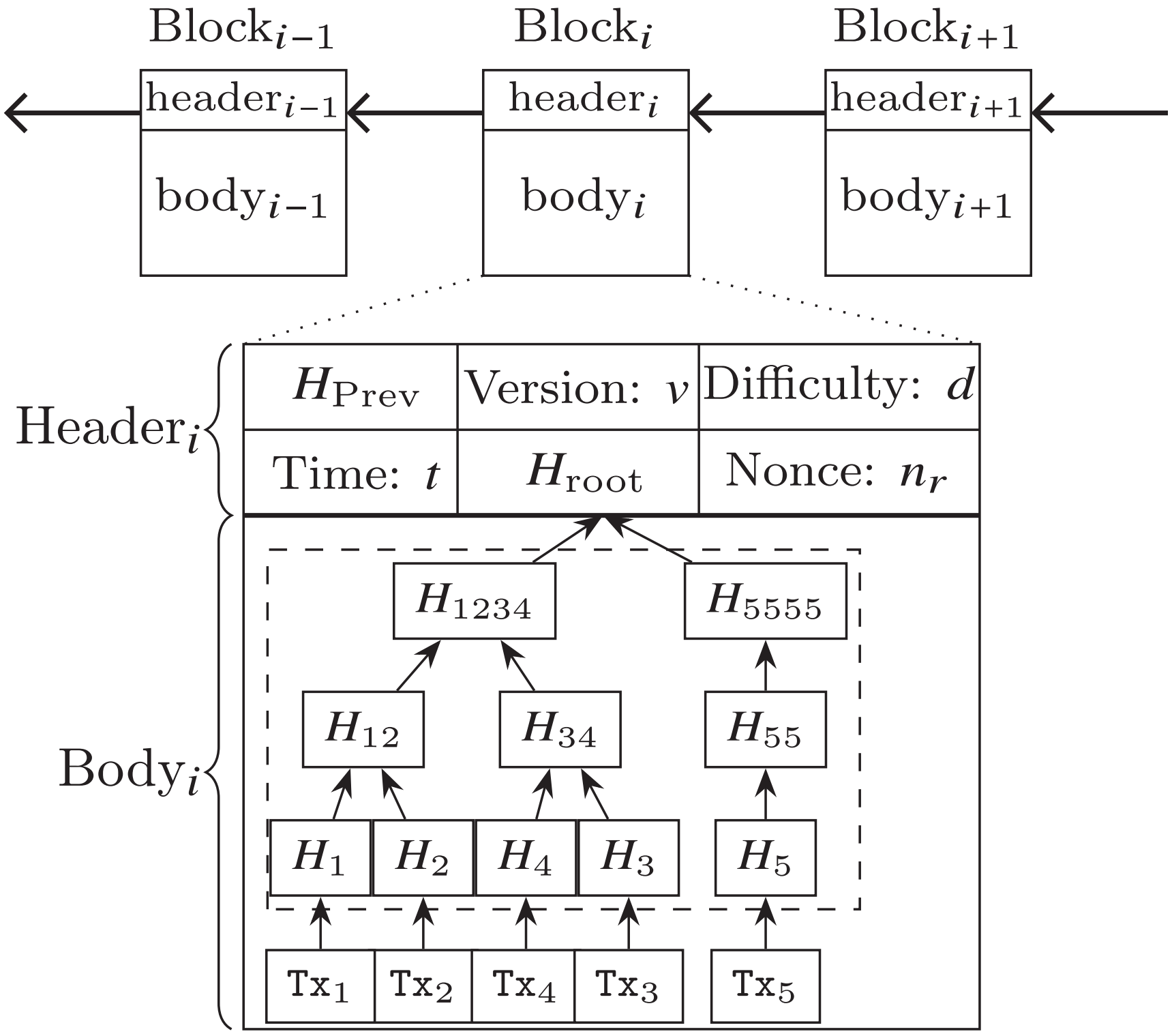

**Figure 1. Data structure of the Bitcoin blockchain**

Bitcoin supports a set of scripts that enable the auto-enforcement of some special financial affairs other than direct electronic cash exchange. This procedure can be considered as the prototype of smart contracts. In the early years, some researchers implemented zero-knowledge contingent payment[19] to achieve a fair exchange of electronic goods. On this basis, more and more smart contracts[20] were developed and implemented.

However, the scripts in Bitcoin are only applicable in limited scenarios. The first reason for this is that Bitcoin intentionally excludes the opcode for loops to avoid potential non-stop operations that lead to the deadlock of the workflow. Although finite loops can be expressed as several repeated operations, the total length of a script is limited up to 520 bytes, with each opcode occupying 4–5 bytes.[21] Moreover, the script language is relatively difficult to learn for young programmers because of its Forth-like, "old-style" appearance. The potential security issues (e.g., miscellaneous attacks) make this situation even worse, requiring developers to prevent possible attacks as much as possible.

Though it is difficult to develop smart contracts in Bitcoin, several decentralized applications (so-called DApps in recent years) are launched utilizing the scripts. The most popular ones are data storage (evidence keeping), voting, gambling, and online poker games (see Section 4.1). Most of these applications consist of several lines of scripts, and they fully utilize the inherent cryptocurrency to realize automatic value transfer without a TTP. These applications' logic is relatively simpler compared with that in Ethereum, as described below.

Ethereum introduces a new virtual machine structure and supports Turing-complete programming languages, which greatly enrich smart contracts' functionalities. Specifically, Ethereum supports the execution of arbitrary deterministic computer programs in theory. The underlying Ethereum Virtual Machine (EVM) recognizes a low-level language called EVM bytecode. To reduce the learning cost and improve development efficiency, several high-level programming languages have been proposed, e.g., Solidity[22] and Serpent,[23] whose grammar is similar to mainstream programming languages. Contracts written in these high-level languages are compiled into EVM bytecode with appropriate compilers. These specially designed structures and languages greatly promote the development of smart contracts. Numerous DApps with complex logic have been proposed in Ethereum and are still thriving nowadays. Typical applications include lottery, loan, auction, and decentralized finance (see Section 5.1). With high-level Turing-complete languages, these applications can be developed and understood by ordinary users more easily.

Figure 2 shows the workflow of a blockchain that supports smart contracts, where the brown arrows represent the processes of deploying a smart contract through the creation transaction $Tx_{create}$. In the development stage, users may refer to design paradigms and use auxiliary tools to get a prototype. Analysis tools are then implemented to confirm the contract's security and correctness (Definition 13, 14). The blue and red arrows refer to the call from smart contracts and users, respectively. After receiving $Tx_{1,call}$ and $Tx_{2,call}$, the miners (i.e., executors of these transactions) verify and package the transactions into the latest block, i.e., Block$_{i+2}$, following the execution mechanism. After the block is appended to the blockchain, the World State, which contains all the states, is updated accordingly.

From another point of view, the execution mechanism among different platforms varies. In the following, we give a brief introduction of the execution mechanisms in Section 2.1.2.1 and Section 2.1.2.2, respectively, taking Bitcoin and Ethereum as representatives.

*2.1.2.1. Contract execution in Bitcoin*. Smart contracts in Bitcoin refer to the transactions setting script hashes as output addresses (Pay to Script Hash, P2SH), which encode the hashes of scripts into Bitcoin UTXOs (Definition 10). P2SH transactions are the basis for multi-signature (MultiSig)[24] transactions, Lightning Network,[25] and other techniques in the Bitcoin ecosystem. These techniques play a significant role in data communication by allowing messages and instructions delivered and executed consistently through a network without mutual trust.

Figure 3 shows a simplified payment process in Bitcoin, where the time field is omitted. When dealing with a transaction that spends UTXOs from a P2SH transaction $Tx_0$, the miners first verify the sender's signature. They then check whether the script code in the transaction payload matches the corresponding script hash $H_{script}$. Finally, they check whether the other payload data $p_{data}$ makes the script evaluated to be true. If so, the signed redeem transaction $Tx_1$ is valid.

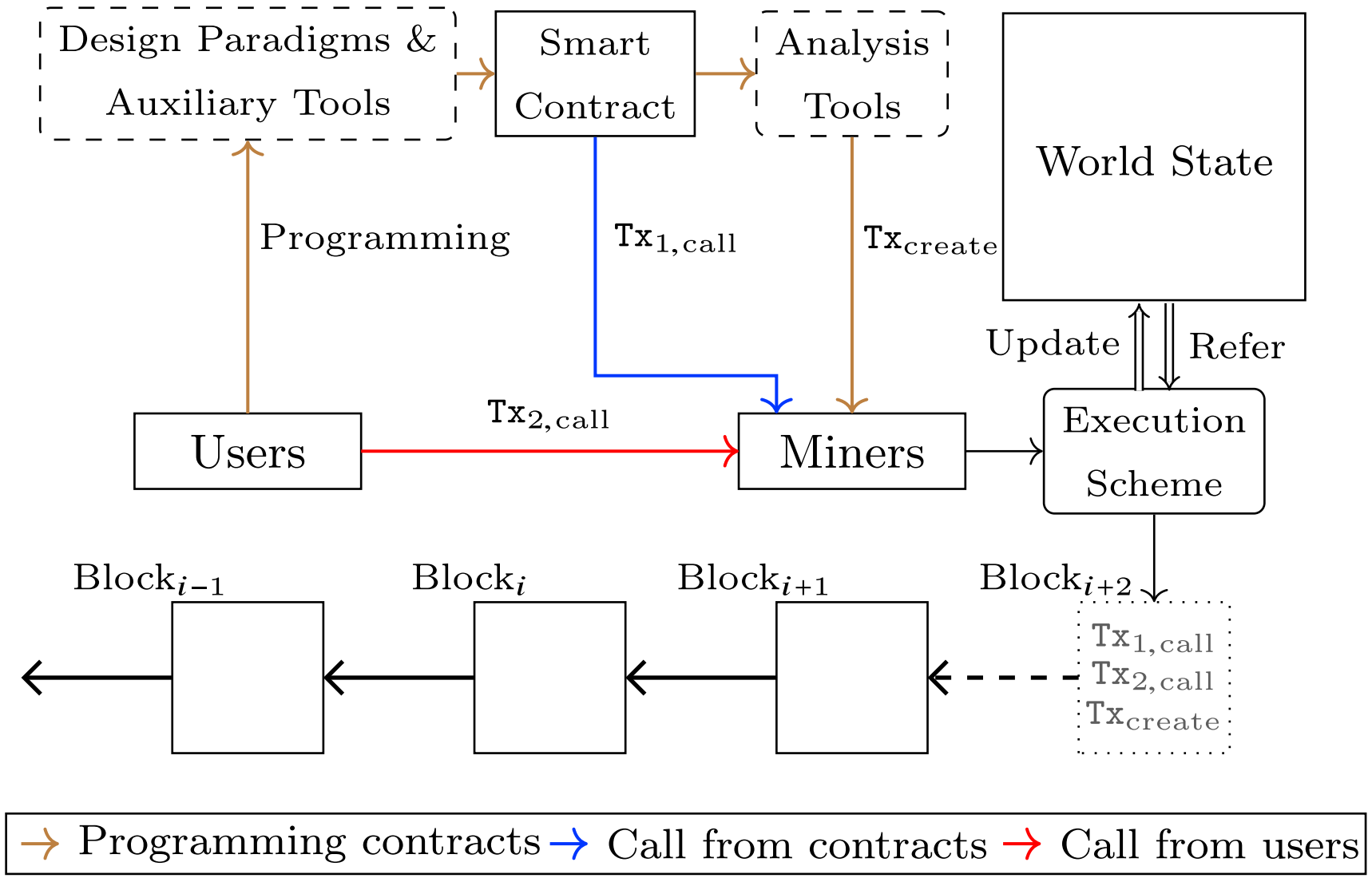

**Figure 2. Workflow of the blockchain supporting smart contracts**

In Bitcoin, it takes approximately 10 min to append a new block to the blockchain. Moreover, it is recommended to wait for at least a sequence of six blocks to make sure that the transactions indeed take effect and cannot be erased or forked, with an overwhelming probability. This introduces a huge delay in the transaction confirmation, which further limits the implementation of Bitcoin smart contracts.

Moreover, since the information on the Bitcoin blockchain is publicly available, the full scripts are exposed to the entire network. Even though Bitcoin is equipped with a pseudonym mechanism, such privacy leakage is still inevitable. Curious readers may refer to the work of Conti et al.[13] for a more detailed survey on privacy issues in Bitcoin.

*2.1.2.2. Contract execution in Ethereum*. The Turing-complete programming languages in Ethereum significantly extend the application scenario of smart contracts. Theoretically, smart contracts in Ethereum can realize any deterministic program. These contracts are executed by the EVM, whose formal definition and execution mechanism are elaborated in Ethereum Yellow Paper.[3]

Ethereum adopts the account model (Definition 11), whereby an account of a smart contract has the same status as that of a user. In other words, a contract account has the same ability to send transactions and trigger or create contracts as that of a personal account.

Transactions are handled by miners who run the EVM. After a transaction is included in the blockchain, the balance and other variables are updated according to the contract rules.

To prevent potential Denial of Service (DoS) attacks (e.g., non-stop execution caused by an infinite loop), Ethereum introduces the gas mechanism. Namely, each operation consumes a certain amount of gas, and the upper bound of gas consumption is set and paid in advance in the transaction. Suppose the execution of a contract function does not terminate before the gas is exhausted. In that case, the contract will be reverted to the initial states before the triggering of this function, and the miners will charge all the consumed gas as execution fees. However, this limits the complexity of smart contracts.

In addition, similar to Bitcoin, smart contracts in Ethereum also suffer from privacy leakage. Several schemes have been proposed to handle such privacy issues and will be discussed later in this review.

## 2.2. Notations

As mentioned in Section 1, the security problems on smart contracts should be carefully settled. Relevant solutions for contract construction and execution usually come with a formal security proof, which might involve mathematical models or cryptographic primitives. To present these schemes in a uniform style, we make an effort to unify the notations in our work, as shown in the following.

Sets are denoted with upper-case calligraphic letters, e.g., $\mathcal{T}$ represents the set of valid transactions for a contract.

For most functions, $F$ is used, along with a subscript denoting the particular usage of this function, e.g., $F_{neg}$, the negligible function with certain security parameter. Some other functions or primitives may be denoted with Greek letters, e.g., $\varphi$ for the primitive that evaluates the witness $\omega$.

Tuples are denoted with upper-case letters, e.g., $U$ denotes an unspent transaction output (UTXO, Definition 10). A dot operation is used to refer to the component inside a tuple, e.g., $U.v$ denotes the value of a UTXO. We use $Tx$ to represent a transaction, and $Tx.in$, $Tx.out$, $Tx.id$, and $Tx.pld$ denote the transaction's input, output, identifier, and payload data, respectively.

Arbitrary-length sequences are denoted as sans serif lower-case abbreviations, e.g., buf, the stack of buffers. Square brackets are used to index individual components, e.g., buf[0], the first item on the buffer stack.

Scalars and variables are denoted with a lower-case letter, e.g., $n$ is often used to represent the number of participants, and $i$, $j$, $k$ are often used as indexes to refer to the members in a set. Moreover, those with special meaning might be Greek, e.g., $\sigma$ denotes a digital signature.

For the names of proposed schemes, we adopt the original text styles in the literature, e.g., OYENTE and Hawk.

Besides, some special representations are used for particular meanings. Hash values are denoted with $H$, whose subscripts may be strings with special meanings, e.g., $H_{root}$, the root hash of a Merkle tree. Time is denoted with $t$. We use ₿$x$ to represent $x$ bitcoins. Smart contracts are usually denoted with $C$, while for those with particular meanings, the typewriter format is used, e.g., *SimpleDAO*, the example contract used to describe a DAO attack. Greek letters such as $\beta$ and $\gamma$ are used to denote payment channels and state channels. Protocols are denoted with $\pi$. The letter $\mathcal{A}$ is used to denote an adversary. Participants in a protocol or contract are denoted as $P$, which often comes with subscripts such as numbers (e.g., $P_1$, the first participant) and letters (e.g., $P_s$, the sender). Address, usually a string in the context of blockchain, is denoted with $a$, and with subscripts indicating the usage of this address, e.g., $a_{\text{mul}}$ is the multi-sig address in the Bitcoin context. Ideal functionalities are denoted with $\mathcal{F}^*$, and the message headers in these functionalities are represented with the small capital letters, e.g., DEPOSIT.

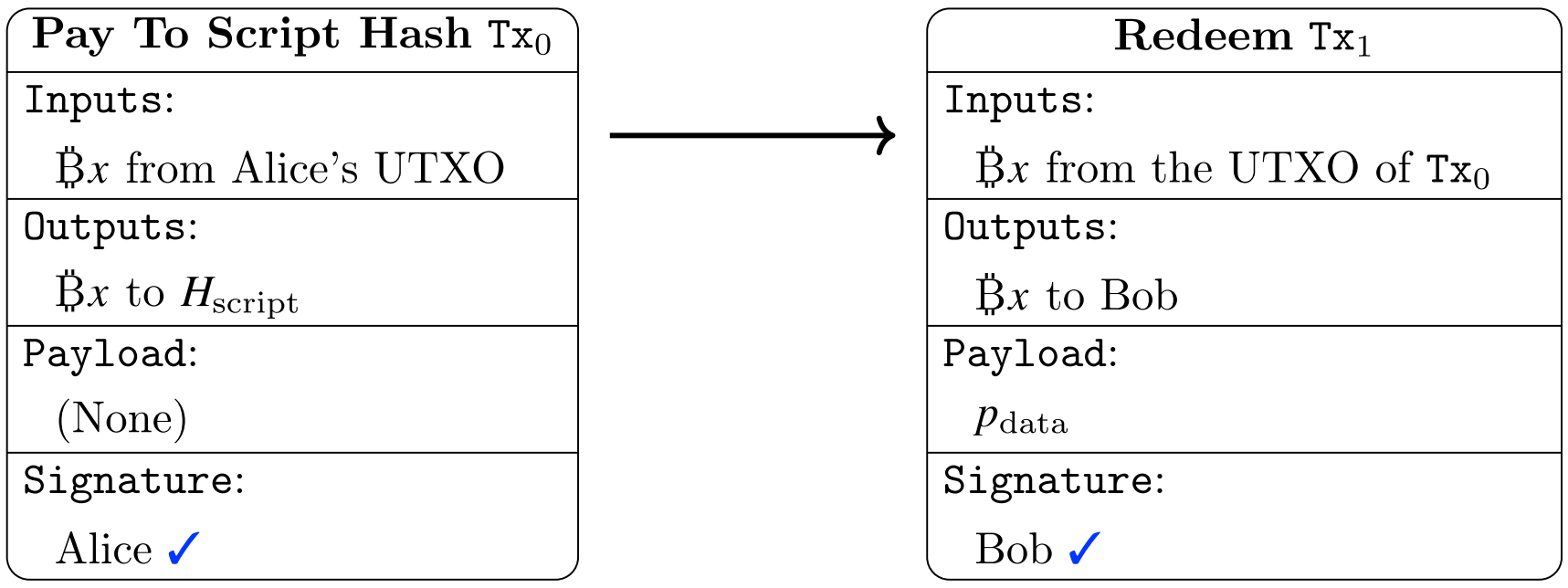

**Figure 3. P2SH transactions between Alice and Bob with the time field omitted. A redeem transaction requires specific payload data from Bob**

As for operations, we use $s \leftarrow 0$ to denote the operation of assigning value 0 to $s$, and $s \leftarrow_{\$} \{0,1\}^{128}$ to denote that $s$ is uniformly picked at random from the set $\{0,1\}^{128}$. The operation $\rightarrow$ is used to denote the concatenation of several nodes that forms a path, as already shown in Figure 1, the path $H_5 \rightarrow H_{55} \rightarrow H_{5555}$. $\langle P, V \rangle$ is used to denote the interaction of two Turing machines $P$ and $V$. The concatenation of strings is denoted with $||$ and the XOR of same-length binary elements with $\otimes$.

Most notations are utilized during the specification of relative schemes and functionalities in Section 4.1, Section 6.2, and the Definitions section, especially when depicting the procedure of the schemes.

## 2.3. Related work

Smart contracts are an essential aspect of the blockchain. Its execution characteristics, efficiency, and security are directly related to the acceptance of this emerging technology. Prior to our work, there have been several surveys on the features of contract platforms, properties of the contracts, and related analysis tools, as shown in Figure 4.

### 2.3.1. Features of platforms

Seijas et al.[26] discuss the languages adopted by the blockchain systems such as Bitcoin, Nxt,[27] and Ethereum, and list the defects of these languages. Furthermore, they point out some promising techniques that may help expand contracts' functionality and enforce their security, such as zero-knowledge proofs (ZKP; Definition 16) and static analysis.

Bartoletti and Pompianu[28] compare six smart contract platforms: Bitcoin, Ethereum, Counterparty,[29] Stellar,[30] Monax,[31] and Lisk.[32] Junis et al.[33] briefly present a basic concept of blockchain and smart contracts. The existing studies only briefly introduce the features of smart contract platforms, and a thorough and comprehensive survey is absent.

### 2.3.2. Properties of contracts

We consider the security, privacy, and performance aspects of smart contracts. There are seven surveys discussing the similar concepts, but not as systematic as ours.

Alharby and van Moorsel[34] investigate 24 papers related to smart contracts. They point out that most research focuses on the problems for DApps and the corresponding solutions. These issues are divided into contract construction, secure execution, privacy, and performance. From the practical perspective, Atzei et al.[12] summarize the vulnerabilities of smart contracts in Ethereum. They categorize four weaknesses in Ethereum to facilitate future development or research on smart contracts (for more details see Section 5.2). Dika[35] analyzes smart contracts in Ethereum from a higher point of view, and categorizes the weaknesses into three levels: blockchain, virtual machine (EVM), and programming language (Solidity). Macrinici et al.[36] collect 64 papers on the issues related to the smart contract applications. They summarize 16 subproblems into three categories: problems on blockchain mechanism, contract programs, and virtual machine. However, they only list the problems but fail to elaborate on the solutions.

Regarding the application scenarios of smart contracts, Bartoletti and Pompianu[28] analyze the smart contracts on Bitcoin and Ethereum up to 2017, and divide the application scenarios into financial, notary, game, wallet, library, and others. They focus on the quantitative statistics in the application layer, aiming to give an impression on the usage of smart contracts. Ayman et al.[37] analyze the programming problems according to the number of questions in the Stackoverflow forum. They conclude the trend in the development of smart contracts based on the statistical results.

### 2.3.3. Analysis tools

Harz and Knottenbelt[38] analyze the languages and security tools designed for smart contracts in 2018, giving a classification and brief introduction. Angelo and Salzer[39] investigate 27 tools for Ethereum contracts from the aspects of open source, maturity, adopted methods, security issues, and others. At the same time, 53 papers related to smart contract security are summarized by Liu and Liu,[40] and are classified from security and correctness aspects. Compared with these two studies, our work covers more up-to-date analysis tools. We describe the smart contract construction schemes, tools, and execution mechanisms, forming a more systematic knowledge of contract construction and execution.

The latest studies of Ante[41] and Almakhour et al.[42] survey the contract analysis tools almost simultaneously. Ante[41] analyzes the smart contract-related literature and provides several statistical and quantitative results, such as the citation statistics, distribution of keywords, and the most concerned smart contract platforms. Almakhour et al.[42] classify the analysis tools into correctness verification tools and vulnerability analysis tools for smart contracts, and provide a detailed description of each tool.

Differently, we start from a higher-level perspective and discuss the topics related to the contract construction and execution schemes with more details. We make a distinct taxonomy of these tools, aiming at providing a road map for the researchers and developers interested in smart contracts.

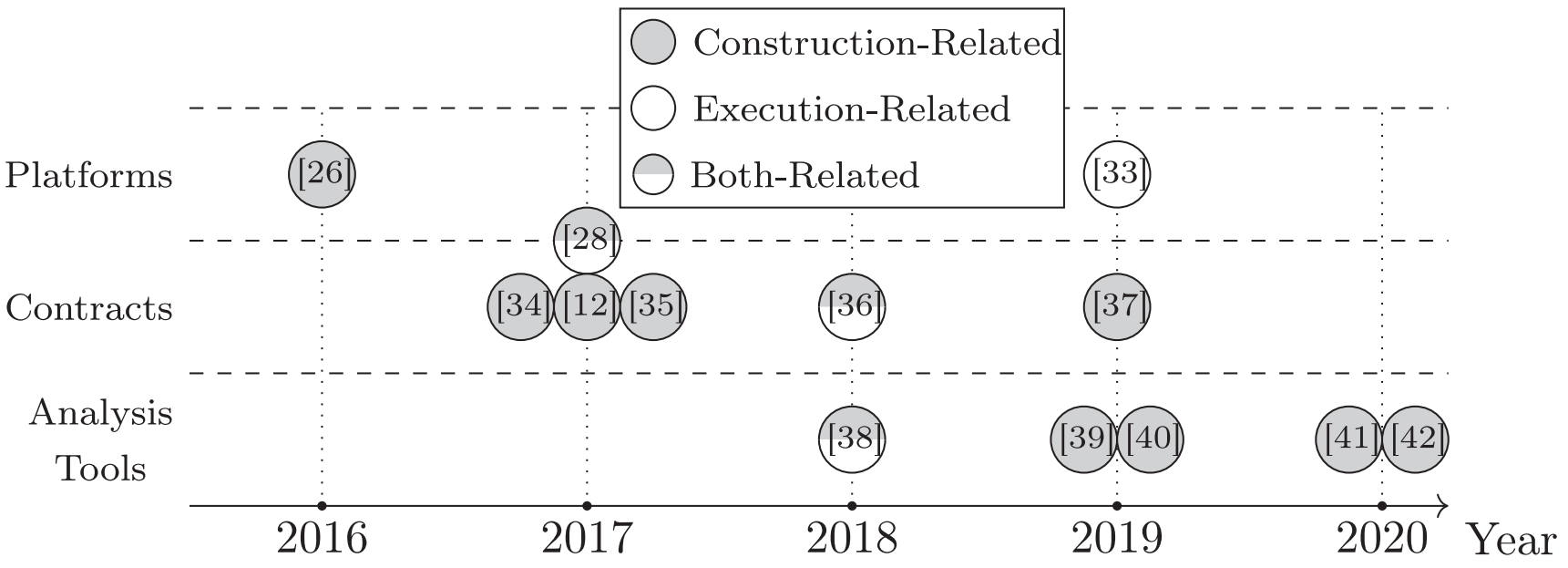

**Figure 4. Related work on construction and execution mechanisms of smart contracts**

## 3. SYSTEMATIZATION METHODOLOGY

In this review, we divide the schemes related to smart contracts into the construction-related and execution-related. As shown in Figure 2, the construction and execution schemes work separately. The construction-related schemes focus on the design paradigms and auxiliary frameworks within the current architecture to facilitate the secure and flexible development of DApps. The execution-related schemes involve the implementation strategies of smart contracts or modifications of the underlying execution mechanism to improve smart contracts' performance and make the contracts applicable to more demanding application scenarios.

We discuss the above two categories in the following. Note that the notion of constructing and designing smart contracts are used synonymously throughout this paper. We discuss the smart contracts in the blockchain context and sometimes use the word contract for short.

### 3.1. Construction of smart contracts

As mentioned in Section 1 and Section 2, smart contracts are crucial and promising building blocks for reliable data communication, as they facilitate the information and value exchanges in a verifiable way. However, due to the demanding requirements on smart contracts, efficiently designing secure and privacy-preserving smart contracts to reach the desired performance remains challenging.

Forms of smart contracts are various, depending on the platforms on which they are running. Therefore, design schemes of smart contracts also rely on the platforms, especially the language they support. Smart contracts written in scripts are mostly used to describe financial transactions, while smart contracts written in Turing-complete languages could theoretically describe any deterministic protocol as a computer program.

We categorize the existing blockchain platforms into two types:[26,34] script-based blockchains represented by Bitcoin and Turing-complete blockchains represented by Ethereum. The former supports only limited expressions of operations (especially no loops), and the latter supports arbitrary functions with Turing-complete programming languages.

Since the smart contracts on these two kinds of platforms have significant differences in the form and execution mechanism, the construction schemes are also quite different. Therefore, we will discuss the smart contract construction schemes separately.

Additionally, the schemes in each category are further divided into two parts, design paradigms[28,37] and design tools,[35,36,38–42] where the design patterns describe some common and useful paradigms, including paradigms for specific applications and best practices for general purpose, and the design tools refer to the tools available in the process of developing smart contracts. Some tools help developers confirm contracts' security to avoid economic losses caused by potential vulnerabilities or bugs. They are called analysis tools and usually take effect after the contract is almost completed. Other tools work during the construction process of contracts, improving the efficiency of development or reducing security issues, and are called auxiliary tools in this review.

### 3.2. Execution of smart contracts

Besides the expressivity of programming languages, the aforementioned two kinds of smart contract platforms also vary in the execution mechanisms. However, from another point of view the mainstream platforms in both categories suffer from several common problems in transaction delay, contract complexity, privacy leakage, and others.

Numerous schemes for contract executions[28,33,36,38] are proposed to solve these problems, aimed at making smart contracts applicable to miscellaneous implementations. Here we divide them into three classes: (1) private contracts with extra tools, (2) off-chain channels, and (3) extensions on core functionalities.

In the first two categories, schemes usually follow the original rules, and schemes in the last category often introduce new functionalities or properties by modifying the underlying execution mechanisms. Solutions in the first category often introduce useful cryptographic protocols or hardware to protect user privacy during the executions of smart contracts. We remark that schemes in the second category are quite prevalent since they do not require extra tools or modifications of the underlying mechanism. Their core idea is to migrate the execution off-chain and only use blockchain for final state settlement or dispute handling, and we call them off-chain channels. Schemes in the last category are designed to add the functionality that the original platforms do not support. Their implementation requires a fork of the existing system, or they even launch a new one instead.

We classify the schemes related to smart contracts in the literature in Table 1. In the "Theory" column, the symbols ●, ◐, and ○ represent that the work has both complete description and formal security proof, has description but no proof, and has neither description nor proof, respectively. Similarly, in the "Realization" column, the symbols of ●, ◐, and ○ denote that the work has open-sourced implementation, has implementation but is not open-sourced, and has no implementation. The word open-sourced here means the source code is released online and available for all users, which is of great significance for

**Table 1. Summary of the construction and executing schemes of smart contracts**

| Class | | Refs.[a] | Year | Keywords | Theory[b] | Realization[c] |
|---|---|---|---|---|---|---|
| Designing contracts with scripts | Design patterns | 20 | 2012 | P2SH transactions & common contracts | ◐ | ● |
| | | 43–45 | 2014–2018 | `OP_RETURN` opcode | ◐ ○ ◐ | ● ● ● |
| | | 46–48 | 2014–2017 | contracts for lottery | ● ● ● | ● ○ ● |
| | | 49 | 2015 | contracts for online poker | ● | ○ |
| | | 50 | 2014 | general fair multi-party protocols | ● | ○ |
| | | 51–53 | 2016 | secure multi-party computation on public blockchains | ● ● ● | ● ○ ○ |
| | | 54–56 | 2015–2018 | probabilistic payment system | ● ● ● | ◐ ○ ○ |
| | | 57–59 | 2016–2019 | scriptless contract | ● ◐ ● | ● ○ ◐ |
| | Design tools | 60–64 | 2014–2018 | security models | ◐ ● ◐ ◐ ◐ | ● ○ ● ◐ ● |
| | | 62–69 | 2017–2019 | languages | ◐ ◐ ◐ ◐ ● ● ◐ ◐ | ● ◐ ● ● ◐ ● ● ● |
| Designing smart contracts with Turing-complete languages | Design patterns | 48,70 | 2017 | contracts for lottery | ● ● | ● ● |
| | | 71,72 | 2018–2019 | lending contracts | ● ◐ | ● ◐ |
| | | 73–77 | 2016–2020 | contracts for e-government | ◐ ◐ ◐ ◐ ◐ | ○ ○ ○ ○ ● |
| | | 78,79 | 2018 | private auction protocol | ● ◐ | ● ● |
| | | 80–84 | 2017–2019 | off-chain computation and storage | ◐ ◐ ◐ ◐ ◐ | ○ ○ ● ◐ ◐ |
| | | 22,85,86 | 2016–2020 | best practices on writing smart contracts | ◐ ◐ ◐ | ● ● ● |
| | | 28,87,88 | 2016–2018 | classification & common patterns | ◐ ◐ ◐ | ● ● ● |
| | | 12,89–96 | 2016–2020 | common vulnerabilities | ◐ ◐ ◐ ◐ ◐ ◐ ◐ ◐ ◐ | ● ● ○ ○ ○ ◐ ● ● ○ |
| | | 97,98 | 2016 | design models | ◐ ◐ | ○ ○ |
| | | 99 | 2016 | interfaces for updating smart contracts | ◐ | ● |
| | Design tools | 100–102 | 2017–2019 | detecting re-entrancy vulnerabilities | ◐ ◐ ◐ | ● ◐ ◐ |
| | | 103–110 | 2017–2020 | detecting gas-related vulnerabilities | ◐ ◐ ● ◐ ◐ ◐ ◐ ◐ | ◐ ◐ ○ ● ◐ ◐ ● ◐ |
| | | 111 | 2018 | detecting trace vulnerabilities | ◐ | ● |
| | | 112 | 2019 | detecting event-ordering bug | ◐ | ◐ |
| | | 113,114 | 2018–2020 | detecting integer bug | ◐ ◐ | ● ● |
| | | 89,115–123 | 2016–2020 | general detection by symbolic execution | ◐ ◐ ◐ ◐ ◐ ◐ ◐ ◐ ◐ ◐ | ● ● ● ● ● ● ◐ ◐ ● ◐ |
| | | 124,125 | 2018–2019 | general detection by syntax analysis | ◐ ◐ | ● ● |
| | | 126–130 | 2018 | general detection by abstract interpretation | ● ● ◐ ◐ ● | ● ● ● ◐ ◐ |
| | | 131 | 2019 | general detection by data-flow analysis | ◐ | ● |
| | | 132 | 2018 | general detection by topological analysis | ◐ | ◐ |
| | | 133,134 | 2018 | general detection by model checking | ◐ ◐ | ○ ◐ |
| | | 135 | 2019 | general detection by deductive proof | ◐ | ● |

(*Continued on next page*)

**Table 1. *Continued***

| Class | | Refs.[a] | Year | Keywords | Theory[b] | Realization[c] |
|---|---|---|---|---|---|---|
| | | 136 | 2018 | general detection by satisfiability modulo theories | ◐ | ○ |
| | | 137–139 | 2018–2019 | general detection by fuzzing test | ◐◐◐ | ●●● |
| | | 140–149 | 2016–2019 | frameworks | ●◐◐●◐●◐◐◐◐ | ○●○◐●●○○○◐ |
| | | 143,150–156 | 2016–2019 | Languages | ●◐◐◐◐◐●● | ◐●●○●○●● |
| | | 90,157–159 | 2017–2019 | basic tools | ◐◐●◐ | ○●●● |
| Execution schemes for smart contracts | Private contracts with extra tools | 160–162 | 2015–2018 | private contracts with multi-party computation | ◐●● | ●◐○ |
| | | 6,140,141 | 2016–2018 | private contracts with zero-knowledge proof | ◐●◐ | ●○● |
| | | 161,163–168 | 2017–2019 | private contracts with trusted execution environment | ●●●●●●● | ◐◐●◐◐●● |
| | Off-chain channels | 25,59,169–173 | 2015–2019 | payment channel network on Bitcoin | ◐●◐◐●●● | ●◐○○○◐◐ |
| | | 174–177 | 2015–2017 | payment channel network on Ethereum | ◐◐◐● | ○●●● |
| | | 178–185 | 2017–2019 | state channel network | ●●●●◐◐◐● | ●○◐○◐●●● |
| | Extensions on core functionalities | 186,187 | 2016–2017 | Bitcoin covenants | ◐◐ | ○◐ |
| | | 188,189 | 2019–2020 | moving contracts across blockchains | ◐◐ | ◐◐ |
| | | 190 | 2018 | proof-carrying smart contracts | ◐ | ○ |
| | | 191 | 2018 | private contracts with one-step proof | ● | ● |
| | | 192 | 2018 | complex contract execution without validation | ● | ◐ |
| | | 193 | 2018 | private execution of arbitrary contracts | ● | ● |
| | | 194 | 2020 | execution of interactive complex smart contracts | ● | ◐ |

[a]If there are multiple references on the same line, there will be multiple marks of ●, ◐, or ○ in the "Theory" and "Realization" columns with the same order.
[b]In the "Theory" column, the symbols ●, ◐, and ○ represent that the work has both complete description and formal security proof, has description but no proof, and has neither description nor proof.
[c]In the "Realization" column, the symbols ●, ◐, and ○ denote that the work has open-sourced implementation, has implementation but is not open-sourced, and has no implementation. The word open-sourced here means the source code is released online and available for all users.

succeeding researchers and developers to learn from these schemes. Besides, if there are multiple references at the same line, there will be multiple ●, ◐, or ○ in the "Theory" and "Realization" columns in the same order.

## 4. CONSTRUCTING SMART CONTRACTS WITH SCRIPTS

Script-based blockchains usually provide simple stack-based opcodes to facilitate a more flexible circulation of cryptocurrencies. For example, a payer could specify a condition under which the payee receives his payment. The primary purpose of such script languages is to facilitate simple financial affairs or demands. Therefore, smart contracts in script-based blockchains are relatively simple and limited compared with those in Turing-complete blockchains.

In this section, we discuss the construction schemes of smart contracts in script-based blockchains. We take Bitcoin as a representative. The reasons are as follows.

(1) Bitcoin is the first and most well-known script-based smart contract platform.
(2) Most state-of-the-art script-based blockchains are derived from Bitcoin, and thereby most construction schemes in Bitcoin could be easily applied to such blockchain systems with slight modifications.
(3) Most relevant studies also focus on the construction of smart contracts in Bitcoin.

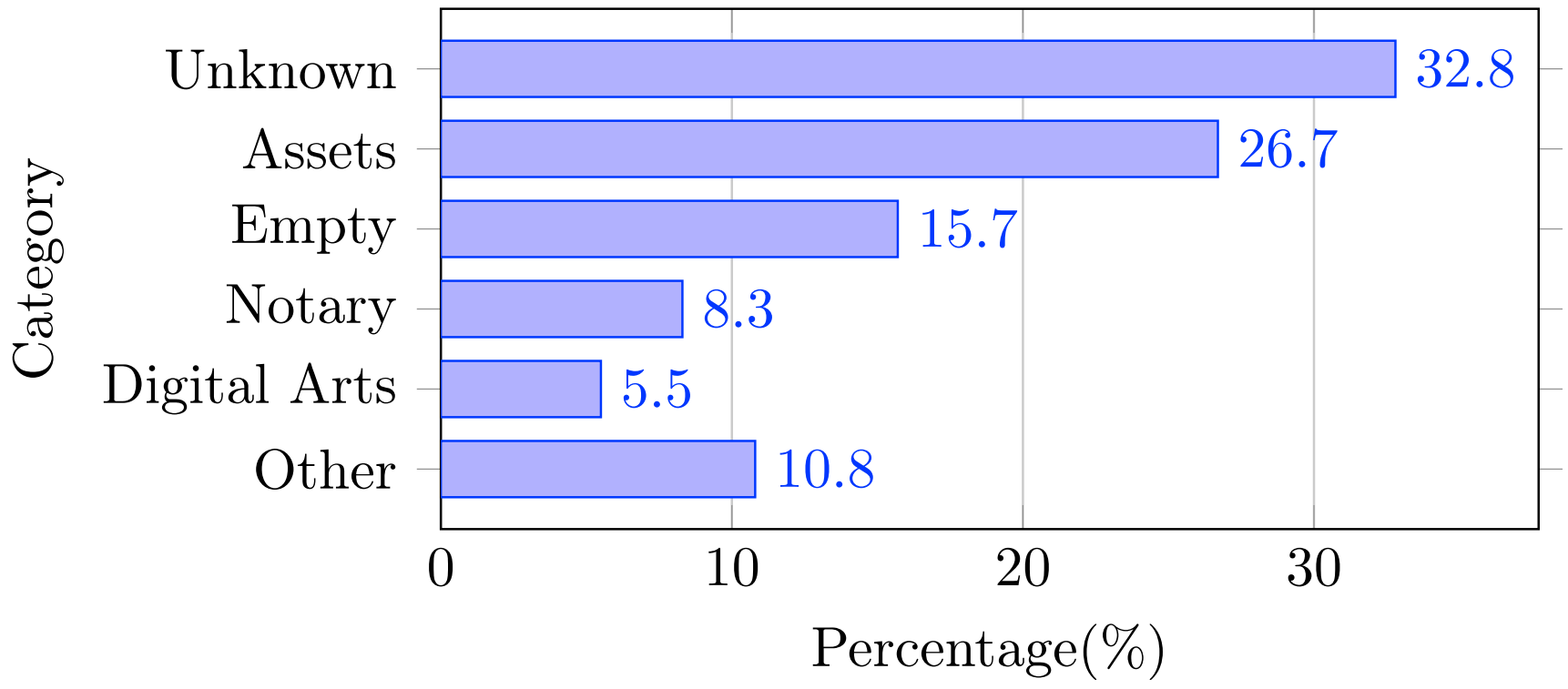

**Figure 5. Usage of the OP_RETURN opcode (until February 2017)**
Figure reprinted with permission from Bartoletti and Pompianu[44].

We divide the schemes related to the contract construction into two categories: design paradigms and tools. Design paradigms here refer to the modular patterns in functionalities, applicable to different scenarios and widely considered secure. Such schemes may help to develop secure smart contracts efficiently (see Section 4.1). Design tools here refer to the solutions aimed at guaranteeing the security of smart contracts (see Section 4.2).

## 4.1. Design paradigms

As mentioned in Section 2.1.2.1, most smart contracts in Bitcoin[19] use P2SH transactions. Before the introduction of the P2SH transaction in BIP16,[195] there were already a few transactions that realized the contract for data storage by modifying (or abusing) the standard ones. With the P2SH solution, transactions are no longer restricted to direct payments. It becomes possible to implement some simple protocols in Bitcoin.

To better support the implementations of smart contracts, design paradigms are researched and proposed in many studies. We divide the design paradigms in Bitcoin into four parts, according to their use and applications.

Firstly, we have collected three data storage schemes proposed in the early years (see Section 4.1.1), taking advantage of the tamper-resistant nature of the underlying blockchain. Such storage schemes may serve as foundations for other applications such as digital forensics and data disclosure.

Secondly, there are eight papers further regarding the blockchain as a public bulletin board based on the storage contracts and implementing the secure multi-party computation protocol (SMPC, Definition 15). SMPC has become a heated research direction in recent years since it enables secure and privacy-preserving communications among participants. Related schemes focus on various features of SMPC, such as fairness and generalization (see Section 4.1.2).

Thirdly, layer-2 protocols are proposed to alleviate the problems of huge confirmation delay and low throughput of Bitcoin. Such protocols optimize both the money and time overhead by moving the calculation off-chain (see Section 4.1.3). Note that we especially move the discussion of layer-2 off-chain channels to Section 6.2, since these schemes are typical and widely studied in the literature.

Finally, to avoid potential privacy leakage due to Bitcoin's public nature, the concept of scriptless contracts is proposed, with which a transaction reveals no information about the contract contents. We have collected three papers on this topic (see Section 4.1.4).

### *4.1.1. Data storage*

The tamper-resistant property of Bitcoin blockchain attracts users to store short messages as witnesses or memorandums on-chain.

The original version of Bitcoin does not support storing data other than transactions. Some users abuse the output field (the hash of the target address in *Tx.out*) in a standard transaction by filling it with meaningful strings, such as sentences and ASCII art.[196] However, the UTXOs in such transactions will never be spent since it is almost impossible to solve the secret keys according to such arbitrary strings. Therefore, such UTXOs will stay in miners' memory pool forever, which results in the loss of hardware capacity.

To avoid such abuse, `OP_RETURN` opcode[43] is introduced, enabling appending additional messages in a transaction. Bartoletti and Pompianu[44] analyzed the usage of this opcode in 2017, and we rephrase the statistical results in Figure 5, where most of the known behaviors are related to digital forensics and data closure. They conclude that at least 22 protocols adopt `OP_RETURN` to provide services such as asset declaration, integrity proof of files, digital copyright, and information recording. Faisal et al.[45] conducted similar work in 2018 and argued that `OP_RETURN` might be utilized by ransomware, serving as a certificate of payment. Their statistical results and classifications are similar to those of Bartoletti and Pompianu,[44] so we omit the details of their work here.

We remark that data storage is the most fundamental application in Bitcoin and other relevant platforms, and it is a prominent feature of blockchain. Such utilization is still prevalent and necessary in today's more complex applications, e.g., launching a complaint with certain pre-stored proofs or commitments when a dispute occurs.

### *4.1.2. Secure multi-party computation*

With the data storage contracts discussed above, Bitcoin can be used as a public bulletin board without a trusted third party, enabling the implementation of SMPC (Definition 15).

Andrychowicz et al.[46] first implement a timed commitment scheme in Bitcoin, whereby a commitment should be opened within a specified time period, otherwise the publisher will be penalized. They further propose a Bitcoin-based SMPC protocol. However, the amount of deposit grows rapidly when the number of participants increases in the multi-party case, and fairness is not guaranteed in practice. That is, the counterparties can abort or claim their deposits by trying to race other transactions on-chain. Thereafter, two respective lottery schemes based on two-party SMPC are proposed by Andrychowicz et al.[46,47]. The former only supports the application of two-party lotteries, while the latter supports arbitrary two-party functions, and the ability to prevent the adversary from generating valid transactions from the published ones (i.e., non-malleability) is strengthened in the latter work.

**Deposit**$(sid, 1^{\lambda})$

Upon invocation by $P_s$:
$F_{\text{receive},s}(\text{DEPOSIT}, sid, ssid, P_s, P_r, \phi, \tau, v)$
if $(\text{DEPOSIT}, sid, ssid, P_s, P_r, \cdot)$ does not exist:
  $F_{\text{record}}(\text{DEPOSIT}, sid, ssid, P_s, P_r, \phi, \tau, v)$ and broadcast

**Claim**$(sid, 1^{\lambda}, \tau)$

Upon invocation by $P_r$ in round $\tau$:
$F_{\text{receive},r}(\text{CLAIM}, sid, ssid, P_s, P_r, \phi, \tau, v, \omega)$
if $(\text{DEPOSIT}, sid, ssid, P_s, P_r, \phi, \tau, x)$ exists and $\phi(\omega) = 1$:
  $F_{\text{broadcast}}(\text{CLAIM}, sid, ssid, P_s, P_r, \phi, \tau, v, \omega)$
  $F_{\text{send},r}(\text{CLAIM}, sid, ssid, P_s, P_r, \phi, \tau, v)$
  $F_{\text{delete}}(\text{DEPOSIT}, sid, ssid, P_s, P_r, \phi, \tau, v)$

**Refund**$(sid, 1^{\lambda}, \tau)$

Upon invocation by $P_s$ or $P_r$ in round $\tau + 1$:
if $(\text{DEPOSIT}, sid, ssid, P_s, P_r, \phi, \tau, v)$ exists:
  $F_{\text{send},s}(\text{REFUND}, sid, ssid, P_s, P_r, \phi, \tau, v)$
  $F_{\text{delete}}(\text{DEPOSIT}, sid, ssid, P_s, P_r, \phi, \tau, v)$

**Figure 6. Ideal claim-or-refund functionality $\mathcal{F}^*_{CR}$[49–53]**
Figure reprinted with permission from Kumaresan et al.[49]

To alleviate the deposit-explosion problem in the multi-party case, Bartoletti and Zunino[48] propose a multi-party lottery contact with fixed deposits, which requires a modification of the Bitcoin mechanism. They then implement the scheme on Ethereum (see Section 5.1.1.1). Kumaresan et al.[49] consider a decentralized online poker protocol. They propose a primitive called secure cash distribution with penalties to guarantee the fair finalization of the poker game. A timed commitment scheme is also used to incentivize honest behavior. Their scheme requires additional opcodes in Bitcoin and thus cannot directly apply to Bitcoin, but it inspires subsequent works on general SMPC protocols.

In addition to the SMPC for specific applications mentioned above, we have collected five studies dedicated to general SMPC protocols. Bentov and Kumaresan[50] introduced an ideal functionality of the fair multi-party protocol in Bitcoin in 2014, which is more general and could also be applied to other script-based smart contracts. Their results are implemented to the decentralized online poker by Kumaresan et al.,[49] with the primitive called secure cash distribution with penalties that guarantees the fair finalization of the poker game. They also utilize the timed commitment to incentivize rational players to behave honestly. Kumaresan et al.[51,52] improve the efficiency of the deposit-based general SMPC protocol they proposed.[49] From another aspect, Kiayias et al.[53] further give the first fair and robust SMPC protocol based on the blockchain.

Note that most SMPC protocols are proposed with formal security proofs. Since SMPC is a combination of several basic schemes (e.g., commitments and the $\mathcal{F}^*_{CR}$), the universally composable (UC) model[197] is considered the most favorable proof model in this context and is frequently used. The general SMPC protocols described above[50–53] use the same ideal claim-or-refund functionality $\mathcal{F}^*_{CR}$ for secure cash distribution. We conclude the contents of $\mathcal{F}^*_{CR}$ in Figure 6, where $\lambda$ is the security parameter, *sid* and *ssid* are two session identifiers, and $\tau$ is the round number. A sender $P_s$ sends a fund of value $v$ to a receiver $P_r$, and $P_r$ should provide a witness $\omega$ such that $\varphi(\omega) = 1$ within the $\tau^{th}$ round to claim the fund, where $\varphi$ refers to the primitive predefined by $P_s$. $F_{send,x}$ (resp. $F_{receive,x}$) is the ideal function that sends (resp. receives) the message to (resp. from) $P_x$, where $x \in \{s, r\}$. $F_{broadcast}$, $F_{record}$, and $F_{delete}$ are the ideal functions that broadcast, record, and delete messages.

The ideal claim-or-refund functionality $\mathcal{F}^*_{CR}$ requires the participants to deposit some money as assurance for their honest behaviors. The funds will be automatically distributed by the smart contracts accordingly. The on-chain scripts serve as an intermediary that no single party could control, which greatly alleviates the hard work for designing SMPC protocols. With this $\mathcal{F}^*_{CR}$, different schemes are designed[49–53] to optimize the efficiency, fairness, and robustness of SMPC.

However, all these blockchain-based SMPC protocols assume that the participants are rational enough to behave accordingly, and this is weaker than the standard malicious adversary model. Besides, due to the low transaction-processing rate, such SMPC protocols may take more time than the traditional ones.

For future research, the efficiency of blockchain-based SMPC is still attractive. The trade-off between best-case and worse-case overhead may be considered: to improve the best-case efficiency to make honest users more comfortable or minimize the worst-case overhead to protect honest users' interest. Although several general SMPC protocols have been proposed, the authors all choose to give a specific implementation of their protocol, and a compiler that converts arbitrary multi-party computation protocols into smart contracts (scripts) is still absent. Therefore, we recommend that researchers and developers pay more attention to the implementation aspects and make these general blockchain-based SMPC protocols really "general" and practical.

#### 4.1.3. Layer-2 protocols

Smart contracts written in scripts are limited in complexity and processing rate. Layer-2 protocols are introduced to overcome these problems. The main idea is to separate the computation from the validation process. That is, the executions of smart contracts are performed off the blockchain, and only necessary steps such as setup, recording, settlement, and dispute resolution are carried out on the blockchain. In this way, the limitation of opcodes and the effect of high transaction delay can be avoided.

Moreover, since only the final results and information used for disputations are publicly available, the privacy during the communications among participants and the contracts is enhanced while taking advantage of the auto-enforcement of smart contracts. Related schemes are classified and summarized in Jourenko et al.[198] (up to 2019) and are divided into three types: off-chain channel, construction schemes for off-chain networks, and off-chain network management. Based on Jourenko et al.,[198] we summarize the design patterns of contracts related to the layer-2 protocol, as described below.

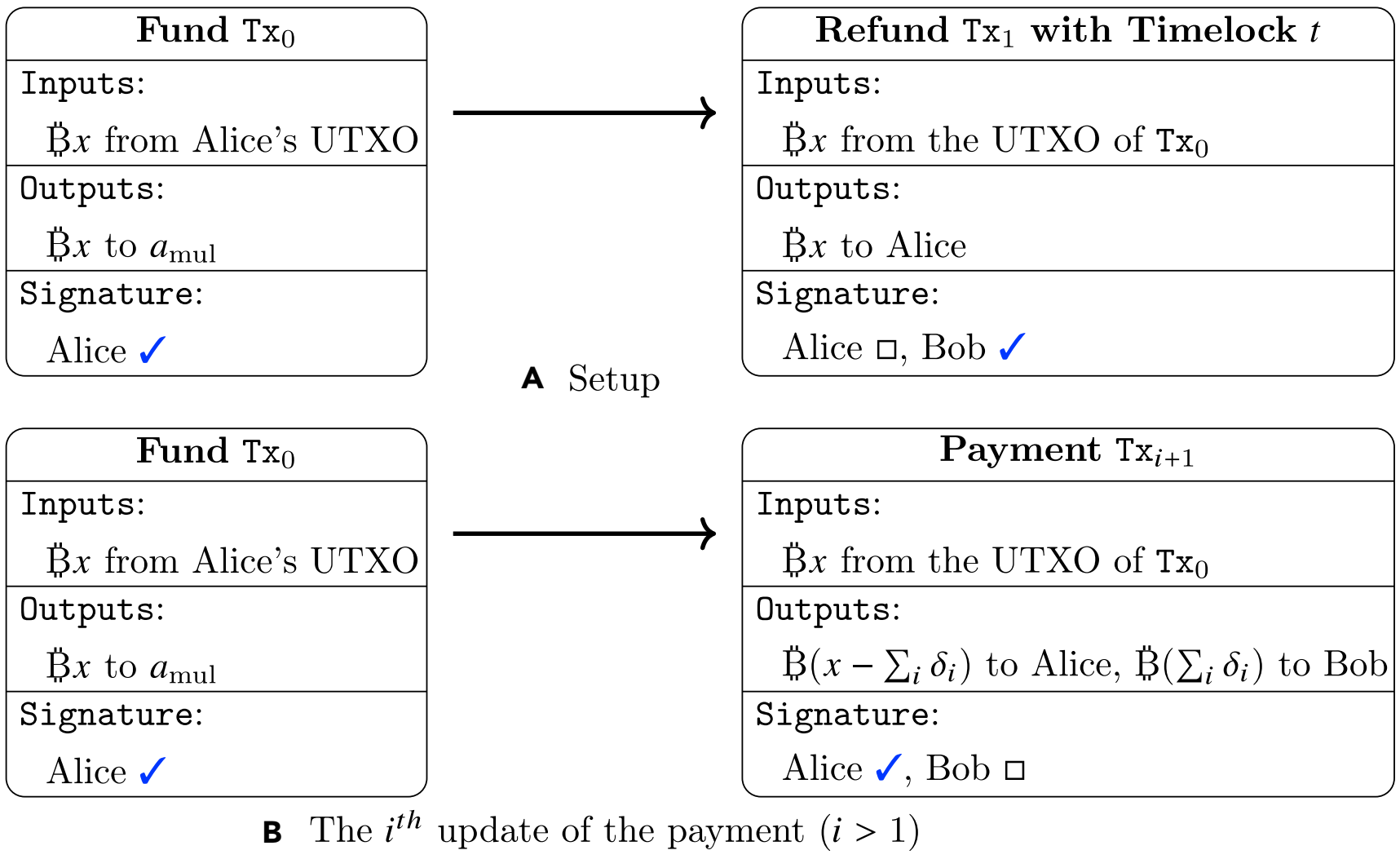

**Figure 7. Micropayment from Alice to Bob[20] (the time and payload field is omitted for simplicity)**
(A) Alice's deposit and refund transactions through a multi-signature mechanism.
(B) Update of the payment transactions.

The payment channel is one of the most significant components among layer-2 protocols. It is first introduced by Hearn[20] and is known as the micropayment channel, a protocol that conducts continuous payments to a recipient. It utilizes the multi-signature opcode `CHECKMULTISIG` provided in Bitcoin to spend the UTXOs in such transactions, requiring more than one signature.

To better illustrate the payment channel, we restate the micropayment channel in Figure 7 according to Hearn,[20] and the time and payload fields are omitted for simplicity. Suppose a channel is established between the payer Alice and the payee Bob. Initially, Alice creates (but does not broadcast) $Tx_0$ that deposits ₿$x$ to a multi-signature address $a_{mul}$, which requires both signatures of Alice and Bob to spend the UTXO. Alice then generates the refund transaction $Tx_1$, which returns the funds from $a_{mul}$ to Alice. There is a timelock $t$ in $Tx_1$: if ₿$x$ is not claimed in $t$, Alice can reclaim it freely with her signature. Alice and Bob jointly sign and broadcast $Tx_1$, and Alice simultaneously broadcasts $Tx_0$. Hence, Alice's ₿$x$ is locked in $a_{mul}$. Whenever Alice wants to pay, she creates the transaction $Tx_{i+1}$ ($i$ denotes the $i^{th}$ update) and sends the signed $Tx_{i+1}$ to Bob. In $Tx_{i+1}$, the fund ₿$x$ in $a_{mul}$ is divided into two parts, ₿$\left(\sum_i \delta_i\right)$ for Bob and the rest for Alice, where $\delta_i$ denotes the value in $i^{th}$ payment and $\sum_i \delta_i$ is the total amount paid to Bob. Finally, Bob signs and broadcasts the latest transaction to finalize the payment, and both Alice and Bob will get what they deserve. Since the inputs of $Tx_i$ all come from the same address $a_{mul}$, only one finalization will take effect. Therefore, Bob has the incentive to broadcast the latest payment to get the coins he deserves.

With properly designed scripts and initial on-chain setup (as described above), Alice could continuously pay Bob in exchange for Bob's service or goods, as long as no dispute occurs and Alice has enough money deposited. Such payment is no longer conducted through blockchain. This strategy avoids the long time delay for each transaction (e.g., waiting for subsequent six blocks in Bitcoin) and saves transaction fees, especially when the payment amount is small and frequent.

Based on the micropayment channel described above, the concepts of payment channel networks and state channel networks are derived. Their main idea and the structure of scripts (or contracts) are similar, and related studies are dedicated to the fairness of the protocol, which is conducted through off-chain communications, so we reserve the introduction of these schemes to Section 6.2.1.

The probabilistic payment system is another research direction in layer-2 protocols. It was first implemented as smart contracts by Pass and Shelat[54] in 2015. The core idea is that the transactions only succeed in a probabilistic manner. In other words, for a fixed success probability $\rho$ and amount ₿$x$ per transaction, on average a recipient will get ₿$\rho x$ in every transaction. It can be applied in lotteries and other situations involving a large number of small payments. In such a system, senders must make two deposits, one for the payment and the other for the penalty. Furthermore, the scheme in Pass and Shelat[54] requires a verifiable trusted service (VTS) to validate probabilistic payments, and anyone can observe the VTS's misbehavior through on-chain information. Moreover, the authors propose a refined scheme that the VTS will never be invoked in the best case. Hu and Zhang[55] adopt a timelocked deposit mechanism in Bitcoin to simplify the initial deployment procedure, which only needs one on-chain transaction to achieve the same functionality as the original scheme in Pass and Shelat[54]. Chiesa et al.[56] propose a concept of decentralized anonymous micropayment. They utilize the fractional message transfer technique to transfer messages in a probabilistic manner and protect privacy in the probabilistic payment system, but they point out that the double-spending in their probabilistic payment system is inevitable and analyzes this attack's effect on their work.

We restate the probabilistic payment system[54] in Figure 8. Similar to that in a micropayment channel, Alice first deposits the escrow ₿$x$ and penalty ₿$y$ to the multi-signature addresses $a_e$ and $a_p$, respectively. Next, Bob generates a random number $r_1$ and sends $F_{SHA-256}(r_1)$ to Alice, along with his address $a_B$, where $F_{SHA-256}$ refers to the SHA-256 hash function. Afterward, Alice generates the payment transaction $Tx_1$ and penalty transaction $Tx_2$ in Figure 8A, where the zero address $a_0$ is used to destroy the funds. $Tx_1$ and $Tx_2$ are sent to VTS after being signed. Note that there is a timelock $t$ in $Tx_1$ and $Tx_2$, similar to that in the micropayment scheme. Next, Alice generates a random number $r_2$. If $r_1 \oplus r_2 < r$ holds, where $r$ is related to the pre-agreed probability, their message transcripts will be sent to VTS, and VTS will sign and broadcast $Tx_1$ if the message is correct, as shown in Figure 8B. Note that during the whole process, when Bob and VTS receive a signature, they shall verify it before other operations. In addition, if Alice spends the ₿$x$ before $Tx_1$ takes effect, the VTS will sign $Tx_2$ so that Alice's penalty deposit will be destroyed.

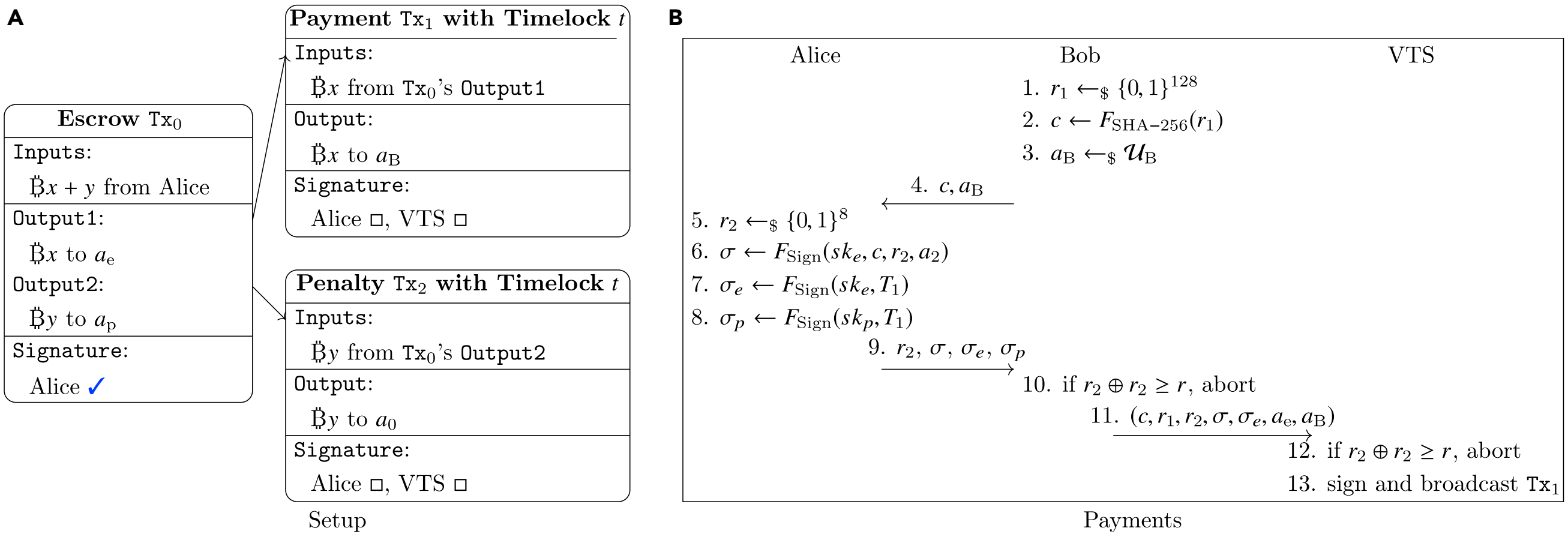

**Figure 8. Probabilistic payment from Alice to Bob**
(A) Deposit and penalty escrow required for setup in the system.
(B) Interactions between Alice and Bob for a probabilistic payment.
Figure reprinted with permission from Pass and Shelat[54].

The probabilistic payment system reduces the number of transactions, decreases transaction fees, and indirectly improves the transaction rate. However, this mechanism is only applicable for limited scenarios since it introduces a third party and assumes the users are rational. Moreover, it is only relevant to the payments rather than general applications. As for future research, the privacy, efficiency, and security issues could be better illustrated, e.g., give formal definitions and proofs of security within the UC model, minimize the worst-case performance, or introduce more practical and realistic assumptions for adversaries.

#### 4.1.4. Scriptless contracts

The traditional P2SH-based smart contracts require publishing the full scripts, which may cause privacy leakage during communication among users and smart contracts. Apart from the layer-2 protocols described above, the scriptless contract is another solution to this problem. It is similar to a standard transaction but achieves the same goal as smart contracts. Since there is no script, the contents of a smart contract would never be disclosed. It is first proposed Banasik et al.,[57] and its original goal is to prevent miners from rejecting packaging P2SH transactions because executing scripts consumes more time and space than standard transactions. Besides, an application instance of selling a factorization of an RSA modulus is provided by Banasik et al.,[57] which requires an additional opcode in Bitcoin.

Scriptless contracts are further summarized by Poelstra,[58] who presents a way to construct a smart contract without exposing scripts. With Schnorr signature[199] (which is not currently compatible with Bitcoin), it is possible to execute scripts only among parties involved in the transaction, and only the final settlement is updated on-chain. Such transactions for scriptless contracts are indistinguishable from the standard ones, so the privacy of contract contents is well protected. However, such contracts' functionality is quite limited, as the final output should be directly verifiable in the same way as a standard transaction.

In the subsequent work, Malavolta et al.[59] present scriptless contracts using ECDSA, which is compatible with both Bitcoin and Ethereum. Their work makes scriptless contracts more practical and applicable without requiring a hard fork.

Some may argue that the hashed timelock contract (HTLC) and revocable sequence maturity contract (RSMC) in Lightning Network[25] (see Section 6.2.1) also include the idea of non-public execution. However, they are only used for efficient payments, whose forms are almost fixed and monotonous, in contrast to the scriptless contracts described above. That is, scriptless contracts are more general and support more kinds of contracts other than off-chain payments.

The privacy-preserving nature of such scriptless schemes makes it an appealing research direction for privacy-sensitive scenarios. For future research, scriptless contracts, especially those compatible with ECDSA, could be adopted in more applications to hide smart contracts' contents. However, this requires a subtle design, and other properties such as fairness and efficiency should be fully considered.

### 4.2. Design tools

We have so far described 18 design patterns for script-based smart contracts, which may help developers start quickly under similar scenarios. From another perspective, due to the lack of readability and limitation in script operations, constructing script-based smart contracts is hard work. Besides, since most smart contracts involve money transfer, vulnerabilities in smart contracts may cause severe economic loss. This makes the security of smart contracts more critical than general computer programs.

In this section, we introduce ten studies on design tools that relieve the burden of contract developers and help them build

**Table 2. Languages for Bitcoin-relevant blockchains**

| Language | Year | Security proof | Open-sourced | Templates available | Description |
|---|---|---|---|---|---|
| Ivy[65] | 2016 | ○ | ● | ● | high-level language, educational purposes only |
| Simplicity[66] | 2017 | ● | ○ | ○ | intermediary representations, verifiable with Coq |
| BALZaC[64] | 2018 | ○ | ● | ● | high-level language, along with a formal model of Bitcoin |
| BitML[67–69] | 2018 | ● | ● | ● | high-level language, process-algebraic language |

● denotes that the language possesses the corresponding property in the column, while ○ means the opposite.

secure smart contracts. We divide these references into analysis tools and languages, where the former helps validate the security of contracts and the latter facilitate the writing of smart contracts.

#### *4.2.1. Analysis tools*

Concerning the security analysis in script-based smart contracts, contracts in Bitcoin are abstracted as timed automata in Andrychowicz et al.,[60] where the states are finite and change chronologically. In this way, the model detection tool UPPAAL[200] for timed automata can be used to ensure that the contract runs as expected.

As for the modeling of smart contracts, Bigi et al.[61] propose an ideal contract model in Bitcoin and provide a security analysis using game theory. The authors analyze the possible behaviors executed by two parties and explain the feasibility of their model. However, the model is limited in the two-party case, which excludes the multi-party situation. More recently, Atzei et al.[62] proposed a formal model of contracts in Bitcoin, which is the base of the high-level language BALZaC[64] (see Section 4.2.2).

There are only a few analysis tools for script-based smart contracts during our survey, partially because smart contracts are less used compared with those in Turing-complete blockchains. Besides, scripts are mostly used to transfer money among users conditionally, and thus vulnerabilities and bugs are less reported. For future research, the various security analysis techniques for Turing-complete blockchains (see Section 5.2.1) may be adopted in script-based platforms. During this migration, most efforts may be consumed by the definition of potential vulnerabilities in scripts. Moreover, multi-party protocols in the script-based platforms are more difficult to model and analyze, which may also be a potential research direction in this context.

#### *4.2.2. Contract languages*

Considering the difficulty in script programming, numerous new languages for smart contracts are proposed to improve the expressivity, readability, and verifiability of the Bitcoin scripts. These languages are shown in Table 2, where ● denotes that the language has the corresponding properties while ○ means the opposite. They can be divided into high-level languages and intermediate representation languages according to their expressivity. High-level languages refer to those that are more expressive and can be directly used by ordinary developers with ease, while intermediate representation languages are useful during the compilation and security analysis.

On the expressivity aspect of contract languages, Ivy[65] is one of the earliest high-level languages designed for Bitcoin. Ivy's syntax is similar to that of common high-level languages, and it adds some specialized keywords for operations in Bitcoin. Some examples are given in its documentation,[65] e.g., LockWithPublicKey and HTLC (the core technique in Lightning Network, for more details see Section 6.2.1). However, the security proof of Ivy itself and the compiled scripts is lacking, so its use is restricted to educational purposes.

As mentioned in Section 4.2.1, Atzei et al.[62,63] propose a formal model of Bitcoin contracts and develop the high-level language BALZaC.[64] They also provide a corresponding analyzer and compiler that compiles BALZaC to standard transactions. Besides, they describe the existing smart contracts with their model, including crowdfunding, timed commitments, micropayment channels, and lotteries. BALZaC enables developers to define Bitcoin smart contracts with a concise syntax. However, it lacks a security analysis.

On the verifiability aspect of smart contracts, O'Connor[66] designs a low-level intermediate representation language, Simplicity, which uses denotational semantics defined in Coq,[201] a popular verification tool. With Coq and other auxiliary tools, Simplicity helps developers easily validate the security of a smart contract. In addition, the author claims that Simplicity also supports static analysis to analyze the efficiency of contract execution.

There are studies that consider both the expressivity and verifiability of the script language in Bitcoin. Bartoletti and Zunino[67] propose a high-level language, BitML, which encapsulates the complex instructions and provides a concise and convenient expression for smart contracts in Bitcoin. They also provide a compiler that converts BitML programs into standard Bitcoin transactions. The correctness of the compiler is proved. Namely, it incurs no additional error or bug during conversion. Other works[68,69] implement several common smart contracts in BitML, such as covenants and timed commitments. However, BitML language is still limited to some extent as there are contracts that could not be expressed by it.

We note that high-level languages are quite attractive for script-based blockchains, which will greatly alleviate the burden of contract programmers, as is done by the Solidity language in Ethereum. However, there will be several problems with the adoption of high-level languages. One of the major challenges is converting the programs written in high-level and script languages to each other without sacrificing correctness. In fact, this problem is also apparent in Ethereum (see Section 5.2.2.2). As we have observed, although BitML is a provable secure high-level language, the expression and syntax are still quite sophisticated compared with modern high-level expressive languages such as JavaScript or Solidity. It may prevent programmers from a quick start. An easy-to-use and provable secure programming language and a compiler that accurately converts the programs into scripts are still needed for script-based blockchains.

**Table 3. Design patterns for Turing-complete smart contracts**

| References | Year | Category[a] | Main contributions | Open-sourced[b] |
|---|---|---|---|---|
| 48,70 | 2017 | specific patterns | multi-party lottery | ●● |
| 71,72 | 2018–2019 | | loan contract | ●○ |
| 78,79 | 2018 | | private auction protocol | ●● |
| 73–77 | 2016–2020 | | smart contracts for government processes and services | ○○○○● |
| 80–84 | 2017–2019 | | off-chain computation and storage | ○○●○○ |
| 22 | 2016–2020 | general patterns | official guidance and examples for solidity | ● |
| 85 | 2016–2020 | | best practices on smart contracts | ● |
| 86 | 2016–2020 | | solidity library | ● |
| 28,87,88 | 2017–2018 | | common patterns for popular applications | ●●● |
| 12,89–94 | 2016–2020 | | vulnerabilities in smart contracts | ●●○○○○● |
| 95,96 | 2016, 2019 | | programming errors found in smart contract courses | ●○ |
| 97,98 | 2016 | | model and schemes for contract design | ○○ |
| 99 | 2016 | | methods for contract update and deletion following existing mechanism | ● |

[a]Specific patterns here mean that the scheme is limited to specific applications, and general patterns denote that the scheme is general for the construction of any contract (even for smart contracts in different platforms).
[b]● (or ○) here indicates that the implementation can (or cannot) be accessed in public online. When there are multiple references on a single line, there will be multiple ● or ○ correspondingly.

## 5. CONSTRUCTING SMART CONTRACTS WITH TURING-COMPLETE LANGUAGES

To extend the limited operations in Bitcoin's script language, Ethereum introduces a new virtual machine structure to support Turing-complete programming languages, greatly extending the application scenarios of smart contracts. The smart contracts are executed in Ethereum Virtual Machine in the form of EVM bytecode. In fact, to facilitate the definition of the execution rules in smart contracts, numerous high-level programming languages are introduced, which can be converted into bytecode by compilers, such as Solidity,[22] LLL (Lisp-Like Language),[202] and Serpent.[23] Such high-level languages are featured with high expressivity and can reduce the difficulty of contract construction.

There are also many derivatives of Ethereum, whose execution mechanisms are almost the same as EVM and support Turing-complete languages. Here we call such systems Turing-complete blockchains. Similar to Bitcoin, most related research is conducted on Ethereum, and the results can be easily transferred to Ethereum's derivative systems. Therefore, we also take Ethereum as the representative to describe the contract construction schemes on Turing-complete blockchains. The schemes here are divided into two parts, design paradigms and tools, as discussed in Section 5.1 and Section 5.2, respectively.

### 5.1. Design paradigms

To reduce errors caused during contract programming, developers are suggested to refer to contract design paradigms, which are carefully designed against common attacks and recognized as safe. We divide the related schemes into two categories: paradigms for specific applications and general purposes. The former refers to the specialized design patterns in some popular application scenarios (see Section 5.1.1) and the latter describes the patterns in general cases (see Section 5.1.2). We remark that both types of paradigms serve as essential building blocks for reliable and verifiable data communication upon blockchain.

We summarize and classify these schemes according to their application scenarios in Table 3. The symbol ● (or ○) in the "Open source" column indicates that the implementation can (or cannot) be publicly accessed online. When there are multiple references on a single line, there will be multiple ● or ○ correspondingly.

#### *5.1.1. Paradigms for specific applications*

In the early stage of smart contracts, lottery, loan, auction, and data storage were the main applications on the market. These contracts are directly related to financial transactions, and their contents and logic are relatively simple. Although such contracts are only designed for specific scenarios, they are still significant and helpful for similar applications. Nonetheless, in recent years smart contracts have attracted governments' attention, and may provide benefits for municipal government processes. In the following, we introduce 16 studies relevant to five common patterns: lottery, loan, auction, e-government, and off-chain computation and storage. We remark that such specific applications are miscellaneous, and thus we only focus on the commonly addressed scenarios in the literature.

Note that although some of the paradigms introduced here might be out of date, their core idea and the revolution are still worth attention. In other words, these schemes might become vulnerable or inefficient during the evolution of smart contracts, and they could be considered as counter-examples in practice.

*5.1.1.1. Lottery*. Traditional lottery schemes require a trusted third party to receive bets from participants and distribute the deposits afterward. However, there exist risks of collusion and absconding for online lottery websites. Due to the anonymity and trustless property of the blockchain and smart contracts, coupled with the inherent cryptocurrencies, smart contracts can replace such TTPs to eliminate these risks and minimize privacy leakage. Lottery thereby becomes one of the most popular applications of smart contracts.

Several lottery contracts implemented in Bitcoin have already been mentioned in Section 4.1.2 as examples of the proposed SMPC protocols. With the Turing-complete languages supported by Ethereum, lottery contracts could be implemented with better performance and less cost. As mentioned in Section 4.1.2, Bartoletti and Zunino[48] devised a multi-party lottery scheme with fixed collateral and implemented it on Ethereum. Their solution comes from a tree of two-party games to determine the final winner among all the participants. The authors also describe a variant that reduces the number of transactions by a set of iteration games between adjacent players, but it cannot guarantee fairness. Miller and Bentov[70] put forward another lottery scheme with zero collateral. They adopt a similar tree of two-party games as that in Bartoletti and Zunino,[48] but their solution requires an extra opcode to take effect in Bitcoin. The authors mention that they implement their scheme on Ethereum.

The lottery schemes can be more easily implemented than those in Section 4.1.2, thanks to the Turing-complete language and account-based model of Ethereum. How to design a fair, collateral-efficient lottery scheme is still an open problem. This might be solved by taking advantage of extra tools such as cryptographic schemes or trusted execution environments (TEE, Definition 17). However, privacy issues are still unaddressed, especially in lotteries and other privacy-sensitive scenarios. Although Ethereum adopts the pseudonym mechanism, the identities may be traced because of the blockchain's transparency. Therefore, off-chain computation and other privacy-preserving techniques (e.g., ZKP) may be considered for future improvements.

*5.1.1.2. Loan*. Similar to the motivations in the lottery schemes, the loan is another popular application of smart contracts. However, in loan contracts participants often want to borrow and lend fiat money since the fluctuation of cryptocurrency market value may cause an undesirable loss for either one of the participants. Moreover, loan contracts have to handle the counter-party risk, whereby a borrower may abscond with funds.

To solve the problems above, Okoye and Clark[71] designed a set of loan contracts named *Ụgwọ* with different methodologies. These contracts ensure the safety of funds for both borrowers and lenders. To manage the unstable market value, the authors adopt an oracle contract (which justifies the current exchange rate) to enable the users to settle the contract with fiat currency (e.g., USD). Moreover, other methodologies, such as mortgage and insurance, are also employed for different security concerns. Norta et al.[72] propose a capital transfer system for users with no access to banks. They propose an Ethereum-based eFiat scheme, where financial institutions provide the exchange rate of fiat currencies (i.e., act as an oracle). They further take the lending process as an implementation example.

The blockchain-based loan schemes usually require an oracle to reflect the cryptocurrency price. However, such an oracle may introduce new vulnerabilities and cause huge economic loss. One example is the attack against the flash loan protocol through oracle manipulation attacks.[203] Therefore, for loan schemes to safely take effect a secure and accurate oracle contract is required, and represents is a promising research direction.

*5.1.1.3. Auction*. In addition to lottery and loan, auction contracts also fully utilize the anonymity and no-TTP features of blockchain. Strain[78] is a private auction protocol whereby participants' identities and bids are hidden. To avoid using an inefficient SMPC protocol, Strain improves the two-party comparison mechanism[204] and verifies the results by using ZKP (Definition 16). Besides, a reversible commitment mechanism is applied to ensure fairness when a malicious termination occurs. However, although the scheme avoids directly publishing the auction details on-chain, it still leaks the order of bids. To solve this problem, Galal and Youssef[79] propose a verifiable secret auction protocol using Pederson commitment[205] and ZKP, and implement their work on Ethereum. Their scheme ensures that participants cannot learn any information about others during the auction process, and anyone is allowed to verify the auction results. However, a formal security proof of their scheme is absent.

We remark that fairness, security, and privacy issues in auction schemes require further consideration in the future. A formal proof of the privacy and security properties is a necessity, and other tools such as cryptographic schemes and hardware could be introduced to improve the state-of-the-art solutions.

*5.1.1.4. E-government*. Blockchain and smart contracts provide convenient verifiability and could make participants collaborate without mutual trust. Ølnes[73] argues that such advantages of smart contracts can be used for applications other than currencies, especially for online government processes (e-government). Ølnes et al.[74] further discuss the benefits and implications of blockchains for e-government applications. With a similar point of view, Hou[75] investigates the blockchain applications of e-government in China. Abodei et al.[76] select Nigeria as the study case and enumerate the blockchain solutions to solve the existing problems in the government's public project processes. However, all the aforementioned discussions involve no practical implementations.

Most recently, Krogsbøll et al.[77] have pointed out that e-government applications are possible with smart contracts if taking care of privacy issues. They give a prototype implementation of governmental services in Denmark, with attractive properties, such as verifiability, that are beneficial for the transparency of governmental affairs.

We remark that e-government might be one of the most promising applications of smart contracts. Researchers and developers may pay attention to the unique requirements exhibited by the government affairs, such as enabling tens of thousands of citizens to participate in a single event efficiently, keeping sensitive personal information private, and preventing malicious manipulation.

*5.1.1.5. Off-chain computation and storage*. Taking smart contracts off-chain is a promising way to avoid the side effects of the high confirmation delay while maintaining the trustless property offered by blockchain. With this idea, Eberhardt and Tai[80] analyze the suitable scenarios of off-chain computation and storage and provide several design patterns for off-chain schemes. The typical scenarios are payment channel network and state channel network.[198] These solutions mainly focus on the execution-related aspects, so we leave the discussion to Section 6.2.

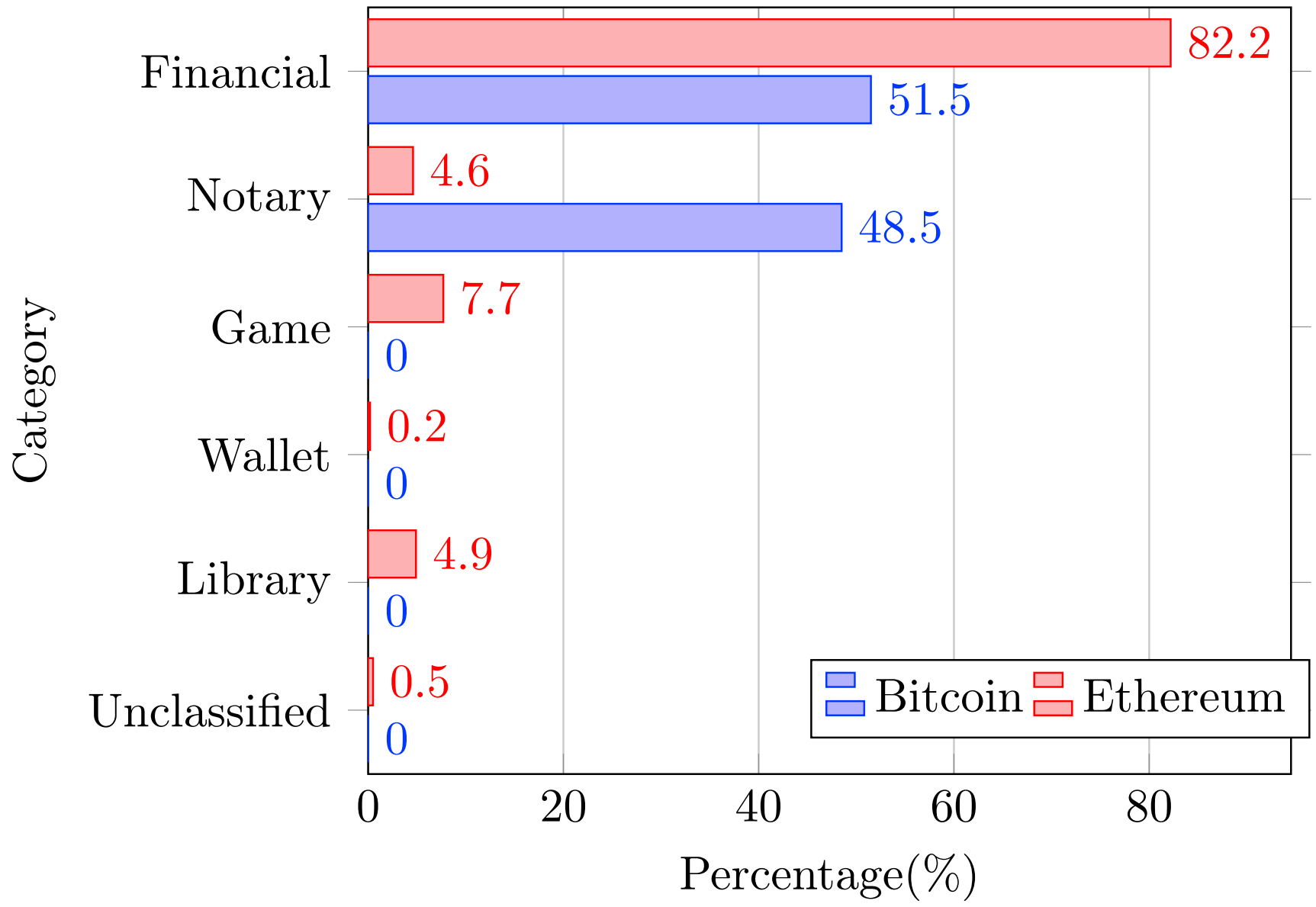

**Figure 9. Distribution of the number of transactions in Bitcoin and Ethereum (until January 2017)**

Figure reprinted with permission from Bartoletti and Pompianu[28].

To attain advantages (e.g., privacy, latency, and transaction fees) from both on- and off-chain implementations, Molina-Jiménez et al.[81] propose a hybrid solution that splits a smart contract into on-chain and off-chain components. However, this comes at the cost of a complex implementation strategy, which depends on concrete applications. The authors provide a proof-of-concept implementation of a simple trading contract[82] to demonstrate the feasibility of their work. Similarly, Li et al.[83] separate a smart contract into the light/public parts and heavy/private parts to avoid high transaction fees and confirmation delay caused by complex on-chain executions. Their solution also suffers from the absence of a general way to make such separation.

As for off-chain data exchange, Norta et al.[84] propose the DataWallet, a data-sharing system. The data requester and provider interact with the help of a smart contract, ensuring the fairness of data exchange.

For future research, a general way to split arbitrary protocols into on- and off-chain parts could be investigated. Moreover, problems such as making these parts collaborate efficiently and preventing potential attacks against the on-chain part are also worth consideration—for example, a fine-grained access control mechanism could be implemented.

#### 5.1.2. Paradigms for general purposes

We have thus far described 16 design paradigms for specific applications. From another aspect, some researchers are trying to give a general pattern that applies to the design of all kinds of smart contracts. Specifically, we have collected three references providing the so-called "best practices" to mitigate common bugs and security risks during development. Another three papers classify smart contracts according to application scenarios. They give patterns for each category, which are referred to as "classification and patterns of common contracts" in this review. There are also eight studies focusing on common errors in the contract construction process, as counter-examples for beginners to learn from, which we call "common vulnerabilities and errors." Finally, there are three "contract design models" describing contract construction experience and methods from a higher perspective. As mentioned in Section 5.1, these paradigms may provide insights for more useful applications on the data communication and others for new incomers.

*5.1.2.1. Best practices*. It is recommended to refer to best practices given by researchers or communities to avoid common vulnerabilities as much as possible. The Solidity document,[22] which serves as official guidance for writing smart contracts, provides tips, requirements, and examples of contract construction. It is considered a must-read document for beginners since it has been verified and reviewed by most developers, and it has been continuously improved and updated. In addition, Consensys Diligence[85] and OpenZeppelin[86] also provide an open-sourced list of best practices and libraries.

We remark that such collecting of best practices is not a straightforward task, since smart contract technology is evolving quickly. Besides, newly found vulnerabilities may continuously influence the forms of best practices. Therefore, the way is long toward enumeration of a fully covered list of the state-of-the-art best practices in contract programming, which will require non-trivial work from the community.

*5.1.2.2. Classification and patterns of common contracts*. Statistics and classification of existing contracts could help developers find their desired reference patterns for their target application scenarios and further avoid contract vulnerabilities that may commonly occur in specific applications.

In terms of classification, Bartoletti and Pompianu[28] first classify the application scenarios of smart contracts. They then list several common contract modes, including token, authentication, oracle, randomness, poll, time constraint, termination, math, and fork check. Their classification covers most smart contracts, and the numeric results are shown in Figure 9,[28] which illustrates the number of transactions for different types of smart contracts on Bitcoin and Ethereum. Note that the zero value in this figure refers to a negligible percentage. According to their results, financial and notary-related smart contracts contribute to the majority of transactions, which is much more apparent in Bitcoin. Although their results only reflect the situation up to January 2017, we infer that this situation might still hold nowadays because of the prevalence of ERC-20 tokens[206] and decentralized finance.[207]

For the security of smart contracts, Wöhrer and Zdun[87,88] give several contract patterns according to some known secure contracts from the perspectives of access control, authentication, contract life cycle,[208] and contract maintenance.

**Table 4. Potential vulnerabilities in smart contracts on Ethereum[12,89,90]**

| Level[12] | Causes[12,89] | Security[90] | Attacks[12,89] |
|---|---|---|---|
| Solidity | call to the unknown | call integrity | the DAO attack |
| | re-entrancy | | |
| | gasless send | – | King of the Ether Throne |
| | exception disorders | atomicity | King of the Ether Throne and GovernMental |
| | type casts | – | – |
| | keeping secrets | – | multi-player games |
| EVM | immutable bugs | – | Rubixi and GovernMental |
| | ether lost in transfer | – | – |
| | stack size limit | – | GovernMental |
| Blockchain | unpredictable state | independence of mutable account state | GovernMental and dynamic libraries |
| | transaction-ordering dependence | | Run |
| | generating randomness | independence of transaction environment | – |
| | time constraints | | GovernMental |

Similar to the best practices part above, we remark that the classification of common usage and design patterns of smart contracts require a continuous update, which turns out to be absent nowadays. Future researchers and developers may contribute to extend the existing works and make a classification of the latest smart contract patterns.

*5.1.2.3. Common vulnerabilities and errors*. Statistics and classification of common vulnerabilities and errors during contract construction may serve as counter-examples, and relevant research may help developers form a good contract pattern.

In terms of common vulnerabilities, Luu et al.[89] were the first to summarize the security problems that commonly occur in smart contracts. These problems are categorized into four types, namely, transaction-ordering dependence, timestamp dependence, mishandled exceptions, and re-entrancy vulnerability. The first three concepts are relatively easy to understand from their name, and the last one refers to a concept unique in the smart contract field. We leave the introduction of re-entrancy to Section 5.2.1, where the analysis tools specially designed for such vulnerability are discussed.

Later, Atzei et al.[12] also summarized the vulnerabilities on smart contracts and divided them into different layers according to the effects of attacks. For example, some of them affect the correctness of execution while others may disturb the underlying execution mechanism. The latest work of Groce et al.[94] makes a classification of 246 defects found in 23 Ethereum smart contracts. They utilize several open-sourced analysis tools (i.e., Slither,[131] Manticore,[118] and Echidna,[139] see Section 5.2.1) along with manual auditing, and find that there are ten defects per contract on average.

In addition to common vulnerabilities, there are studies on the security properties that smart contracts should satisfy. Following the work of Luu et al.[89] and Atzei et al.,[12] Grishchenko et al.[90] presented four security features that smart contracts should meet. The smart contracts with such features could automatically avoid several known vulnerabilities. Their classification is useful for subsequent development and contributes to the design of related vulnerability detection tools.

We integrate the above results in Table 4. The blanks in the table indicate that the corresponding vulnerabilities are not discussed in these papers. We remark that Mense and Flarscher[91] and Dika and Nowostawski[92] do similar work that provides a taxonomy of the security issues in the literature. Moreover, both studies give the severity level of each vulnerability.

With regard to common errors, Delmolino et al.[95] enumerate four common problems found in their smart contract courses. Although they adopt Serpent[23] as the programming language, many problems are universal for all languages when designing smart contracts. The authors summarize four types of errors, namely design errors on state machines, the absence of cryptographic protocols, unreasonable incentive mechanisms, and vulnerabilities inherent in Ethereum. These errors are of educational meaning for all developers, especially beginners. Angelo et al.[96] also give a summary of their smart contract course but focus more on the teaching process rather than the design skills.

From a practical perspective, Pérez and Livshits[93] investigate six frequently mentioned vulnerabilities in Ethereum and find that the actual exploitation of these vulnerabilities is relatively rare: the hacked amount only takes up to 0.27% of the total ETH marked as vulnerable. The authors explain that this is because most vulnerable ETH is held by several contracts that are not exploitable in practice (e.g., the exploitation requires a malicious majority).

The common vulnerabilities and errors mentioned above are essential to the subsequent research and development of smart contracts, especially the analysis tools that will be discussed in Section 5.2.1. Inspired by Pérez and Livshits,[93] future research may pay attention to the actual influence of the vulnerabilities on deployed smart contracts. Nevertheless, such classification requires constant updates to reflect the latest discoveries.

*5.1.2.4. Design models*. Design models are often built from a higher point of view, and can be seen as a design philosophy orthogonal to the aforementioned best practices. Both design models and best practices can help new developers to prepare for the construction of smart contracts.

Clack et al.[97,98] first discuss the basis, design method, and research direction of smart contract templates, then discuss the basic requirements of smart legal contracts, i.e., the contracts serving legal purposes. They also propose a design model for the storage and release of contracts from a higher level. Marino and Juels[99] summarize the available methods for contract updates without modifying the existing execution mechanism. Developers are often required to reserve certain interfaces at the beginning, and it is good practice to consider such updating demands.

Due to the immutability of blockchains, specifications of smart contracts cannot be easily updated. Therefore, effectively updating the deployed smart contracts to eliminate detected vulnerabilities is one of the prominent research directions

**Table 5. Analysis tools for smart contracts**

| | | Refs. | Year | Tool[b] | Main Methods | Input[c] | Open-sourced[d] |
|---|---|---|---|---|---|---|---|
| Targets | | | | | | | |
| Specific-purpose | re-entrancy attacks | 100 | 2017 | ECFChecker | modular reasoning | Solidity | ● |
| | | 101 | 2018 | ReGuard | fuzzing | Solidity, EVM bytecode | ○ |
| | | 102 | 2019 | Sereum | taint analysis | EVM bytecode | ○ |
| | gas-related | 103 | 2017 | GASPER | symbolic execution | EVM bytecode | ○ |
| | | 104 | 2018 | GasReducer | pattern matching | EVM bytecode | ○ |
| | | 105 | 2018 | – | symbolic execution | Solidity | ○ |
| | | 106 | 2018 | MadMax | decompilation and logic-based specification | EVM bytecode | ● |
| | | 107 | 2019 | GASTAP | symbolic execution | Solidity, EVM bytecode | ○ |
| | | 108 | 2019 | GASOL | symbolic Execution | Solidity, EVM bytecode | ○ |
| | | 109 | 2020 | Syrup | symbolic execution and SMT-solving | EVM bytecode | ● |
| | | 110 | 2020 | GasChecker | symbolic execution | EVM bytecode | ○ |
| | trace vulnerability | 111 | 2018 | MAIAN | symbolic execution | EVM bytecode | ● |
| | event-ordering bugs | 112 | 2019 | ETHERRACER | symbolic execution and fuzzing | EVM bytecode | ○ |
| | integer bugs | 113 | 2018 | OSIRIS | symbolic execution and taint analysis | Solidity, EVM bytecode | ● |
| | | 114 | 2020 | VERISMART | CEGIS-style verification | Solidity | ● |
| Techniques[a] | | | | | | | |
| General-purpose | symbolic execution | 89 | 2016 | OYENTE | symbolic execution | Solidity, EVM bytecode | ● |
| | | 115 | 2018 | ETHIR | symbolic execution | Solidity, EVM bytecode | ● |
| | | 116 | 2019 | SAFEVM | symbolic execution and SMT-solving | Solidity, EVM bytecode | ● |
| | | 117 | 2018 | Mythril | symbolic execution and SMT-solving and taint analysis | EVM bytecode | ● |
| | | 118 | 2019 | Manticore | symbolic execution | Solidity | ● |
| | | 119 | 2018 | TEETHER | symbolic execution and constraint solving | EVM bytecode | ● |
| | | 120 | 2019 | sCompile | symbolic execution | EVM bytecode | ○ |
| | | 121 | 2019 | SMARTSCOPY | summary-based symbolic evaluation | application binary interface | ○ |
| | | 122 | 2018 | SECURIFY | abstract interpretation and symbolic execution | Solidity, EVM bytecode | ● |
| | | 123 | 2020 | VERX | symbolic execution and predicate abstraction | Solidity | ○ |
| | syntax analysis | 124 | 2018 | SmartCheck | syntax analysis | Solidity | ● |
| | | 125 | 2019 | NeuCheck | syntax analysis | Solidity | ● |
| | abstract interpretation | 126,127 | 2018 | EtherTrust | abstract interpretation | EVM bytecode | ● ● |
| | | 128 | 2018 | Vandal | abstract interpretation logic-driven analysis | EVM bytecode | ● |
| | | 129 | 2019 | Gigahorse | abstract interpretation | EVM bytecode | ○ |
| | | 130 | 2020 | eThor | abstract interpretation | EVM bytecode | ○ |
| | ~ | 131 | 2019 | Slither | data-flow analysis and taint analysis | Solidity | ● |
| | ~ | 132 | 2018 | SASC | topological analysis, syntax analysis, and symbolic execution | Solidity | ○ |
| | | 133 | 2018 | – | model-checking | Solidity | ○ |

(*Continued on next page*)

Table 5. *Continued*

| | Refs. | Year | Tool[b] | Main Methods | Input[c] | Open-sourced[d] |
|---|---|---|---|---|---|---|
| model checking | [134] | 2018 | Zeus | symbolic model checking and abstract interpretation SMT-solving | Solidity, C#, Go, Java, etc. | ○ |
| ∼ | [135] | 2019 | – | deductive proof | Why3[209] | ● |
| ∼ | [136] | 2018 | – | SMT-solver | Solidity | ○ |
| fuzzing | [137] | 2018 | ContractFuzzer | fuzzing | application binary interface, EVM bytecode | ● |
| | [138] | 2019 | ILF | fuzzing | Solidity | ● |
| | [139] | 2020 | Echidna | fuzzing | Solidity, Vyper[210] | ● |

[a]∼ means that the taxon is trivial for a single object.
[b]If the corresponding tool has no name, the blank is filled with –.
[c]The input only refers to the form of smart contracts, so the analysis specification required in some tools is not included.
[d]● here means that the corresponding tool is open-sourced and can be referenced online, while ○ implies the opposite.

nowadays. Other design models for security and privacy issues are also of great significance.

### 5.2. Design tools

Design tools discussed here are used to help developers build smart contracts more efficiently, and usually come in the form of useful software rather than boring instructions. They may reduce the security concerns of developers or simplify the development process intuitively or interactively. Here we further divide the existing design tools into analysis tools and auxiliary tools. Analysis tools are used to perform security analysis when the contracts are almost completed to find potential vulnerabilities (see Section 5.2.1). On the other hand, auxiliary tools usually take effect in contract development, facilitating the development to some extent (see Section 5.2.2).

Prior to our work, we have observed six studies surveying the contract design tools. Dika[35] conducts a detailed comparison of the analysis tools in 2017 from the aspects of the efficiency, accuracy, and types of supported vulnerabilities. Harz and Knottenbelt[38] further analyze the related languages and security tools. They provide a brief introduction and classification of these languages and tools. Later on, Grishchenko et al.,[126,127] Angelo and Salzer,[39] Liu and Liu[40] review the smart contract-related security tools from several different aspects.

Based on the discussions above, we summarize, compare, and analyze the existing contract languages and vulnerability detection tools in detail (up to August 2020). Our work on the contract design tools can be viewed as a further extension to the surveys mentioned above.

#### *5.2.1. Analysis tools*

A smart contract is a piece of executable computer program deployed on a blockchain. Therefore, traditional code analysis methods can be naturally extended to the field of security analysis. Since there are many vulnerabilities unique to smart contracts, many tools are specifically designed for these threats. Specifically, we have collected 15 tools focusing on detecting certain vulnerabilities that frequently appear and may cause severe consequences. There are another 26 tools aimed at general security analysis, which simultaneously detect multiple potential vulnerabilities, check user-defined properties, and remind the developers of potential risks. The former tools usually work with higher accuracy and help mitigate the pressure caused by specific attacks. The latter remind developers of the vulnerabilities and risks that are overlooked during development.

We list the analysis tools in Table 5 according to the type and number of vulnerabilities they can detect. We divide them into six classes in this paper: (a) re-entrancy attacks related; (b) gas consumption related; (c) trace vulnerability related; (d) event-ordering bugs related; (e) integer bugs related; and (f) general detection tools. The first five classes of tools focus on specific vulnerabilities, and we group them according to their targets. The last type of tools achieves general detection of vulnerabilities, and we group them according to the main techniques they utilize, including symbolic execution, syntax analysis, abstract interpretation, data-flow analysis, topological analysis, model checking, deductive proof, satisfiability modulo theories, and fuzzing test.

We note that there might be several specially designed domain-specific languages for the analysis tools to support user-defined conditions in the process of security analysis. Such languages are not included in Table 5, because they usually are accompanied by their corresponding tools. Moreover, contents in the column "Input" represent the form of smart contracts (e.g., Solidity or EVM bytecode) that the tool accepts as input. The symbol ● in the column "Open-sourced" means that the corresponding source code can be referenced online, while ○ implies the opposite. We remark that specific- and general-purpose tools are illustrated separately. We group the tools capable of detecting multiple vulnerabilities into the general-purpose category, even though some of them can detect the vulnerabilities targeted by the specific-purpose ones.

*5.2.1.1. Re-entrancy attacks related*. It is the re-entrancy attack that made users suffer huge economic losses in the infamous DAO event.[11] In a re-entrancy attack, an attacker utilizes the fallback function to steal money from the smart contracts designed in a non-standard manner.

The fallback function stands for a predefined function that has no name or parameters. It is used to handle exceptional requests, such as calling a function that does not exist. Inspired by Atzei et al.[12] and Rodler et al.,[102] we depict the fallback mechanism and re-entrancy attack in Figure 10. Ideally, when a contract $C_A$ calls a public function in another contract $C_B$, it waits until the executed function is finished before returning to the remaining part of $C_A$. However, in some cases (e.g., when the `call` function is invoked to transfer money), after the invoked function is finished, the EVM would call the fallback function in $C_A$ before returning to the expected part in $C_A$, as shown in the left column of Figure 10. If $C_B$ calls the contract $C_A$ in the fallback function, then it comes to a situation of recursive invocation, as shown in the right column of Figure 10. When the latter invocation starts, the former is not finished, and thus contract $C_B$ re-enters $C_A$.

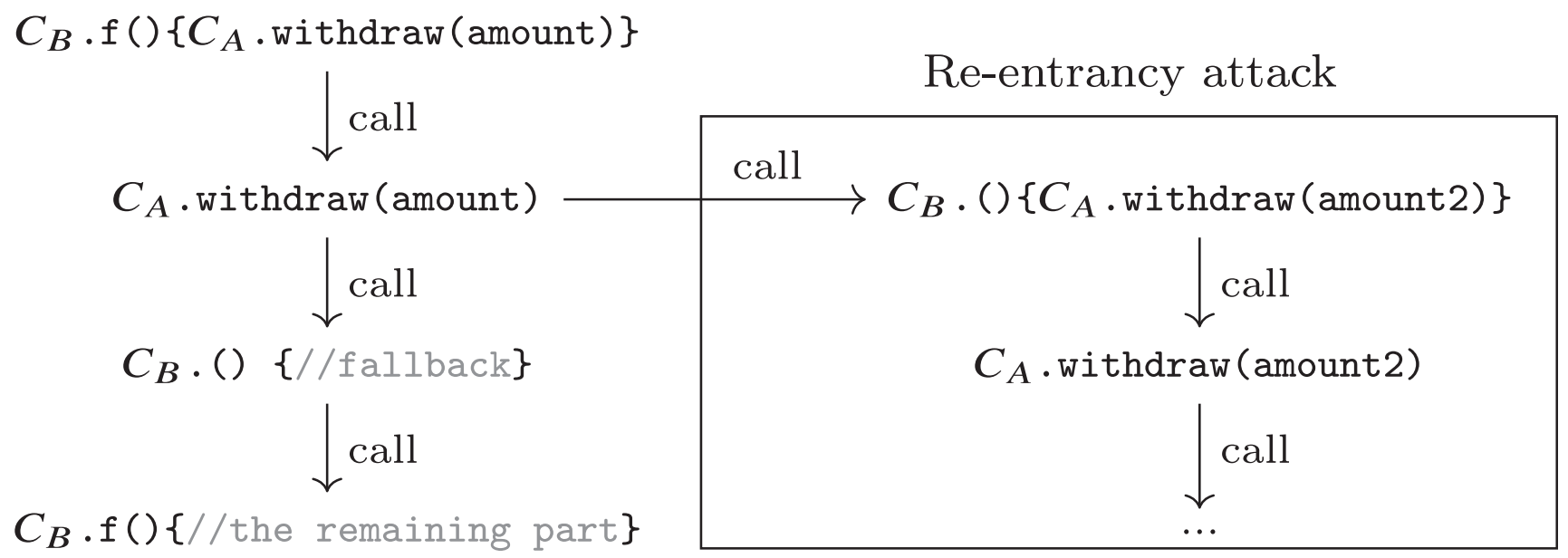

**Figure 10. Fallback mechanism and the re-entrancy attack**
Figure reprinted with permission from Atzei et al.[12] and Rodler et al.[102].

Such a re-entrancy is undesirable and may cause unexpected consequences, especially when the contract $C_A$ is not carefully designed to handle such misbehavior. As depicted in Figure 11 (inspired by Atzei et al.[12] and Rodler et al.[102]), when the contract `Attacker` ($C_B$ in Figure 10) sends some value to contract `SimpleDAO` ($C_A$), the fallback function is invoked after the value is processed by $C_A$. We can see that the fallback function of $C_B$ calls the `withdraw` function of $C_A$, and gets into lines 3–6. However, when the process gets into line 5, a "call" operation in $C_B$ (the `msg.sender`) is invoked, and when the "call" is finished, the fallback function is invoked again, before the process of getting into line 6. That is, the value stored in $C_A$ is transferred but the balance record is not adjusted. When the fallback function of $C_B$ calls the `withdraw` function again, the condition at line 4 will still be true, and the value will be transferred repeatedly before the credit record is adjusted.

We mention that the re-entrancy attack is not influenced by the out-of-gas exception discussed in Section 2.1.2.2, because the `call.value()` function at line 5 only returns false upon failure and does not invoke the reversion procedure.

Researchers have proposed various analysis tools to defend against such attacks based on different methods and aspects. Grossman et al.[100] propose the concept of Effective Callback Freedom, which requires that the invocation of the callback (same to fallback) function should not affect the states or behaviors of the original program. They claim that this concept can be effectively used to detect re-entrancy attacks in Ethereum, and they integrate their idea into the online detector named ECFChecker.

ReGuard[101] is another tool for detecting re-entrancy vulnerabilities, utilizing the fuzzing test (see Section 5.2.1.6). ReGuard accepts Solidity or EVM bytecode as input. It parses the input into an intermediate representation and subsequently converts it into a C++ program. The fuzzing engine is then employed to generate random inputs. Finally, based on the fuzzing results and the re-entrancy automaton proposed by the authors, the final detection report is generated.

From the perspective of sustainability, Rodler et al.[102] point out that existing tools could only detect potential vulnerabilities before the smart contracts are deployed. Namely, deployed contracts are protected from attacks. To solve this problem they propose the Sereum scheme, which extends EVM by introducing taint analysis and an attack detector. Sereum monitors the EVM bytecode instructions at runtime and utilizes a writelock mechanism that locks the states when calling outside functions. This scheme fundamentally prevents the re-entrancy attack from the underlying execution layer.

The re-entrancy attack is one of the most harmful events for smart contracts. It enables an unauthenticated adversary to steal money from the contract account, sacrificing other users' interests. Various analysis tools have been designed against such attacks as mentioned above, utilizing different techniques. We note that most general tools also include the re-entrancy attack as an analysis target (see Section 5.2.1.6). However, most tools require manual checks to exclude false-negative results.

For future research, researchers can consider the trade-off between efficiency and accuracy of the detection. Moreover, we recommend automation to be augmented in the future. Such automatic tools could give several suggestions for possible fixes of the detected vulnerabilities, or even directly correct them and prove that the modified programs achieve the same purpose as before.

*5.2.1.2. Gas consumption related*. To prevent DoS attacks, Ethereum introduces the gas mechanism to limit the number of operations in a contract. However, this may cause new problems. For instance, it increases users' economic cost since the gas should be paid in advance for the contract execution. There are several tools for optimizing gas consumption, using various methods and techniques. We summarize the relationship among these tools in Figure 12, where the arrows between tools refer to the dependency. The gray names refer to the tools that are not specifically designed for gas consumption.

Chen et al.[103] point out that the non-standard design of smart contracts may lead to unnecessary gas cost. By collecting and analyzing smart contracts on Ethereum, the authors enumerate seven patterns that may cost more gas than expected. Based on symbolic execution (see Section 5.2.1.6), they further develop an analysis tool GASPER that can detect three of them. GASPER accepts EVM bytecode as input and detects three kinds of abuse patterns. The subsequent work[104] further lists 24 gas abuse modes and develops a tool named GasReducer. The input that GasReducer accepts is also EVM bytecode. After code disassembling and pattern matching, it recognizes all 24 abuse patterns. Moreover, it automatically replaces the costly operations with cheaper instructions that accomplish the same functionality. Finally, by recalculation and verification of the optimized codes, GasReducer ensures that the outcome contract still works as expected. GasChecker[110] is another branch from GASPER. GasChecker enriches the gas-wasting patterns from seven to ten

```
contract SimpleDAO { // contract A
  ... // some other contents
  function withdraw(uint amount) public {
    if (credit[msg.sender] >= amount) {
    msg.sender.call.value(amount)(); // A calls
         B
    credit[msg.sender] -= amount;
    }
  }
  function() public payable {
    // fallback function
  }
}

contract Attacker { // contract B
  ...// some other contents
  SimpleDAO public dao = SimpleDAO(addr);
  function sendToDAO() {
    ... //send some value to dao
  }
  function() { // fallback function
    dao.withdraw(amount);
  }
}
```

**Figure 11. Simplified contract example to conduct a re-entrancy attack**

Figure reprinted with permission from Atzei et al.[12] and Rodler et al.[102].

and can detect all these patterns through a parallel version of symbolic execution upon EVM bytecode, which greatly improves the performance when the workforce increases.

Similar to Chen et al.,[103] Marescotti et al.[105] propose two algorithms for calculating the maximum amount of gas that may be consumed by smart contracts, where the latter can be seen as a simplified version of the former. These two algorithms also use the symbolic execution method as Gasper. Different to Gasper, by assuming a one-to-one correspondence between the EVM and Solidity gas consumption path, both algorithms can directly analyze the contracts written in Solidity. However, their solution only calculates gas consumption statistics but does not provide optimization suggestions.

Grech et al.[106] also analyze the gas consumption problem in Ethereum and propose an analysis tool, MadMax. based on Vandal's decompilation technology and logic-driven model (see Section 5.2.1.6). MadMax decompiles EVM bytecode into a control-flow graph and then uses a logic-based specification to detect predefined gas-related vulnerabilities. The authors also give suggestions for developers who write smart contracts with high-level languages, which is not accomplished in Marescotti et al.[105]

By improving and integrating existing tools, Albert et al.[107] propose Gastap. First, they improve the Oyente tool (see Section 5.2.1.6) to generate a control-flow graph with more information. They then improve the EthIR tool[115] (see Section 5.2.1.6) so that it can convert the control-flow graph into a rule-based representation. Based on these two improvements, along with other auxiliary tools for calculation, Gastap finally gets the upper bound of gas consumption of each function in the contract, as is done in Marescotti et al.[105] The Gasol[108] tool further extends Gastap to give out suggestions on the gas consumption optimization, provides an editor plugin for the convenience of programmers, and can automatically optimize the gas-expensive fragments of code (if needed). The automatic optimization is not supported by Gastap. Orthogonal to the techniques used in Gasol, Albert et al.[109] propose another way to find the cheapest alternative instructions for a smart contract through super-optimization[211] and Max-SMT encoding, which takes EVM bytecode as input. They first extract a so-called stack functional specification by symbolic execution, then use the Max-SMT encoding method to feed the optimization task into the SMT-solver, and finally get a set of optimized operations with the same functionality as that of the inputs. The authors also present an open-sourced implementation named syrup.

Gas-related problems have attracted enormous research. The gas consumption is directly relevant to user experience, since everyone needs to pay for it when calling smart contracts. Moreover, the gas consumption may also influence the throughput of blockchain. For instance, there is an upper bound of 8 million gas[212] per block in Ethereum, and when the upper bound is achieved, no transactions will be further included in this block. Therefore, gas-related research also helps to improve the throughput of blockchain.

For future research, improvements on the efficiency and accuracy aspects could be further investigated. As supplementary to prior works, research directions such as gas-wasting patterns and novel analysis techniques with automation are also worth consideration.

*5.2.1.3. Trace vulnerability*. As Nikolić et al.[111] point out, most existing security analysis tools only focus on a single call of a contract but ignore the problems that may occur when called multiple times. For the latter case, they find a new type of vulnerability named trace vulnerability. Contracts containing such vulnerabilities may: (1) be destroyed by any user; (2) be unable to withdraw funds; (3) transfer funds to any address. To fix this problem, the authors propose the Maian tool based on symbolic execution. Maian accepts EVM bytecode as input, along with user-defined analysis targets. This could confirm the existence of trace vulnerabilities.

The security analysis considering multiple calls of a contract is intuitively more complex, and we find that the trace vulnerability is hardly mentioned in other literature. Therefore, we note that future research could be further conducted with other detection techniques for better efficiency and accuracy.

*5.2.1.4. Event-ordering bugs*. The event-ordering bugs evolve from the transaction-ordering dependence described in Luu et al.,[89] where the original concept describes the case when the order of transactions influences the final states in the contracts. Kolluri et al.[112] define the notion of event-ordering bug. Namely, when users call the same function in a contract, the order of calling may lead to inconsistent results among the miners, some of which may be undesirable. To eliminate such risks, the authors combine the methods of symbol execution and fuzzing test and propose the EtherRacer tool, which directly works on EVM bytecode. To achieve better performance and accuracy, optimizations are made to avoid resource explosion during symbolic execution. With the fuzzing test, EtherRacer directly provides a counter-example to intuitively explain the existence of bugs, i.e., two sets of inputs of different orders that result in distinct outputs.

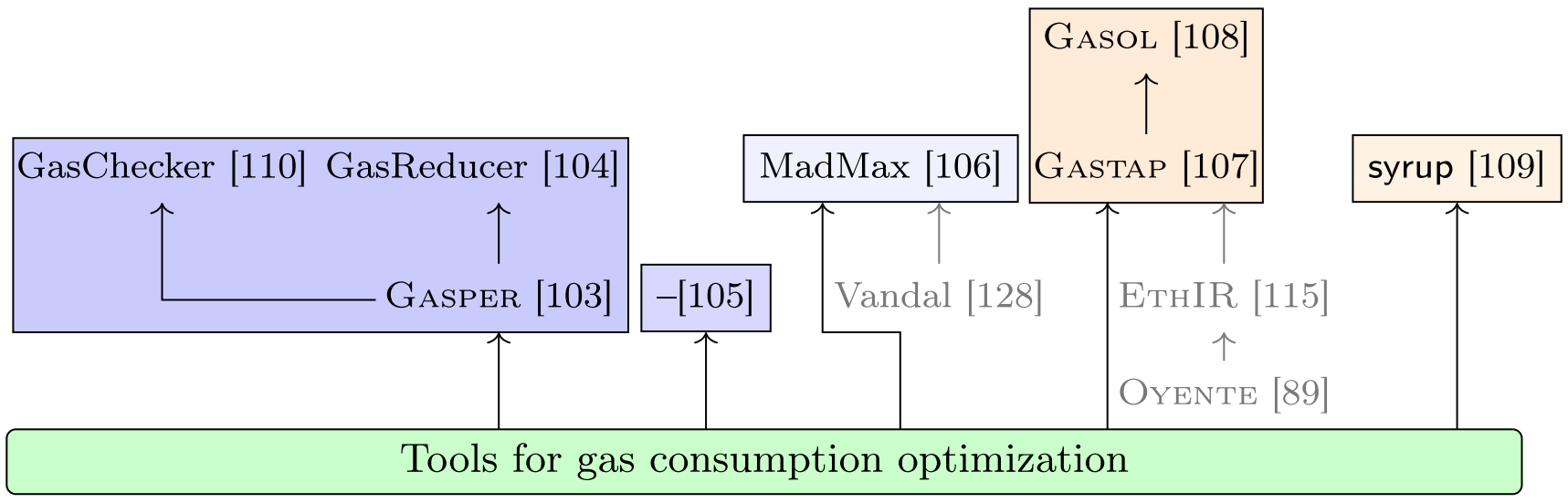

**Figure 12. Relationships among tools for gas consumption optimization**

The event-ordering bugs are among the earliest mentioned vulnerabilities in Ethereum, and they make the state change of smart contracts unpredictable. We remark that there might be applications designed in this way on purpose, such as on-chain games, to obtain randomness. In other words, how to avoid or utilize such event-ordering bugs is worth consideration in the future. Moreover, as indicated by Kolluri et al.,[112] how to avoid the resource-explosion problem during symbolic execution can also be a promising direction.

*5.2.1.5. Integer bugs*. Integer bugs refer to the bugs that are related to integer arithmetic in smart contracts. Torres et al.[113] first sort the integer bugs in smart contracts into: (1) arithmetic bugs that include integer overflow, underflow, and divided by zero; (2) truncation bugs that occur when converting longer integers into shorter ones; and (3) sign-related bugs that occur during the conversion between signed and unsigned integers. They then propose the Osiris tool for these integer bugs, based on symbolic execution and stain analysis methods. Osiris accepts Solidity or EVM bytecode as input. Compared with Zeus[134] and other general tools that detect multiple vulnerabilities (see Section 5.2.1.6), Osiris can detect more integer bugs with the same dataset and has a lower false-positive rate. Torres et al. further point out that Zeus could not detect integer bugs, so the claimed zero false negative is not accurate.

Based on the works above, VeriSmart[114] further improves the integer bug detection's accuracy and efficiency. The authors point out that arithmetic-related vulnerabilities account for more than 90% of the reported vulnerabilities. Moreover, existing tools usually come with inevitable false-positive or false-negative reports, making manual checks necessary. To address these two problems, the authors develop the VeriSmart tool that detects all known arithmetic bugs with almost negligible false alarms. VeriSmart accepts contracts written in Solidity as input. It draws on the idea of the counter-example-guided inductive synthesis (CEGIS) framework,[213] by constantly searching candidate invariants to check whether the contract meets security requirements. VeriSmart manages to avoid expensive operations through a well-designed decision procedure, which also helps to improve the scalability of the tool.

Furthermore, the authors claim that with appropriate improvements VeriSmart can be used to detect other vulnerabilities.

Integer bugs may result in critical consequences, since most smart contracts involve numbers and calculations. Therefore, in this area the detection accuracy is more important, and false positives can be tolerated as long as false negatives are few enough. We mention that improvements in the automation and efficiency aspects are also attractive for future research.

*5.2.1.6. General analysis*. Apart from the tools specifically designed to detect specific vulnerabilities, many tools are designed to detect multiple vulnerabilities at once. We classify 26 general detection tools according to the main technique they adopt. Among them, as the mainstream technology used in software analysis, symbolic execution also occupies a dominant position in the research of security analysis tools of smart contracts. Moreover, the fuzzing test technique is currently attracting more attention in contract analysis. These tools are now introduced in terms of the main technique they adopt.

*5.2.1.6.1. Symbolic execution*. Symbolic execution is commonly used to analyze the security of computer programs. Informally, it uses symbolic values to find out the value or range of the inputs that trigger the execution of each part of the program and then helps to determine whether the program works according to the developer's expectation.

On the one hand, symbolic execution has higher accuracy than other methods such as taint analysis or data-flow analysis. On the other hand, the consumption of memory grows rapidly as the size of the target program grows, which is called a memory explosion. Related research mainly focuses on the accuracy rate, operation efficiency, and calculation cost in the process of analysis.

Oyente[89] is the first symbolic-execution-based tool for smart contract validation. It detects four types of vulnerabilities categorized by the authors. The input of Oyente is EVM bytecode. Oyente is improved in Albert et al.[115] so that the results can be used to generate control-flow graphs and forms the basis of the EthIR[115] tool. The control-flow graph generated by EthIR contains both control-flow and data-flow information of the input EVM bytecode. In addition, EthIR also generates the corresponding rule-based representation for further analysis. Utilizing Oyente and EthIR, Albert et al.[116] designed the tool SAFEVM, which employs the above tools to convert the Solidity program or EVM bytecode into rule-based representation and further converts it into a special C program.[214] Thereafter, existing analysis tools are applied to verify the security of the converted C program.

Mueller[117] combines symbolic execution with other technologies such as SMT-solver and taint analysis and proposes the analysis tool Mythril,[215] which works with a symbolic execution backend LASER-Ethereum.[216] Mythril accepts EVM bytecode as input. Compared with OYENTE, it has better support from the community and is under constant optimization. By the time of writing (August 2020), Mythril has evolved into a security analysis tool supporting smart contracts on various platforms derived from Ethereum and can be used to analyze multiple common bugs and vulnerabilities. Manticore[118] is another widely used and flexible analysis tool based on symbolic execution and satisfiability modulo theories. It supports user-defined analysis by providing several API for the access of the core engine. Moreover, it can infer concrete inputs for a given program state and supports various computer programs in traditional environments (e.g., x86, ARM, and WASM) other than Ethereum.

As mentioned earlier, symbolic execution only works well to analyze short contracts due to the efficiency issues, especially the memory explosion problem. Krupp and Rossow[119] optimize the procedure of symbol execution with the help of control-flow graph. They propose TEETHER, focusing on the automatic detection and utilization of vulnerabilities in smart contracts. According to the EVM bytecode input, TEETHER generates a control-flow graph and sorts the critical paths related to the fund transfer. Then, by constraint solving, symbolic execution results are used to enumerate possible attacks against these critical paths. In this way, TEETHER provides the witness of detected vulnerabilities while making the detection more automated. Chang et al.[120] adopt a similar idea of partial analysis and propose the tool sCompile, which only works on the critical part of a smart contract involving fund transfer. Given the EVM bytecode, sCompile first generates the corresponding control-flow graph and then analyzes whether the transfer-related path meets the predefined security properties. Paths are ranked according to the results and some predefined rules. Thereafter, sCompile performs the symbolic execution on the higher-ranked paths and finally generates an analysis report. This partial analysis solution improves the scalability of symbolic execution.

To solve the same scalability problem, Feng et al.[121] propose the idea of summary-based symbolic evaluation and the corresponding tool SMARTSCOPY. SMARTSCOPY not only supports analyzing large contracts efficiently but also generates counter-example attacks for the detected vulnerabilities. To reduce the space and time overhead for large contracts, SMARTSCOPY symbolically evaluates the methods indicated by the application binary interface (ABI)[217] of contracts, and summarizes the impact of each method on the blockchain. It then conducts the range splitting and pruning procedure and finally gets to the symbolic execution. The authors argue that SMARTSCOPY can detect the newly defined BatchOverflow bugs that other tools previously overlooked.

In the aforementioned solutions for scaling symbolic execution,[119–121] less important paths are skipped or ignored during the analysis. Although this improves efficiency, it may lead to false-negative results. Tsankov et al.[122] propose an analysis tool SECURIFY, which combines abstract interpretation with symbolic execution. SECURIFY guarantees to traverse all possible paths in a contract, reducing the false-negative results caused by the incomplete symbolic execution. Its input is EVM bytecode, along with a security model defined by a domain-specific language. Through steps of decompiling, semantic fact inferring, and security pattern checking, it determines whether a contract meets the predefined properties in the security model. Moreover, the security models in SECURIFY are separate from the analysis tool itself. Therefore, by optimizing the security model, the accuracy of vulnerability detection could be further improved.

From another point of view, to improve the efficiency of analysis, VERX[123] adopts the concept of delayed predicate abstraction. Its main idea is to combine symbolic execution and abstraction methods: symbolic execution is used in the individual execution of transactions while abstraction is conducted between transactions. The delayed abstract process reduces the infinite state space brought by unlimited transactions to limited space. VERX takes contracts written in Solidity as input, along with the security requirement, and it outputs the predicate of whether a contract meets the given properties.

Symbolic execution is one of the most popular analysis techniques for smart contracts. Problems such as resource explosion and low efficiency are discussed in many studies. This technique can be utilized to detect multiple vulnerabilities within one invocation. Besides, there are six existing tools that support the analysis of user-defined properties. It dramatically improves the flexibility of security analysis. Combined with the constantly updating vulnerabilities and bugs in Section 5.1.2.3, future research may focus on making analysis tools compatible with newly found vulnerabilities. A general compiler is also needed to convert the descriptions of vulnerabilities into some languages that the analysis tools could recognize.

*5.2.1.6.2. Syntactical analysis*. Syntactical analysis is a method to analyze computer programs by parsing them into a tree and analyzing the relation of each component.

Tikhomirov et al.[124] first summarize four potential problems in the smart contract programming process: (1) security-related issues; (2) function-related issues; (3) execution-related issues; and (4) development-related issues. They design a static analysis tool, SmartCheck, which can find such problems. SmartCheck converts the Solidity source code into an XML parse tree,[218] then uses XPath queries[219] to find the matched patterns. As the authors point out, SmartCheck cannot guarantee accuracy or work without manual checks. Nevertheless, this method provides an efficient way to detect potential vulnerabilities.

Later in 2019, Lu et al.[125] proposed the tool NeuCheck, which manages to improve the processing speed and complement several detected vulnerabilities that are not fully covered by the existing tools (e.g., SmartCheck). NeuCheck takes Solidity source code as input, parses it into an XML parse tree by a syntactical analyzer, and utilizes an open-sourced library to complete the analysis.

The syntactical analysis is relatively lightweight and efficient compared with symbolic executions, although the accuracy cannot be guaranteed. Future research could improve accuracy by enriching the syntax patterns. Automation is also needed and could be realized by offering possible suggestions for contract correction.

*5.2.1.6.3. Abstract interpretation*. The basic idea of abstract interpretation is to verify whether a program meets certain specific properties according to the approximation of the program's semantics. Related research mainly focuses on the tool's usability and accessibility, and abstract interpretation in these studies usually comes with other tools, such as Horn clause resolution and control-flow graph.

Based on their work[90] that defines the full semantics of EVM bytecode, Grishchenko et al.[126] propose the tool EtherTrust, which first abstracts the EVM bytecode as a series of Horn clauses and then uses the resolution of such clauses to verify the contract's reachability. The authors also give a security analysis of the reliability of this tool. Moreover, they clarify the mechanism and details of EtherTrust in their later work.[127]

After the birth of EtherTrust, Schneidewind et al.[130] claimed that all existing tools such as EtherTrust,[127] Zeus,[134] NeuCheck,[125] and MadMax[106] all fail to provide provable soundness as contract analyzers. From this point of view, they propose eThor, an analyzer that is provably sound and tested to be practical through a large-scale experimental implementation. The soundness is proved against the semantic of EVM bytecode defined by Grishchenko et al.[90] This tool can verify user-defined properties described by the *HoRSt* language, which reduces the difficulty of defining Horn clause-based abstractions of the contract properties.

Vandal[128] is an analysis framework with the idea of abstract interpretation. It creatively converts (decompiles) the input EVM bytecode into a logical relationship and uses the logic-driven methods to verify the correctness and security of such a logical relationship. Using this framework, users can easily define security requirements and conduct security analysis. The authors of Vandal also compare Vandal's performance with the aforementioned Oyente[89] and Mythril[215] in their work. They show that Vandal and Mythril find more types of vulnerabilities than Oyente, and Valdal is the most efficient among these three. Based on the Vandal framework, Grech et al.[106] further propose the Madmax tool to find gas-related vulnerabilities in Ethereum contracts.

Grech et al.[129] extend Vandal and propose Gigahorse, which outperforms Vandal in both precision and completeness, capable of decompiling over 99.98% of deployed contracts compared with Vandal's 88%. Gigahorse firstly decompiles EVM bytecode into a new intermediate representation, which specifies the data- and control-flow of the input bytecode. It then infers the functions and variables through a set of carefully designed heuristics. The authors mention that Gigahorse could be a better dependency for the previously discussed Madmax.[106]

As indicated by Schneidewind et al.,[130] there is a trend that an analysis tool should carry it with a formal proof of its performance (e.g., soundness and completeness). In abstract interpretation, the smart contracts are parsed into formal descriptions, and it is especially suitable to generate formal proofs based on these descriptions. We remark that this might be a promising future research direction.

*5.2.1.6.4. Data-flow analysis*. In data-flow analysis, the variables' runtime information is collected to check whether a program meets the expected property during the execution. Slither[131] is an analysis tool based on this method. It can automatically detect vulnerabilities and help developers to have a better understanding of the developed smart contracts. Slither converts a Solidity contract to a control-flow graph and further compiles it to an intermediate representation named SlithIR. Utilizing data-flow analysis and taint analysis techniques, the tool can analyze large contracts infeasible for symbolic execution.

As mentioned above, the data-flow analysis consumes fewer resources and thus supports large contracts. This is an attractive property, since large smart contracts are more probably to suffer from vulnerabilities, and the overhead for the analysis of such contracts directly influences the efficiency of contract development. We remark that data-flow analysis is combined with other technologies in several tools,[115,129] and more combinations could be expected in the future to improve the efficiency of vulnerability detection.

*5.2.1.6.5. Topological analysis*. Topological analysis for smart contracts is mainly based on the topological structure graphs that illustrate the relations among multiple smart contracts. Using the Solidity-parser,[220] Zhou et al.[132] analyze the calling and dependency relationships in and between contracts, based on the Solidity source code. They propose a method to form a topological graph for developers to analyze the structure of their contracts. In addition this method also uses symbolic execution, syntax analysis, and other methods to find potential logical vulnerabilities in a contract.

The topological analysis is usually combined with other methods to obtain the final results, such as syntax analysis and symbolic execution.[132] Zhou et al.[132] mention that future research could be focused on the automatic generation of test cases that are used to indicate the topological relationships among functions and contracts, and how to improve the accuracy and alleviate the burden of manual check should be addressed in the future.

*5.2.1.6.6. Model checking*. Model checking is a method that verifies whether a system meets certain properties by modeling it into a finite state machine. Regarding the security analysis of smart contracts, related studies mainly focus on model construction, accuracy, and efficiency.

On model construction, Nehai et al.[133] propose a model-checking-based solution to check the security of contracts. They build a three-layer model for Ethereum smart contracts: the kernel layer, application layer, and environment layer. These three layers correspond to the blockchain, smart contract, and execution environment, respectively. They propose a method that compiles the Solidity source code into the NuSMV input language,[221] which is friendly to model checking. Finally, they check whether the implementation meets the given user-defined security properties. If not, the model-checking approach would return a counter-example, which helps refine the implementation.

On the aspects of accuracy and efficiency, Kalra et al.[134] combine the methods of symbolic model checking, abstract interpretation, and constrained horn clauses, and propose the Zeus tool. They claim that the tool achieves zero false-negative results, with a lower false-positive rate than existing tools such as Oyente. In addition, Zeus works faster than Oyente. Theoretically, Zeus accepts smart contracts written in various high-level languages as input (e.g., Solidity, C#, Go, and Java), and thereby can be extended to support platforms other than Ethereum, such as Hyperledger Fabric.[7] It converts a smart contract into an intermediate representation by a specially designed compiler and inserts checkpoints into the intermediate representation according to user-defined rules. Finally, with verification tools based on the constrained horn clauses, Zeus verifies the security of smart contracts (Definition 13).

In addition to the procedure of model checking itself, several auxiliary components such as contract model, intermediate representation, and corresponding compilers are also worth consideration in future research.

*5.2.1.6.7. Deductive proof*. The aforementioned model checking only works on small-scale contracts, as the number of states in the model grows with the contract size. Nehai and Bobot[135] propose a deductive proof method, where they compile the Solidity contracts into programs in Why3 language.[209] They then use the Hall-logic-based detection tools brought by Why3 to analyze the properties of a contract. They also provide a compiler that compiles the contracts written in Why3 into EVM bytecode so that developers can directly write contracts in Why3.

The scheme discussed above is an implementation of Why3, a widely used programming language designed for provable secure programs. We remark that in the future, other languages and tools with attractive properties such as provability could be adopted in this area to provide analysis alternatives.

*5.2.1.6.8. Satisfiability modulo theories*. Satisfiability modulo theories (SMT) involve a formula whereby the parameters are functions or predicate symbols, and the goal of SMT is to determine the satisfiability of this formula. It usually comes with other methods, serving as an auxiliary method. However, Alt and Reitwießner[136] have used it as the main technology to complete the security analysis. They assert that SMT can be directly integrated into the Solidity compiler to enable users to conduct security analysis while compiling and meanwhile provide counter-examples to the vulnerabilities. However, the authors only provide the idea and simple use cases but do not provide a complete plan or implementation of this idea.

SMT is relatively lightweight and often used along with other techniques,[109,117,118] and the utilization of this technique alone is less touched. For future research, the utilization of SMT could be future considered to support the analysis of large-scale contracts, and combinations of SMT and other techniques are also worth consideration to achieve better performance (either accuracy or efficiency).

*5.2.1.6.9. Fuzzing test*. The fuzzing test has become a prevalent technique for bug detection in recent years. The core idea is to generate random data as input and monitor the abnormal behaviors of the target program under these inputs. A large number of random inputs are used. Bugs unreachable in the normal cases could be found by random collisions in this way. Existing research on smart contract analysis mainly focuses on the completeness and efficiency of the analyzing process.

ContractFuzzer[137] is a typical example based on the fuzzing test to detect multiple types of vulnerabilities in smart contracts. It first generates the random inputs according to the ABI of a smart contract and then records the execution results of these inputs. Thereafter it performs security analysis with predefined test oracles, which describe the characteristics of specific vulnerabilities. Evaluation results show that ContractFuzzer has a lower false-positive rate than Oyente, with a higher false-negative rate under certain circumstances.

Regarding the completeness and efficiency of the fuzzing test, He et al.[138] argue that tools such as ContractFuzzer[137] could not reach some paths in depth, thereby failing to find related vulnerabilities and causing false-negative results. On the other hand, tools based on symbolic execution can reach deep paths but consume enormous resources. Combining these two methods, the authors propose the concept of Imitation Learning-based Fuzzer (ILF), which learns the procedure of the symbolic execution-based tools, imitating the behavior of the symbolic execution paradigm. Thereafter, the test set is generated for the Solidity contract. In this way, the fuzzing test can be used to efficiently find more vulnerabilities.

To support user-defined analysis, the tool Echidna[139] is proposed. It checks user-defined properties and assertions, and further estimates the worst-case gas consumption for the contracts. Echidna accepts contracts written in Solidity and Vyper[210] and is adopted by the auditing service in Groce et al.[94]

The fuzzing test technique has become prevalent in recent years. However, its accuracy and soundness cannot be proved, since it relies on a random collision to find the unexpected behaviors. We remark that the generation of counter-examples could be considered in the future, and suggestions for corrections are also attractive for efficient programming.

#### 5.2.2. Auxiliary tools

Apart from the analysis tools discussed above, there are many other tools for auxiliary purposes, such as frameworks, languages, and basic tools. The "framework" refers to a set of tools or schemes that simplify or facilitate the development of smart contracts. "Language" indicates the new smart contract languages, such as high-level languages with high expressivity (e.g., Solidity) and intermediate representations used during the process of compilation or analysis. "Basic tool" refers to the basic tool for other high-level tools (e.g., the analysis tools discussed in Section 5.2.1, and the high-level languages and frameworks in Section 5.2.2.1 and Section 5.2.2.2). Such studies are much more fundamental and could not be applied to the development procedure directly.

Table 6 summarizes the auxiliary tools discussed in this paper, where the ● (resp. ○) in the "Open-sourced" column represents that the corresponding tool is open-sourced (resp. not). When there are multiple references on the same line, there will be multiple ● or ○ in the same order.

*5.2.2.1. Frameworks*. Frameworks refer to the auxiliary tools available during the contract construction process, simplifying the development of smart contracts. We have collected ten studies on contract frameworks. Some of them help developers achieve the privacy goal or analyze the security properties of a contract. Others provide developers with simpler tools or familiar languages, reducing learning costs and enabling a quick start for beginners. Note that these goals are orthogonal and might be achieved simultaneously by a single scheme.

The public information on the blockchain causes privacy issues during the executions of smart contracts. Kosba et al.[140] propose the Hawk framework. Developers can write Hawk contracts and set privacy portion $\varphi_{priv}$ and public portion $\varphi_{pub}$ in the contract, where $\varphi_{priv}$ helps to hide the private input of users while $\varphi_{pub}$ refers to the data that are allowed to be publicly disclosed. Accordingly, a standard contract will be generated automatically in the framework, along with cryptographic protocols that ensure the correctness of the contract (Definition 14) and the privacy of the users. Therefore, developers with little knowledge of the complex cryptographic schemes can efficiently construct a privacy-preserving smart contract. Hawk implements zero-knowledge succinct non-interactive argument of proof (zk-SNARK)[223,224] and other cryptographic schemes to ensure the privacy of the smart contract. The authors also give a security proof within the UC model.[197] However, Hawk has not yet been open-sourced (up to August 2020). Eberhardt and Tai[141] adopt a similar idea to describe off-chain computation using zk-SNARK and propose a toolbox named ZoKrates. ZoKrates includes a special high-level language to describe off-chain computations and a compiler that converts the contract to a ZKP protocol. Compared with Hawk this toolbox is open-sourced, but a formal security proof is absent.

**Table 6. Auxiliary tools for smart contract construction**

| Refs. | Year | Type | Main contributions | Open-sourced[a] |
|---|---|---|---|---|
| [140] | 2016 | Framework | Hawk framework for writing and compiling privacy-preserving smart contracts | ○ |
| [141] | 2018 | Framework | ZoKrates framework for writing and executing off-chain contracts | ● |
| [142] | 2016 | Framework | a framework for verifying smart contracts by translating both Solidity and EVM bytecode to F$^\star$ | ○ |
| [143] | 2018 | framework and high-level language | a quantitative game-theoretic framework for analysis and a corresponding contract programming language. | ○ |
| [144] | 2018 | Framework | FSolidM framework modeling smart contracts as finite state machines, and a GUI for creating contracts | ● |
| [145] | 2019 | Framework | VeriSolid framework extending FSolidM, supporting specifying and verifying the desired properties of a contract | ● |
| [146] | 2019 | Framework | a framework defining business smart contracts in the state machine model, with a programming called TLA+[222] | ○ |
| [147] | 2019 | Framework | a framework using the Event-B formal modeling for easier verification | ○ |
| [148] | 2019 | Framework | Takamaka framework for writing and executing smart contracts in Java | ○ |
| [149] | 2019 | Framework | FEther framework for verifying smart contracts in Coq | ○ |
| [150] | 2016 | high-level language | a functional programming language (Idris) library for smart contracts | ● |
| [151] | 2017 | high-level language | Findel, a domain-specific language for financial purposes | ● |
| [152] | 2018 | high-level language | Lolisa, a (nearly) alternative language for Solidity with stronger type system, designed to facilitate symbolic execution and formal proof in Coq[201] | ○ |
| [153] | 2018 | high-level language | Flint, an inherently safer language for smart contracts with additional restrictions and mechanisms | ● |
| [154] | 2019 | high-level language | Featherweight Solidity, a calculus for Solidity, supporting precise definition of the behavior of smart contracts | ○ |
| [155,156] | 2018–2019 | intermediate representation language | Scilla, an intermediate language suitable for formal analysis and verification | ● ● |
| [157] | 2017 | basic tools | a formal definition of EVM in Lem language, which can be combined with multiple theorem provers to verify smart contracts | ● |
| [158] | 2018 | basic tools | an extension of Hirai[157] with a sound program logic at the bytecode level | ● |
| [90] | 2018 | basic tools | a semantic framework of EVM bytecode and its formalization in F$^\star$ | ○ |
| [159] | 2018 | basic tools | a formal definition of EVM using $\mathbb{K}$ framework, which can be combined with multiple theorem provers to verify smart contracts | ● |

[a]● (resp. ○) here represents that the corresponding tool is open-sourced (resp. not). When there are multiple references on the same line, there will be multiple ● or ○ in the same order.

Other than frameworks that simplify the design of smart contracts, there are also frameworks aimed at facilitating the security analysis, based on verifiable languages, game theory, finite state machine model, and other techniques. Smart contracts under these frameworks are more suitable for security analysis, making the analysis results more accurate and convincing, as discussed in the following.

F$^\star$[225] is a specially designed language targeted for security analysis. Bhargavan et al.[142] propose a framework that translates smart contracts into F$^\star$ to further conduct the security analysis. Concretely, it compiles (resp. decompiles) the Solidity code (resp. EVM bytecode) into F$^\star$ and evaluates the equivalence between these two results. Such equivalence reveals the correctness of the functionality (on the source code layer) and the runtime security (on the bytecode layer). However, the authors only provide several simple examples, and a complete scheme for the compilation and decompilation is not provided.

Smart contracts usually involve multiple parties with conflicting interests, whereby game theory works well. Chatterjee et al.[143] describe smart contracts as a two-party game and further propose a quantitative stateful game-theoretic framework. The core technique is the refinement of abstraction, which is used to avoid state space explosion during the modeling of two-party games. A simplified contract language without loop instructions is also proposed to support concurrent instructions of the parties. It is used in the game-theoretic model, and the authors claim that it can be translated into Solidity. However, the corresponding compiler is not provided in their work.

Besides, the state machine can also be applied to describe the state transitions in smart contracts. Mavridou and Laszka[144] regard the smart contracts as finite state machines and propose the FSolidM framework to design smart contracts efficiently. Besides, they provide four plugins that can be used to detect known vulnerabilities (e.g., re-entrancy and transaction-ordering dependence) on the generated smart contracts. In their later work[145] the authors propose the VeriSolid framework, which extends the supported Solidity expressions and updates the code generator. Formal operational semantics are also provided in the latter work. VeriSolid enables developers to describe the security requirements and verifies whether the generated contract meets such targets. Xu and Fink[146] also apply the state machine model on smart contracts but focus on the smart contracts in the business field. They propose the Temporal Logic of Actions (TLA) model, written in TLA+ language,[222] to describe the properties that a contract must satisfy. Finally, the TLC (Temporal Logic model Checker)[146] is also proposed to ensure that the contract meets the user-defined properties. However, the state machine model may fail to describe certain smart contracts because of the space explosion problem, and this could be a research direction for future optimization.

Apart from those popular and well-known models, there are another two models useful in contract designing frameworks. Banach[147] proposes a framework using the Event-B model,[226] which is designed to describe and verify the discrete event system. With this model and framework, smart contracts can meet specific properties in the designing layer, and in turn the security analysis becomes relatively easier. However, this work needs further improvement on completeness and syntactic complexity to be brought into practice.

There are also studies introducing general programming languages into the context of smart contracts. Spoto[148] proposes the framework Takamaka that implements Java as the programming language. In Takamaka, Java programmers can easily develop smart contracts with familiar tools. Takamaka specially designs a Storage class for smart contracts and a gas-computing mechanism related to Ethereum. To avoid malicious programs or abused functions in Java, it also maintains a whitelist of permitted functions for smart contracts. The framework also allows clients to use Java Virtual Machine (JVM) to run smart contract bytecode, taking advantage of JVM's excellent bytecode execution rate and garbage collection mechanism. However, this also makes it incompatible with the state-of-the-art Ethereum. In other words, it requires a hard fork to take effect. Moreover, the framework is still under development and not yet released.

Frameworks for the construction of special smart contracts are quite useful and may relieve programmers of security concerns. The state-of-the-art solutions usually include new languages and compilers that simplify the design procedure or facilitate security analysis. However, some existing frameworks are in the proof-of-concept stage only while others may lack formal proofs on their security or privacy abilities. We note that with a formal language and other rigorous methods such as cryptographic schemes, a framework with certain provable properties may be introduced in the future.

*5.2.2.2. Contract languages*. Researchers for contract language are devoted to improving the security of smart contracts (Definition 13), including avoiding common errors and making it more suited for security analysis. Specifically, four schemes restrict the language's functionalities so that errors will be less probable to occur. Another four studies focus on the semantic abstraction and (or) modifications on the original language to facilitate vulnerability detection. In this section, we summarize the contract languages proposed in these studies.

Most languages for contract programming, e.g., Solidity and Serpent, are used for procedural (or imperative) programming, where the states in the contract are changed by multiple statements, while in functional programming programs are composed of multiple functions. Pettersson and Edström[150] argue that functional programming can help to avoid many common errors during construction. Besides, it is easier to apply detection tools in contracts written in functional languages. Accordingly, the authors propose an Idris[227] library (Idris is a functional programming language), taking advantage of its dependent type. They further design a compiler from Idris to Serpent to illustrate the feasibility of their conclusion.

In addition to functional programming languages, Biryukov et al.[151] propose a domain-specific language to improve the security of contracts under specific scenarios. Findel[151] is a language for financial contracts. By separating the contract description and execution methods, financial contract developers can simply pay attention to the contract contents without concerning about how the contracts are executed. The authors argue that Findel can be used in common financial derivatives, and they conduct a test of this language on Ethereum.

Lolisa[152] is an alternative language that adopts a stronger static type system aimed at making the contract more suitable for security analysis (particularly symbolic execution). For the specification of Loisa, a formalization of the syntax and semantics of Solidity is proposed by Yang and Lei.[152] It is claimed that Lolisa includes nearly all expressions of Solidity, and thus Lolisa and Solidity could be converted to each other easily. Moreover, it is more convenient to do symbolic execution on contracts written in Lolisa. The authors also propose an interpreter that converts Lolisa to Coq[201], a specially designed language for semantic verification. Taking Lolisa as the intermediary, the security of smart contracts in Solidity can be easily analyzed and proved with the help of Coq. The subsequent work FEther[149] illustrates the detail of the verification of smart contracts with the combination of Lolisa and Coq. A debugger is also provided for programmers to debug target contracts in Coq.

Based on a similar idea to Lolisa, Crafa et al.[154] define the formal semantics in the core of Solidity and propose a high-level language named Featherweight Solidity. The behavior of the Solidity contract could be accurately defined with it. In this way, analysis tools can directly analyze contracts in Solidity without converting them to EVM bytecode. In addition, the analysis results show that there are defects in the type system of Solidity. Specifically, they conclude that runtime type errors are unavoidable in current systems.

Apart from the semantic abstraction and definition of the original language, other studies focus on the design of intermediate languages that are useful in contract compilation and verification. Sergey et al.[155,156] point out that most high-level languages sacrifice verifiability to improve their expressivity. In other words, the security verification of contracts in these languages is more complex than those in low-level languages (e.g., bytecode). To address this problem, they propose an intermediate language Scilla. High-level languages can be compiled into Scilla for security analysis before being further compiled into EVM bytecode. Its design is based on the automaton model, and it separates the modules of communication, calculation, and state transition. Such modulation enables the analysis tool to work in a focused manner. The contract compiled into Scilla will be further converted to Coq[201] to utilize the powerful analysis function of Coq. However, the scheme has not yet been fully implemented.

Adding security mechanisms to the original programming language also helps avoid potential security issues since Solidity only focuses on the expressivity and completeness aspects but does not consider the convenience of security analysis. Schrans et al.[153] propose a high-level language, Flint, which: (1) adds a permission mechanism to limit undesirable function calls; (2) optimizes the fund-related operations to ensure safe transfer; (3) introduces an unmodifiable property to limit the modification of key states; and (4) uses the same application binary interfaces as Solidity to ensure that contracts written in these two languages can interact with each other. These features reduce the difficulty of designing contracts and corresponding analysis tools. The authors also provide an analysis tool that performs syntax analysis on Flint contracts and a compiler that compiles the contract into EVM bytecode.

As we have mentioned in Section 3.1 and Section 4.2.2, a key difference between Bitcoin and Ethereum is their support for high-level languages. An expressive contract language could greatly save learning efforts for new programmers. On the other hand, several schemes aim to facilitate the security analysis of smart contracts through some intermediate representations and tools. All of these languages require a proper compiler to generate EVM bytecode correctly. We remark that a formal proof of the correct compilation could be considered to illustrate the soundness and completeness for future research. Moreover, user-friendly interfaces are also important to facilitate the development of smart contracts.

*5.2.2.3. Basic tools*. Basic tools refer to certain basic theories used to develop higher-level tools (e.g., security analysis tools, high-level languages, frameworks). Such research usually involves the formal definition of the underlying infrastructure such as virtual machine and bytecode.

There are three studies that adopt different tools to define EVM formally. Such formal definitions are the key to designing contract analysis tools. Hirai[157] gives a formal definition of EVM in the Lem language that can be combined with analysis tools (or framework), such as Coq[201] and Isabelle/HOL,[228] to check EVM's property. The author further proves some security properties of EVM with Isabelle/HOL. Amani et al.[158] extend Hirai's work and propose the sound program logic for EVM so that the correctness of a smart contract can be verified with Isabelle/HOL. They also prove the correctness of their program logic in their work. Based on Hirai,[157] Hildenbrandt et al.[159] propose KEVM, another formal definition of EVM under the $\mathbb{K}$ framework.[229] KEVM is unambiguous, readable, and executable, which can be used as a theoretical foundation for formal analysis of smart contracts. As an example, the authors briefly describe a KEVM-based gas analysis tool and a domain-specific language for analyzing the application binary interfaces of smart contracts. Furthermore, with specially designed interpreters, KEVM passes the official test suite[230] provided by the Ethereum community. The results show that KEVM is more efficient than the EVM defined by Hirai[157] and can detect more vulnerabilities.

There are also studies formally defining EVM bytecode, which contribute to designing high-level languages and analysis tools. Grishchenko et al.[90] argue that the semantics in Ethereum Yellow Paper[3] are incomplete and do not follow the standard rules of definition. They are the first to define the small-step semantics of EVM bytecode. They use the same $F^{\star}$ language as the preceding work,[89] and their complete semantic framework of EVM serves as the theoretic foundation of $F^{\star}$-based analysis tools.

As we have observed, there are six studies giving formal definitions of EVM and bytecode semantics utilizing different specification languages. However, the definitions of EVM and bytecode semantics are illustrated independently, and their relationship and correspondence are not described formally. Therefore, for future research, composing these definitions or universally defining these basic tools to support the development of higher-level theories could be further investigated.

## 6. EXECUTING SMART CONTRACTS

In 2017, Buterin[231] gave a speech on the design challenges of smart contract mechanisms, focusing on the security and application aspects. In this section, we list and group the studies on execution mechanisms of smart contracts, most of which aim at improving privacy and performance. Before introducing these studies, we first summarize the deficiencies in existing contract execution mechanisms in the following, taking Bitcoin and Ethereum as representatives. Such defects may hinder several useful applications from being adopted, which are sensitive to the delay and privacy of on-chain transactions. We remark that these deficiencies mainly occur in the public blockchains (Definition 1), while if these issues are handled properly both public and consortium blockchains will benefit.

(1) *Forced disclosure of smart contract contents*. Although pseudonyms are used in most blockchain systems to protect users' privacy, due to the inherent execution mechanisms of blockchains the contents of smart contracts are still forced to be disclosed in public so that miners can execute the contracts and all nodes can agree on the final states. To fully guarantee anonymity, some blockchain systems even directly remove the smart contract functionality, e.g., Zerocash,[232] But this is not an ideal strategy, since smart contracts are quite attractive for several application scenarios if the privacy concerns are eliminated.
(2) *Low processing rate*. In most blockchains a transaction must be verified, executed, and packaged by all miners

before taking effect, but this duplicated and resource-consuming strategy puts huge limitations on the processing rate. Technically, the processing rate is subject to basic parameters of a blockchain, such as the block size and time interval between blocks. Although it is not difficult to modify these parameters, new problems[233] may arise afterward. For example, a bigger block size would consume more space, and a smaller interval may result in frequent and undesirable forks. Therefore, existing smart contract platforms are not suitable for applications with high demands on the processing rate.

(3) *Limited contract complexity*. To ensure the liveness (Definition 4) of a blockchain, mechanisms to prevent DoS attacks are introduced. For instance, Bitcoin only supports a limited set of operations, which essentially prevents the occurrence of endless loops. Ethereum introduces the gas mechanism that limits the number of instructions in a single transaction. On the one hand, these protective mechanisms guarantee the liveness while on the other hand, they also limit the possibility of executing complex smart contracts on-chain. In other words, the applications with sophisticated operations and heavy overheads are not applicable in the existing blockchains.

As mentioned in Section 4, script-based blockchains are only suitable for implementations of simple financial-related contracts. Such systems are used more as a public ledger than a smart contract platform in practice. Therefore, most related studies focus on the improvement of Turing-complete blockchains and are divided into three categories as follows.

(1) *Private contracts with extra tools*. To avoid revealing sensitive information in a contract, auxiliary tools such as ZKP (Definition 16) and TEE (Definition 17) are adopted in 13 studies. These schemes are aimed at solving the forced content disclosure problem mentioned above, and at the same time may help to mitigate the problems of processing rate and complexity, as we discuss in Section 6.1.
(2) *Off-chain channels*. In these schemes, the executions of contracts are moved off-chain to avoid the low transaction-processing rate and high confirmation delay. Meanwhile, such methods also protect the privacy of contract contents to some extent. We group the existing ten off-chain schemes into payment channels and state channels in this review, as we discuss in Section 6.2.
(3) *Extensions on core functionalities*. The core functionalities here refer to the way of processing and executing smart contracts in a blockchain. The improvements on such functionalities usually require a (hard) fork (Definition 6), or even an alternative blockchain would be proposed, either increasing the functionalities a smart contract could achieve or improving the performance of the blockchain. Some are less relevant to smart contracts, e.g., SegWit,[234] side-chain,[235,236] and Sharding.[237] Such schemes are omitted in this paper, and we focus on the nine improvements and extensions on contract execution mechanisms.

We summarize the schemes discussed in this section in Table 7 according to the above classification. In the "Theory" column, ● denotes that the scheme has both description and its corresponding security proof while ◐ denotes that the scheme has only description without a security proof; in the "Implementation" column, ● means that the scheme is implemented and open-sourced while ◐ means that the scheme is implemented but not open-sourced, and ○ denotes that the scheme is not implemented. We remark that the smart contracts discussed below refer to those written in Turing-complete languages if not otherwise claimed.

## 6.1. Private contracts with extra tools

In most mainstream public blockchains, smart contracts are stored on-chain so that everyone can see and validate their content. In other words, to process a transaction that calls a smart contract, all miners will operate accordingly to the transaction and contract and finally agree on the execution results. Such a mechanism may cause privacy issues, e.g., a business company may be curious about its competitor's daily sales and big orders. This will prevent the implementation of certain business contracts.

To handle the privacy issues in smart contracts, cryptographic schemes or hardware tools are introduced, and we call such solutions private contracts with extra tools. In the following, we group them into private contracts based on SMPC (Definition 15), ZKP (Definition 16) and TEE (Definition 17), and discuss them in Section 6.1.1, Section 6.1.2, and Section 6.1.3, respectively.

### *6.1.1. Secure multi-party computation*

The goal of SMPC (Definition 15) is similar to that of smart contracts, both involving multiple parties that do not trust each other and generate correct execution results. To some extent, SMPC could be viewed as a special form of smart contracts, apart from the fact that SMPC is almost conducted off-chain. When combined with smart contracts, SMPC could improve the privacy of contract contents and alleviate the problems caused by the high latency and low throughput in a blockchain. Related studies focus on the privacy, fairness, and correctness of blockchain-based SMPC, as in the following.

Some researchers are dedicated to improving data privacy with the help of SMPC. Enigma[160] is a privacy-preserving computation platform without a trusted third party. It records the hash of crucial data on-chain to guarantee integrity and utilizes SMPC to conduct private computation. In this way, no one gets any extra information except its inputs and outputs, according to the privacy of SMPC.

In SMPC, fairness is another non-negligible problem. Choudhuri et al.[161] regard blockchain as a tamper-resistant public bulletin board that anyone can write on. They adopt the witness encryption (WE) method to solve the fairness problem in SMPC. According to their scheme, all participants will get their desired outputs, or no one gets them at the end of the protocol.

From another aspect, Sánchez[162] combines SMPC with proof-carrying code[238] and proposes Raziel to guarantee the correctness of smart contract codes and their executions. In Raziel, SMPC guarantees the correctness and privacy of contract executions, and proof-carrying code ensures the correctness and verifiability of the contract codes. Raziel also adopts non-interactive ZKP to prevent the proof-of-code from revealing extra information.

**Table 7. Execution schemes of smart contracts**

| Classification | | Refs. | Scheme name[a] | Year | Keywords | Theory[b] | Realization[c] |
|---|---|---|---|---|---|---|---|
| Smart contracts with auxiliary tools | secure multi-party computation | 160 | Enigma | 2015 | blockchain-based multi-party computation | ◐ | ● |
| | | 161 | – | 2017 | fairness of multi-party computation | ● | ◐ |
| | | 162 | Raziel | 2018 | private contracts through multi-party computation | ● | ○ |
| | zero-knowledge proof | 6 | Quorum | 2016 | public and private contracts | ◐ | ● |
| | | 140 | Hawk | 2016 | integrating zk-SNARK into contracts | ● | ○ |
| | | 141 | ZoKrates | 2018 | zk-SNARK toolkit for private contracts | ◐ | ● |
| | trusted execution environment | 161 | – | 2017 | multi-party computation through TEE | ● | ◐ |
| | | 163 | – | 2018 | contract privacy in Hyperledger Fabric | ● | ◐ |
| | | 164 | PDO | 2018 | private data object | ● | ● |
| | | 165 | Ekiden | 2019 | general private contracts based on TEE | ● | ◐ |
| | | 166 | FastKitten | 2019 | complex Bitcoin contracts | ● | ◐ |
| | | 167 | ELI | 2019 | Enclave-Ledger interaction | ● | ● |
| | | 168 | Teechain | 2019 | settlement of Lightning Network | ● | ● |
| Off-chain channels | payment channel network | 169 | – | 2015 | duplex payment channel network (BTC) | ◐ | ○ |
| | | 25 | RSMC and HTLC | 2016 | Lightning Network (BTC) | ◐ | ● |
| | | 170 | – | 2016 | comparison of payment channel network (BTC) | ◐ | ○ |
| | | 171,172 | –, Bolt | 2016–2017 | anonymous payment channel network (BTC) | ● ● | ○ ◐ |
| | | 173 | Fulgor and Rayo | 2017 | concurrency of payment channel network (BTC) | ● | ◐ |
| | | 59 | AMHL | 2019 | payment channel against wormhole attacks (BTC) | ● | ◐ |
| | | 174,175 | –, Sparky | 2015–2016 | Lightning Network (ETH) | ◐ ◐ | ○ ● |
| | | 176 | Raiden | 2017 | advance payment channel network (ETH) | ◐ | ● |
| | | 177 | PERUN | 2017 | virtual payment channel (ETH) | ● | ● |
| | state channel network | 178 | – | 2018 | state channel for online poker | ● | ● |
| | | 179 | – | 2018 | general state channel | ● | ○ |
| | | 180 | Sprites | 2017–2019 | worse-case time optimization | ● | ◐ |
| | | 181 | – | 2019 | multi-party virtual state channel | ● | ○ |
| | | 182 | – | 2018 | conflicts in state channel | ◐ | ◐ |
| | | 183 | – | 2018 | protecting honest party | ◐ | ● |
| | | 184 | – | 2019 | state assertion channel | ◐ | ● |
| | | 185 | Pisa | 2019 | state channel outsourcing | ● | ● |
| Extensions on core functionalities | extending opcode | 186,187 | Bitcoin covenants | 2016–2017 | limiting the spending of UTXOs | ◐ ◐ | ○ ◐ |
| | | 188,189 | – | 2019–2020 | moving smart contracts cross-chain | ◐ ◐ | ◐ ◐ |
| | improving security | 190 | PCSCs | 2018 | proof-carrying smart contracts | ◐ | ○ |
| | improving efficiency and privacy | 191 | Arbitrum | 2018 | one-step proof of execution | ● | ● |
| | | 192 | YODA | 2018 | complex contracts without validation | ● | ◐ |
| | | 193 | Zexe | 2018 | private execution of arbitrary contracts | ● | ● |
| | | 194 | ACE | 2020 | parallel execution of interactive smart contracts | ● | ◐ |

[a]– denotes the proposed scheme is not named in the literature.
[b]● denotes the scheme has description and security proof, ◐ means the scheme has description without security proof.
[c]● denotes the scheme is implemented and open-sourced, ◐ means the scheme is implemented but closed-sourced, ○ represents the scheme is not implemented.

Smart contracts in the format of SMPC protocols are more flexible and not constrained by the inherent execution mechanisms of blockchain. These schemes are usually applied to the application layer that does not rely too much on the execution mechanisms. They only utilize the existing blockchain functionalities to ensure the safety or fairness of the protocol. For example, participants may upload some witnesses or proofs to make sure everyone behaves honestly. Most procedures are conducted off-chain. In this way, the contract contents are only disclosed among participants, making the procedure more privacy-preserving than normal contracts.

The trade-off between best- and worst-case complexity is a heated topic regarding the SMPC-related schemes. Moreover, efficiently compiling general functions into SMPC-based smart contracts and off-chain protocols is still an open problem worth consideration in the future.

We remark that the SMPC-based solutions require all participants online to complete the computation protocol, which may be unrealistic in some cases. Moreover, existing SMPC schemes are featured with high computational and communication complexity, which can be new obstacles for smart contracts to be widely adopted. In comparison, the solutions with ZKP described below could slightly reduce the overhead of communication.

#### 6.1.2. Zero-knowledge proofs

ZKP (Definition 16) is a popular and relatively mature cryptographic technique to protect users' privacy. Several efficient non-interactive solutions are proposed in recent years, which are quite suitable for blockchain and smart contracts. For example, zero-knowledge succinct non-interactive arguments of knowledge (zk-SNARK)[223,224], whereby the correctness of computation can be verified in a non-interactive way with fewer required resources, are widely used in blockchain. Related studies mainly focus on improving the transparency of the ZKP to enable developers to write private contracts even when they do not know the detail of ZKP protocols.

Quorum[6] is a typical example to protect the privacy of contract contents using ZKP. It is derived from Ethereum, while the smart contracts are divided into public and private contracts. The public ones are the same as those in Ethereum and the private ones interact among contract participants using zk-SNARK and update corresponding states without revealing extra information, thus avoiding privacy leakage during the contract execution procedures.

Rather than propose an alternative system, two studies attempt to automatically integrate the ZKP protocol into the Ethereum smart contracts within the design process. The Hawk[140] framework is a typical example that applies zk-SNARK. It automatically generates smart contracts and the corresponding protocols that guide the users to protect their legitimate rights and interests during the contract execution. Eberhardt and Tai[141] also present a toolkit named ZoKrates based on zk-SNARK for private contracts. Different from Hawk, the contracts constructed in ZoKrates are mostly executed off-chain. ZoKrates comes with a high-level domain-specific language, which is used to describe the off-chain computation. The authors also provide a compiler to generate transactions that submit the final results on-chain. Using the specialized language and compiler provided by ZoKrates, developers could easily and implicitly write a private contract without understanding zk-SNARK.

We remark that the schemes mentioned above are irrelevant to smart contracts' core execution mechanisms (e.g., virtual machine). These schemes apply ZKP to the underlying mechanisms to protect privacy. We leave relevant solutions to Section 6.3.3.

Although the ZKP-based execution schemes effectively hide the contents of smart contracts, they inevitably introduce more communication and storage overhead, increasing miners' burden for validating and executing the transactions. Besides, ZKP schemes such as zk-SNARK require a trusted setup, and how to remove such a setup remains a challenge.

From this point of view, Bulletproofs[239] that works without a trusted setup arises. This is more consistent with the idea of blockchain that removes the trusted third party, and it has become a hot research direction for ZKP application on the blockchain. However, we remark that it is more fundamental and independent of the execution scheme itself. Inspired by Eberhardt and Tai,[141] we point out that future research may also focus on a more practical perspective by making such complex protocols transparent for developers and provide some formal proofs about the correctness and privacy of the generated contracts and protocols.

To sum up, although ZKP-based private contract schemes avoid the massive multi-round communication introduced by SMPC, problems in the aspects of efficiency, storage, and trusted setup are still extant. The contract execution schemes based on TEE (Definition 17) increase efficiency by adding new assumptions on the hardware security, i.e., assuming TEE's reliability, as discussed below.

#### 6.1.3. Trusted execution environment

Taking advantage of the TEE (Definition 17) hardware, such as Intel SGX[240] and ARM TrustZone,[241] the privacy of contract contents can also be guaranteed. Such TEE usually provides proofs of correct executions, revealing limited information about the communications among the users and contracts. Relevant studies mainly focus on its practical applications, such as execution efficiency and weakening the dependence on TEE's specific types.

TEE solves some problems that are difficult for traditional cryptographic schemes, such as privacy and fairness. Several studies are discussing ways to protect contract contents with TEE. Brandenburger et al.[163] introduced SGX into Hyperledger Fabric to enable trusted private executions of smart contracts. In their scheme, efficiency is sacrificed to some extent due to the employment of extra hardware. However, since the efficiency of Hyperledger Fabric has been greatly improved compared with other public blockchains (e.g., Ethereum), such loss is acceptable. Almost simultaneously, Bowman et al.[164] proposed the so-called private data objects (PDO), which utilizes TEE to execute contracts and update the state. However, PDO is designed for consortium blockchains, and thus many security threats are excluded from consideration in their work.

TEE could also be used to solve the problem of fairness in SMPC. Choudhuri et al.[161] propose a solution to achieve fairness in SMPC that combines TEE with blockchain. However, their scheme only achieves one-time SMPC. That is, each invocation requires a new setup so it is only suitable for some special contracts, for example, one-time lottery and voting. Enclave-Ledger interaction (ELI)[167] is a general blockchain-based SMPC solution, which converts the multi-step computation into a protocol involving three parties: a public ledger (or blockchain), TEE, and a host application. The scheme only uses the blockchain as an underlying component but puts no requirement on the mechanism of blockchain.

To address the privacy and fairness issues in Bitcoin, Teechain[168] is proposed to prevent malicious behaviors during the establishment and settlement of Bitcoin Lightning Network.[25] In Teechain, TEE serves as the trusted root to ensure all parties' legitimate rights and interests during the execution of the off-chain payments.

Actually, the introduction of TEE brings forth a new security assumption, i.e., assuming the adopted TEE hardware is secure. Ekiden[165] tries to weaken this assumption to some extent. In Ekiden, executions of transactions are moved into TEE, and the TEE provides the proof of the correct execution. In this way, no one except the participants knows the content of a contract, and each participant only knows the inputs and outputs of the private computation. Since it does not put limitations on the specific types of TEE, Ekiden avoids trusting a single TEE provider. Besides, Ekiden also manages to optimize the processing rate in TEE. While keeping the contents private, it handles thousands of transactions per second, nearly 100 times more than that in Ethereum.

TEE could also be used to support complex smart contracts that are infeasible otherwise. FastKitten[166] is a typical TEE-based scheme that extends Bitcoin to support arbitrary smart contracts by focusing on the efficiency of off-chain contract execution. In FastKitten, smart contracts are executed in an operator's TEE, where the operator works as a miner so that it obtains no trust from the others and will not learn anything about the contract. Checkpoints are introduced to improve the efficiency of validation. Anyone can start from any checkpoint to calculate the final state in the blockchain. Besides, FastKitten adopts the mechanisms of challenge-response and deposit-penalty, where the former is used to identify malicious behaviors and the latter is used to charge a penalty if someone misbehaves. In this way, rational participants will behave themselves and honest ones can always get their deserved money. FastKitten also describes the process of off-chain transactions and further provides a formal security proof.

By assuming the reliability of TEE hardware, the efficiency loss of contract executions caused by the complex and heavy cryptographic schemes is avoided. However, this strategy increases the cost of contract executors (or miners), since they have to update their hardware and validate the proof of correct executions before packaging the transactions. In addition, the assumption of TEE being always secure might not be realistic because there can be bugs and vulnerabilities in the TEE equipment, which will bring in new points of attack. For example, the latest study[242] shows a subversive attack for SGX. Moreover, there are concerns that device providers may insert a backdoor into their products.

Therefore, designing execution schemes that are irrelevant to specific hardware types is a problem worthy of consideration. For concrete applications, the balance between on- and off-chain overhead could be considered since on-chain affairs take more time. Off-chain operations put other requirements such as bandwidth and storage on the executors. Moreover, compilers to automatically generate TEE-based protocols given general function descriptions are still absent, which could be a promising direction for future research.

### 6.2. Off-chain channels

To resist various attacks against blockchain (e.g., DoS attacks), smart contracts inevitably have bottlenecks in transaction-processing performance. Therefore, to improve the efficiency, one solution is to execute contracts in off-chain channels and only release the settling transactions on the blockchain. This also hides the contract contents and the details of users' behaviors during data communication. Off-chain channel schemes mainly describe the interactive protocols among multiple parties, ensuring that none of them bears the unnecessary loss, and the misbehaving ones receive penalties accordingly. We divide the state-of-the-art off-chain schemes into payment and state channels, and conclude that related research mainly focuses on the security, fairness, efficiency, and feasibility issues. In the following, Section 6.2.1 interprets payment channels and Section 6.2.2 discusses the state channels.

#### *6.2.1. Payment channels*

We have discussed the micropayment channel protocol in Section 4.1.3, which derives many advanced payment channel network (PCN) schemes. Most of these schemes are aimed at achieving fair and efficient payments off-chain. In the following, we will introduce PCN schemes in Bitcoin and Ethereum, respectively, in Section 6.2.1.1 and Section 6.2.1.2. The main difference between the two categories is due to the transaction model (Definitions 10, 11) and contract language, and the design of PCN schemes in Ethereum is more convenient because of the general-purpose language.

*6.2.1.1. Payment channel networks in Bitcoin*. Since Bitcoin only supports limited operations in smart contracts, off-chain protocols must be carefully designed. The initial design of the micropayment channel only supports one-way payment (see Section 4.1.3). In other words, a receiver cannot use the same channel to send money back to the sender. For this issue, five solutions for bidirectional channels are proposed. Specifically, two studies try to realize the functionality of such off-chain payments and the other three schemes improve the preceding schemes for privacy or security issues, as in the following.

Based on the micropayment channel[20] scheme, Lightning Network[25] is proposed. It is composed of many bidirectional payment channels and enables the duplex transfer of funds between two users without a direct channel.

To establish bidirectional channels, the RSMC[25] has been introduced. Informally, in RSMC, two parties generate a deposit transaction that collateralizes the funds to a multi-signature address. They then sign the refund transactions as a refund commitment, which includes an expiration time. The party that broadcasts the refund transaction first is restricted to claim the refund after the expiration time, while the other party can immediately get a refund, thereby preventing the deliberate refund operation. The refund commitments are then signed and broadcast. Thereafter, both parties update the payment value off-chain. This will involve multiple public and private key pairs and addresses. Each time a new transaction is generated, the two parties have to send the old key pair to the other, which is regarded as an agreement to accept the new transaction and give up the old one. During the update process, if either party

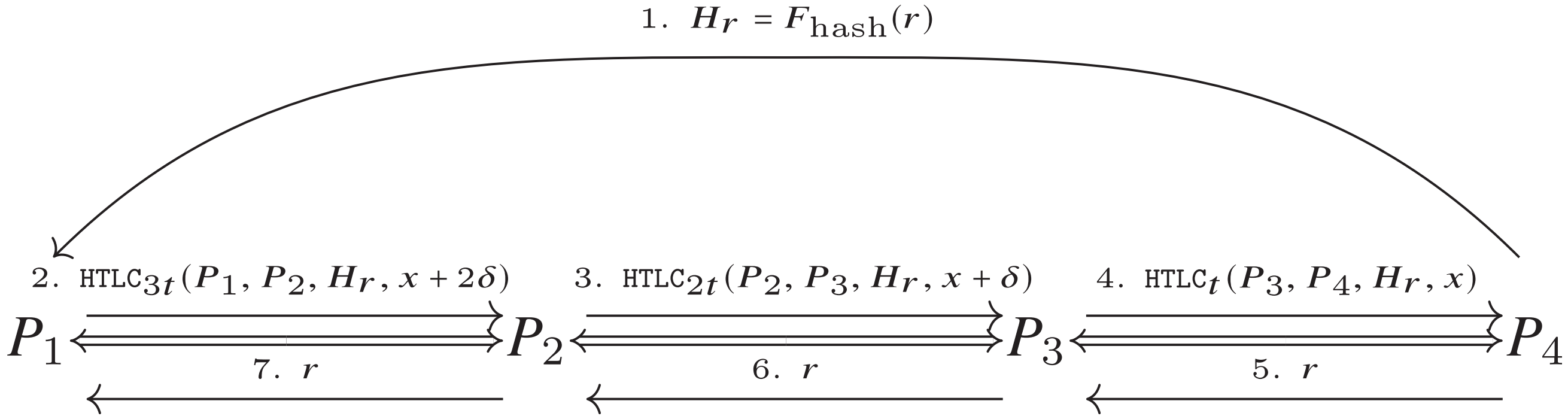

**Figure 13. Using a hashed timelock contract (HTLC) to establish a Lightning Network**[25]

broadcasts the old refund commitment, the other can get all the deposited funds.

In payment channel networks, there are cases when two users without a direct channel want to perform an off-chain payment. In this situation, an HTLC is used. We show a brief description of HTLC in Figure 13,[25] where there is a path from $P_1$ to $P_4$ ($P_1 \to P_2 \to P_3 \to P_4$). To acquire ₿$x$ from $P_1$, $P_4$ selects a random number $r$ and sends its hash value $y = F_{\text{hash}}(r)$ to $P_1$. Next, $P_1$ generates an HTLC contract $HTLC_{3t}(P_1, P_2, y, x+2\delta)$, where the subscript $3t$ stands for the timelock (as discussed in Section 4.1.3), and $\delta$ is the fee that each intermediary charges. $P_2$ (resp. $P_3$) sends their HTLC to $P_3$ (resp. $P_4$) in a similar way, where the lock time decreases to ensure that the intermediaries behave honestly. $P_4$ uses $r$ to claim the ₿$x$ in $HTLC_t$, and passes the $r$ to $P_3$, and $r$ is passed through the channel in the same manner. All rational intermediaries have the motivation to behave honestly or they will lose their money.

The Lightning Network is the combination of RSMC and HTLC and is one of the most popular PCN solutions. It is implemented in many languages, such as C language[243] and Scala language.[244]

We remark that all steps in Figure 13 are off-chain (we note that the initialization step is not illustrated in this figure), and the only on-chain procedures occur during the settlement and dispute phase. Therefore, the Lightning Network reduces the frequent small transactions on-chain, avoids the transaction delay, and somehow improves the blockchain's throughput.

Almost simultaneously, Decker and Wattenhofer[169] proposed another duplex micropayment channel scheme, which uses a diminishing timelock to prevent the other party from aborting. In Figure 14A,[169] we give an illustration of the one-way micropayment channel described by Hearn,[20] and in Figure 14B[169] we show the method to update the channel with the reducing timelock. The branch with a smaller time lock $t$ can be confirmed earlier in the system so that the values in the bidirectional channel are updated. With this method, the duplex micropayment channel is shown in Figure 14C.[169] Firstly, both parties set the maximum and minimum value of $t$, additional branches per node $n$ ($n = 1$ Figure 14), and maximum invalidation depth $d$ ($d = 3$ Figure 14). Thereafter, the two parties use the duplex micropayment channel pair in the first line to conduct the payments. When the fund in one of the channels is exhausted, both parties update the channel, reset the initial funds, and decrease the value of $t$ in order, as shown in the second and third lines. Similarly, by connecting channels end by end, such duplex payment channels can also be extended into a PCN.

As depicted in Figure 14, most procedures in the duplex micropayment channel are also conducted off-chain. This scheme serves a purpose similar to that of the Lightning Network, improving throughput and reducing transaction fees. The two bidirectional payment channel schemes[25,169] are compared and analyzed in McCorry et al.[170] in terms of on-chain privacy, operational overheads, and outsourcing.

From another point of view, the above two PCN schemes cannot fully protect users' privacy. For example, the hash lock in HTLC could be used to track the participants on the same path. Heilman et al.[171] adopt blind signature[245] and realize a fair, anonymous, and off-chain exchange of BTC with vouchers issued by an untrusted third party. This scheme is compatible with both Lightning Network and the duplex micropayment channels described above. Green and Miers[172] propose an opcode `OP_BOLT` to achieve anonymity in three forms of micropayment channels (one-way, bidirectional, and PCN). However, this solution is only applicable to platforms that are born with anonymity (e.g., ZeroCash[232]) or other cryptocurrencies that support coin mixing. Intuitively, the solutions of both Heilman et al.[171] and Green and Miers[172] require a soft fork (Definition 7) to take effect.

Some researchers point out that the aforementioned PCN solutions could not handle the concurrency of off-chain payments well, which might cause problems such as transaction blocking or conflicting. Malavolta et al.[173] propose two protocols for the concurrency issues in PCN, namely, Fulgo and Rayo. Fulgo comes with formal provable privacy within the UC model and is compatible with Bitcoin's script language. When a conflict occurs, all the conflicting transactions will be canceled and resent after a certain delay to prevent permanent blocking. Rayo is another scheme guaranteeing that at least one of the payments will be completed, yet it to some extent sacrifices privacy compared with Fulgo. The authors also propose an advanced multi-hop HTLC, which introduces randomness to the timelock and combines ZKP (Definition 16) to avoid privacy leakage of the routing information.

In terms of the fairness in PCN, based on Fulgo and Rayo,[173] Malavolta et al.[59] propose a new PCN protocol that is secure against wormhole attack, whereby an adversary that controls multiple intermediaries can exclude the honest nodes in the path by directly passing the random preimage $r$ to other corrupted nodes.

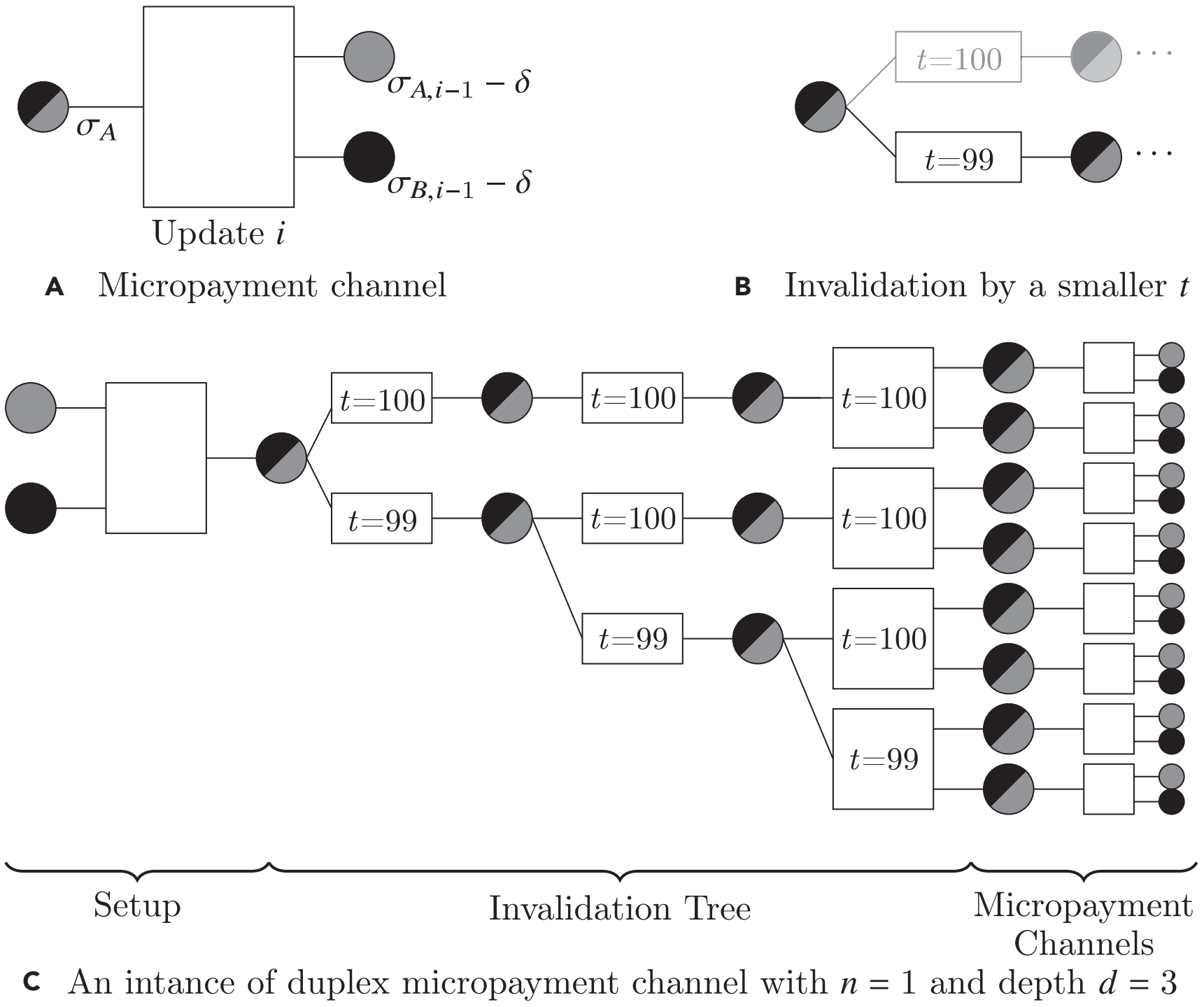

**Figure 14. Duplex micropayment channel with maximum time $t_{max} = 100$, minimum time $t_{min} = 99$, additional branches per node $n = 1$, maximum invalidation depth $d = 3$**
Figure reprinted with permission from Decker and Wattenhofer[169].

Their solution assigns a random number to each intermediary so that the adversary can no longer conduct the original attack, whereby honest participants have sufficient incentives to serve as intermediaries. This solution improves the fairness of PCN and preserves the interests of honest participants. Moreover, the authors provide a formal security proof within the UC model.

As we have observed, research on PCN with formal security proofs is becoming a new trend. Apart from the security issues, other problems such as fairness[59] and rebalancing[246,247] may arise in practice. Exploring and addressing potential vulnerabilities and problems in PCN would be a promising research direction. This could improve user experience and attract more users to get involved in PCN applications.

*6.2.1.2. Payment channel networks in Ethereum*. The aforementioned payment channel networks are also useful in Turing-complete blockchains such as Ethereum, and the deployment of PCN in these systems is more convenient. Related research can be divided into studies in the theoretical and application aspects, as discussed in the following.

From the theoretical aspect, Tremback and Hess[174] propose a general model for payment channels in which smart contracts in Turing-complete languages are used (called smart conditions in their work). However, the implementation and security proof of their model are not provided. The authors also propose a routing protocol that finds a suitable path for the payment, but as the concept of routing is beyond the scope of this review, we omit the description here.

From the application aspect, Peterson[175] implements the Lightning Network on Ethereum with the Solidity language. Raiden[176] is another advanced version of the Lightning Network that improves the throughput of the blockchain system. Moreover, Raiden tries to add more functionalities other than the processing of standard transactions, such as updating variables in smart contracts. However, it is still under development and only supports the standard off-chain payment for now.

Similar to the PCNs already mentioned, earlier PCN solutions in Ethereum also suffer from the problems of feasibility and privacy. As pointed out by Dziembowski et al.,[177] although such schemes support efficient off-chain payment, they require the intermediaries to be always online so that they can forward the necessary messages along the path and monitor possible misbehaviors of others. Besides, from the privacy perspective the intermediaries will know the identities of the sender and receiver and the amount of the transaction. To address these problems, the virtual payment channel PERUN[177] is proposed. It builds a virtual payment channel between two users based on the established payment channels, as shown in Figure 15. In this way, the intermediaries are only involved during the establishment and settlement of the virtual channel. Only essential affairs and events are handled by the contracts automatically, and other procedures in the non-malicious case are conducted off-chain, similar to the Lightning Network. Besides this, Dziembowski et al.[177] give the security proof of the scheme within the UC model and give a proof-of-concept implementation on Ethereum.

As shown in Figure 15,[177] there is a payment channel between Alice and Ingrid, denoted as $\beta_A$. Similarly, $\beta_B$ represents the payment channel between Bob and Ingrid. There are two corresponding smart contracts $C_A$ and $C_B$ on the blockchain. In $\beta_A$, the amounts deposited by Alice and Ingrid are $y_A$ and $y_I$, respectively, and the amounts deposited by Bob and Ingrid in $\beta_B$ are $z_B$ and $z_I$, respectively. With $\beta_A$ and $\beta_B$, a virtual payment channel $\gamma$ is built. Deposit in $\beta_A$ will be partially frozen when creating $\gamma$, namely, $y_A$ for Alice and $y_I$ for Ingrid, respectively. Similarly, $z_B$ and $z_I$ will be frozen for Bob and Ingrid in $\beta_B$, respectively. To prevent these deposits from being frozen permanently, the three participants will set an expiration time when establishing the channel. After the expiration time, or when the channel $\gamma$ is settled on-chain, they can freely retrieve their deposits.

We remark that PCN in Ethereum also suffers from contract vulnerabilities and other issues. Formal proofs within standard frameworks and exploration of possible problems are two promising research directions in this field.

### 6.2.2. State channels

Inspired by PCN, the updates of variables in smart contracts can also be conducted off-chain, which is the key point of state channel networks.[248,249] A state channel updates the states in smart contracts according to the predefined functions and algorithms in an off-chain way. Similarly to PCN, only the establishing and settling transactions are conducted on-chain. Related research mainly focuses on the generalization, usability, efficiency, and privacy of state channel networks.

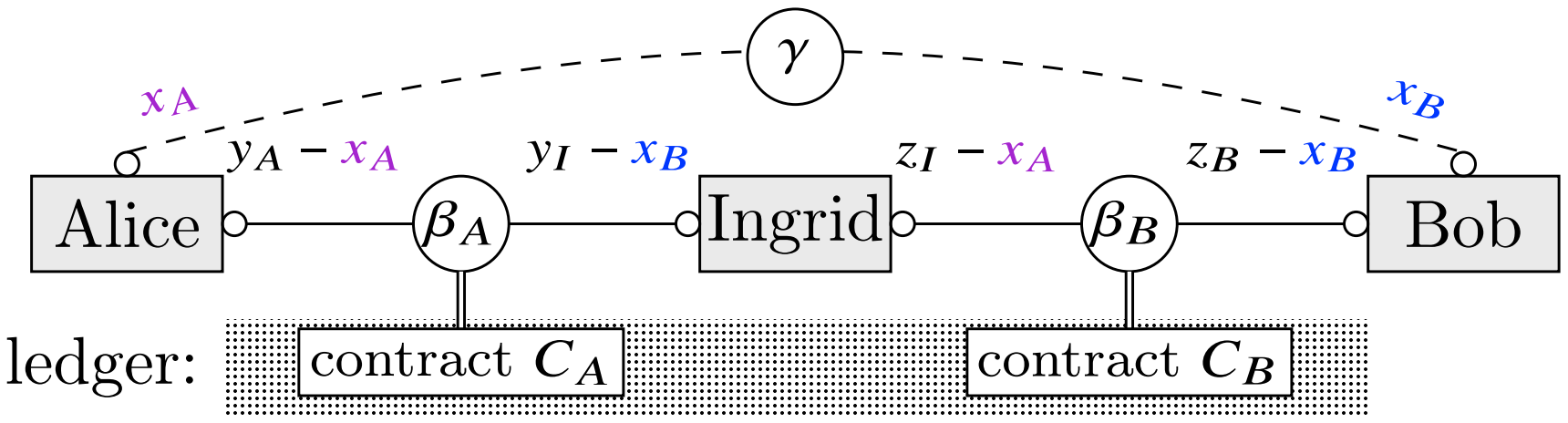

**Figure 15. Virtual payment channel in PERUN**
Figure reprinted with permission from Dziembowski et al.[177].

One typical application of the state channel network is online poker. Bentov et al.[178] design and implement an efficient online poker contract on Ethereum through a functionality named secure cash distribution with a penalty, which transfers the money from losers to winners as soon as the game is finalized. The authors also give security proof within the UC model. The online poker scheme is merely a special case of state channel, and this scheme can also be used for bidirectional payment channels and other applications involving smart contracts.

The online poker scheme mainly implements the ideal multiple sequential cash distribution functionality $\mathcal{F}^*_{\text{MSCD}}$. We rephrase it in Figure 16.[178] The *sid* and *ssid* are session identifiers. The *n* participants are denoted $P_1$,$P_2$,$\cdots$,$P_n$. $\mathcal{A}$ is the adversary, whose corrupted nodes are represented as $\{P_a\}_{a\in\mathcal{C}}$. The set of honest participants is $\mathcal{H}$, where $\mathcal{H} = \{1,\ldots,n\}\ \mathcal{C}$, $h = |\mathcal{H}|$. The deposit and penalty amount is *d* and *q*, respectively. $\overrightarrow{b}$ is the fund distribution vector, and $b_i$ represents the funds that $P_i$ deserves. The number of functions is *m*. The ideal function that receives (resp. sends) messages from (resp. to) *P* is denoted by $F_{\text{receive,P}}$ (resp. $F_{\text{send,P}}$), and $F_{\text{broadcast}}$ is the ideal broadcast function. The last field in the messages with headers SETUP, ADDMONEY, RETURN, PENALTY, EXTRA, and REMAINING is the value of the fund.

As shown in Figure 16, there are three phases in $\mathcal{F}^*_{\text{MSCD}}$. In the deposit phase, it accepts deposits of value *d* from honest parties, and the penalty of value *hq* from adversary $\mathcal{A}$. In the execution phase, each participant takes part in the multi-party functions several times with input payload $p_i, i\in\{1,\cdots,n\}$. The execution results include the change *Z* on contract state *S* and that on fund distribution vector $\overrightarrow{b}$. In this phase, participants can also increase their deposits. When the execution phase is completed, or the adversary $\mathcal{A}$ aborts, the participants enter the claim phase. If $\mathcal{A}$ does not abort, each honest participant $P_i$ will receive his deposited fund *d* and his deserved fund $b_i$; else, the honest ones can share the adversary's deposited penalty *hq* and the additional penalty $q_i$ (if any). Finally, $\mathcal{F}^*_{\text{MSCD}}$ returns the remaining funds to $\mathcal{A}$.

In summary, this ideal functionality first gets deposits from all participants, then executes the desired functions and transfers messages among users, and finally updates the states and distributes the deposited amounts accordingly. We mention that this functionality could be adopted in many scenarios, since most smart contracts are designed to handle money among multiple participants, to implement some prefixed functions.

In addition to online poker, state channels can also facilitate the application of payment channels. Miller et al.[180] specifically describe the model of state channels within the UC model and construct an improved payment channel, Sprites, that reduces the worst-case time to settle a transaction. It constructs a global preimage manager (`PM`) contract to verify the hash preimage instead of transferring it among participants, as is done in the Lightning Network. The receiver directly submits the preimage of the hash lock to `PM` for verification, thereby reducing the time cost brought in by the transfer of preimage. Sprites also supports the dynamic deposit and withdrawal of the fund, which greatly improves the usability of the payment channels. However, it does not consider privacy issues. The ideal functionality $\mathcal{F}_{\text{State}}$ of a state channel given in their work under the UC model is rephrased in Figure 17.[180] We mention that this model is relatively general, and thus could be adopted for future research on state channels with few modifications.

As shown in Figure 17, $\mathcal{F}_{\text{State}}$ initializes the variables, receives auxiliary input message *m* from contract *C*, appends it to the stack buf and aux$_{\text{in}}$, and sets the pointer *ptr*. During the $\tau^{\text{th}}$ round of execution, it receives the payload data $p_{\tau,i}$ ($i = 1,\cdots,n$) from each $P_i$ within $O(\Delta)$ time, and forwards these messages to the adversary $\mathcal{A}$. After receiving all messages, it updates the function $F_{\text{contract}}$ with inputs, including the contract state *S*, inputs $p_{\tau,i}$ ($i = 1,\cdots,n$), and other data in the stack. If there is any non-empty output *o* in $\mathcal{F}_{\text{State}}$, it will be handled according to the output rule *C.output*.

Inspired by the concept of virtual payment channel in PERUN,[177] Dziembowski et al.[179] define the general state channel, supporting the off-chain execution of arbitrary smart contracts. Similar to PERUN, a higher-level channel is built upon two existing channels with a common third party. These higher-layer channels are called virtual state channels. The users only need to interact with their common third party, rather than the blockchain, to open and close the higher channel. Conflicts are resolved by this third party first. If this fails, contracts on the blockchain are then invoked. For better understanding, this general state channel is illustrated in Figure 18 given by Dziembowski et al.,[179] whose concept is partially similar to that of HTLC in Figure 13 and PERUN payment channel in Figure 15. Every channel can be connected, regardless of which lower channels they are built on.

In Figure 18, five state channels are recorded on-chain. On this basis, $\gamma_1$ (resp. $\gamma_2$) is the first-layer virtual state channel between $P_1$ and $P_3$ (resp. $P_4$ and $P_6$). Furthermore, $\gamma_3$ is a higher-level virtual state channel, and $\gamma_4$ is built upon $\gamma_3$ and $\gamma_2$. The authors also provide the state channel and virtual state channel functionality, along with a formal security proof. Meanwhile, Coleman et al.[250] propose the Counterfactual framework to build a general state channel that enables the update of arbitrary smart contracts. In their framework, developers no longer have to design specific state channels for

**Deposit phase**$(sid, \mathcal{H})$:

Upon invocation by any $P_j$ or $\mathcal{A}$:

Initialize $flg = \perp$.

$F_{\text{receive},\text{P}_\text{j}}(\text{SETUP}, sid, ssid, j, d)$ for all $j \in \mathcal{H}$.

$F_{\text{receive},\mathcal{A}}(\text{SETUP}, sid, ssid, hq)$ where $h = |\mathcal{H}|$.

**Execution phase**$(sid)$:

Upon invocation by any $P_j$:

Initialize $flg = 0$ and $\vec{b} \leftarrow \vec{0}$.

For $id = 1, 2, \cdots$, sequentially do:

1. If $F_{\text{receive},\text{P}_\text{j}}(\text{EXIT}, sid, ssid)$:
   $F_{\text{broadcast}}(\text{EXIT}, sid, ssid, j)$,
   go to the claim phase.
2. If $F_{\text{receive},P_j}(\text{ADDMONEY}, sid, ssid\|id, b_j)$:
   If $F_{\text{receive},P_k}(\text{ADDMONEY}, sid, ssid\|id, b_j)$ for each $k \neq j$:
   $F_{\text{broadcast}}(\text{ADDMONEY}, sid, ssid\|id, b_j)$.
   $\vec{b} \leftarrow \vec{b} + (\cdots, 0, b_j, 0, \cdots)$.
3. Initialize state $S = \perp$.
   $F_{\text{receive},P_j}(\text{FUNCTION}, sid, ssid\|id, g^{(id)}$ for all $j \in \mathcal{H}$.
   $F_{\text{broadcast}}(\text{FUNCTION}, sid, ssid\|id, g^{(id)})$.
4. Parse $g^{(id)} = \{g_k^{(id)}\}_{k \in \{1,\cdots,m\}}$. For $k = 1, \cdots, m$, sequentially do:
   (a) $F_{\text{receive},P_j}(\text{INPUT}, sid, ssid\|id\|k, j, p'_j)$ for all $j \in \mathcal{H}$.
   (b) $F_{\text{receive},\mathcal{A}}(\text{INPUT}, sid, ssid\|id\|k, \{p'_a\}_{a \in C})$.
   If no such message was received:
   update $flg = 1$ and go to the claim phase.
   (c) Compute $(Z, \vec{b}', S') \leftarrow g_k^{(id)}(p'_1, \cdots, p'_n; S, \vec{b})$.
   (d) $F_{\text{send},\mathcal{A}}(\text{OUTPUT}, sid, ssid\|id\|k, Z, \vec{b}')$.
   (e) If $F_{\text{receive},\mathcal{A}}(\text{CONTINUE}, sid, ssid\|id\|k)$:
   $F_{\text{send},P_i}(\text{OUT}, sid, ssid\|id\|k, Z, \vec{b}')$ for all $P_i$.
   (f) If $F_{\text{receive},\mathcal{A}}(\text{ABORT}, sid, ssid\|id\|k)$:
   set $flg = 1$, and go to the claim phase.
   (g) $\vec{b} \leftarrow \vec{b}'$.

**Claim phase**$(sid, flg, d, \vec{b}, \mathcal{H}, C)$:

Upon invocation by any $P_j$ or $\mathcal{A}$:

$F_{\text{receive},\mathcal{A}}(\text{EXTRA}, sid, ssid, \{q_i\}_{i \in \mathcal{H}}, \sum_{i \in \mathcal{H}} q_i)$, $q_i = 0$ if not received.

If $flg = 0$ or $\perp$, $F_{\text{send},P_r}(\text{RETURN}, sid, ssid, d + \vec{b}_i)$ for $i \in \mathcal{H}$.

If $flg = 0$, $F_{\text{send},\mathcal{A}}(\text{RETURN}, sid, ssid, hq + \sum_{a \in C} b_a)$.

If $flg = 1$, $F_{\text{send},P_i}(\text{PENALTY}, sid, ssid, d + q + \vec{b}_i + q_i)$ for $i \in \mathcal{H}$.

$F_{\text{send},\mathcal{S}}(\text{REMAINING}, sid, ssid, \Sigma_{a \in C} \vec{b}_a)$.

**Figure 16. Ideal functionality $\mathcal{F}^{*}_{\text{MSCD}}$ for multiple sequential cash distribution with penalties**
Figure reprinted with permission from Bentov et al.[178].

their application. However, the scheme lacks formal security proof, and the framework is still under development.

These state channel network schemes[178–180,250] only support two-party smart contracts. In other words, contracts that involve more parties are not applicable in such state channels. Dziembowski et al.[181] propose a multi-party virtual state channel that retains the advantages of the virtual state channel.[179] Specifically, a state channel could be opened and closed without interacting with the blockchain in the best case, and such processes are almost instantaneous and zero-cost. Regarding the worst case, they reduce the time for conflict resolution from $O(n\Delta)$ to $O(\Delta)$, where $\Delta$ is the maximum time delay for on-chain settlement. When multiple parties are involved, the potential security threats and conflicts become more complicated. To alleviate this concern, Dziembowski et al.[181] apply the UC model and give a formal security proof in their work. The multi-party virtual state channel requires all participants to stay in a common state channel network, as shown in Figure 19.[181] Here, only necessary operations are conducted through on-chain multi-party contracts, such as deploying the contracts, resolving the disputes, and settling the channel. Other procedures are moved off-chain to gain efficiency, such as transferring funds, exchanging messages, and updating the balances. This could be a basis for future improvements and modifications.

Specifically, in Figure 19 five parties from $P_1$ to $P_5$ are connected by four on-chain channels. $P_1$, $P_3$, $P_4$, and $P_5$ jointly build the multi-party virtual state channel $\gamma$ that excludes $P_2$. The `mpVSCC` between them refers to the instance of the multi-party virtual state channel contract. The $x/y$ at the end of each channel indicates the contract's initial/final amount.

On the security aspect, Close and Stewart[182] argue that state channels applicable for arbitrary contracts may face the following three conflicts: (1) conflicts related to external states, such as exchange rates and temperatures; (2) concurrency conflicts occurring when participants perform conflicting operations almost simultaneously and could not agree on the order of operations; and (3) silent conflict when one participant suddenly loses its response, causing the suspension or abortion of off-chain operations. The ForceMove[182] framework, which puts constraints on the applications of the state channel essentially to avoid the first two conflicts, is proposed to solve these problems. Besides, ForceMove

**Initialize**(aux$_{\text{in}}$, $ptr$, S, buf):

Upon invocation by $C$:

  aux$_{in}$ $\leftarrow\perp$

  $ptr \leftarrow 0$

  $S \leftarrow \emptyset$

  buf $\leftarrow \emptyset$

**Input**($C.input(m)$):

Upon invocation by $C$:

  append $m$ to buf and aux$_{\text{in}}$, $j \leftarrow |\text{buf}| - 1$

  within $\Delta$: set $ptr \leftarrow \max(ptr, j)$

**Proceed**($\tau$):

Upon invocation by $C$:

  for each party $P_i$:

    $F_{\text{receive},P_i}(p_{\tau,i})$

    if $p_{\tau,i}$ is not received within $O(\Delta)$: $p_{\tau,i} \leftarrow\perp$

    $F_{\text{send},\mathcal{A}}(i, p_{\tau,i})$

  after receiving all inputs:

    $(S, o) \leftarrow F_{\text{contract}}(S, \{p_{\tau,i}\}, \text{aux}_{\text{in}}[ptr])$

    if $P_i$ for all $i = 1, \cdots N$ are honest:

      $F_{\text{send},P_i}(S)$ within time $O(1)$

    else:

      $F_{\text{send},P_i}(S)$ within $O(\Delta)$

    if $o \neq\perp$:

      invoke $C.output(o)$ within $O(\Delta)$

**Figure 17. Ideal functionality of state channel $\mathcal{F}_{\text{State}}$**
Figure reprinted with permission from Miller et al.[180].

introduces a new operation to ensure that the last conflict is resolved smoothly.

Although the state channel helps avoid fees caused by frequent transactions, an honest party must submit numerous proofs to make things work as expected when a dispute occurs. It is pointed out by McCorry et al.[183] that such proofs may involve a large amount of data, which will result in high transaction fees. That means an honest party has to pay for the misbehavior of malicious parties. To protect the interests of honest parties, Buckland and McCorry[184] propose the concept of state assertion channels. In their scheme, honest parties only submit the hash value of the final state for finalization, avoiding high transaction fees. It adopts the concept of optimistic smart contracts, where the update is accepted without validation, and parties who disagree with it can submit their proofs for invalidation. When a wrong update is verified, the provider will be rewarded as their incentives.

From another perspective, honest parties in a state channel must always be online in case the counter-party submits an older version of the states, which is sometimes unrealistic. McCorry et al.[185] propose the Pisa solution to outsource this work to a third party and give a formal security proof of their scheme. Moreover, a proof-of-concept implementation based on a simplified version of Sprites[180] is also provided by McCorry et al.[185] Compared with other outsourcing solutions, such as Monitor[251] and WatchTower,[252] it only takes $O(1)$ storage space for the third party (which is $O(N)$ in Monitor, where $N$ is the number of transactions generated off-chain). It directly applies to Ethereum, while WatchTower is not compatible with platforms such as Bitcoin and Ethereum.

State channels are attractive for off-chain execution of smart contracts, and relevant schemes with formal security proofs have been discussed earlier. However, a concrete implementation and a compiler that correctly converts smart contracts into state channels are absent and remain a topic for future research.

### 6.3. Extensions on core functionalities

Off-chain channels discussed above mainly focus on the off-chain protocols while retaining the original execution mechanisms of the underlying blockchain. In this section, we discuss several schemes that extend the core functionalities of the smart contract platform. Specifically, Section 6.3.1 introduces the extensions on opcodes that add the functionalities of smart contracts could achieve, Section 6.3.2 introduces the schemes that enhance the security of deployed smart contracts, and Section 6.3.3 describes the solutions that improve the efficiency and privacy of contract execution. All these extensions and alternatives are aimed to make smart contracts more applicable for universal adoptions on data communication and value exchanges.

#### *6.3.1. Extension on opcodes*

By adding new opcodes, more appealing functionalities could be achieved in smart contracts, making them better meet daily use.

The covenant in Bitcoin refers to a mode that the future transfer of the fund is restricted according to certain user-defined rules. This functionality enriches the application scenario of Bitcoin. From this point of view, Möser et al.[186] extend Bitcoin with an opcode that enables the so-called covenant mode, making it possible to track the flow of a specific payment. It also enables the vault transaction, which takes more time to take effect than a standard one. Within this time, the owner possessing the recovery key can invalidate the vault transaction, avoiding the economic loss caused by the private key theft and enhancing the security of the private key. O'Connor and Piekarska[187] propose another opcode that only involves computational operations and leaves out the transaction data, realizing the same functionality as a covenant. They also introduce an opcode that realizes the vault mode. Intuitively, both solutions above require a soft fork (Definition 17) on Bitcoin.

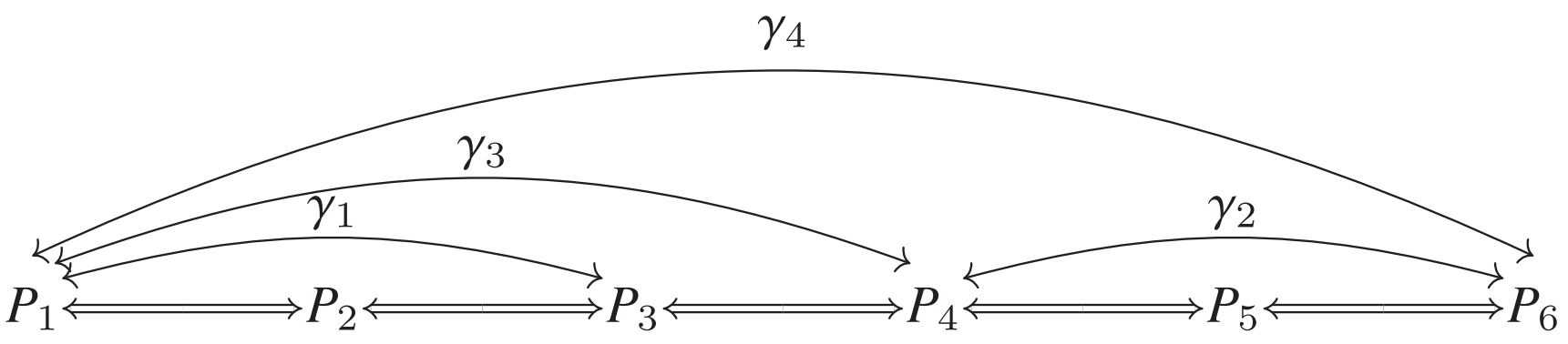

**Figure 18. Construction of virtual state channels**
Figure reprinted with permission from Dziembowski et al.[179].

There are also demands to move smart contracts across blockchains to achieve a better performance (which is relevant to the target platform) or simply as a backup. Fynn et al.[188] propose an opcode `OP_MOVE` in EVM and a corresponding keyword in Solidity to realize the cross-chain movement of smart contracts. Such movement could only be done between two blockchains with the same execution environment. In fact, Westerkamp[189] has already proposed a similar moving protocol, whose solution does not involve the modification of opcodes but requires a large gas overhead for the migration.

We remark that such extensions on opcodes might bring in unexpected vulnerabilities, e.g., the `OP_LSHIFT` opcode could be exploited to crash a Bitcoin node.[21] Therefore, theoretical and practical analysis is required before deployment. As for future research, opcodes that might facilitate the application of smart contracts could be further investigated. For example, safe math algorithms (especially the cryptographic schemes such as commitments) could be integrated into one opcode for developers' convenience. Moreover, inspired by `OP_MOVE`,[188] opcodes that communicate with other blockchains could be considered for efficient value- and data exchange across blockchains.

#### 6.3.2. Improvements on security

As mentioned in Section 1, smart contracts are difficult to update due to the tamper-resistant nature of blockchain. When a bug or vulnerability is found in a deployed contract, users and developers can do nothing to remedy the situation. To mitigate this risk, Dickerson et al.[190] propose the concept of proof-carrying smart contracts (PCSCs) based on the idea of proof-carrying codes. Its implementation involves modifying the underlying consensus and execution mechanism. Namely, the blockchain only maintains the key properties of the deployed contracts. The creator firstly uploads some key properties of the contract to the blockchain as a commitment. Thereafter, the miners check that such key features remain unchanged before and after the update operation. In this way, the upgrade of smart contracts could be realized without harming security.

The update of smart contracts has been a continual problem for a long time and is a promising research direction. Inspired by Dickerson et al.,[190] we remark that auxiliary schemes and tools, such as chameleon hash,[253] are worthy of consideration.

#### 6.3.3. Improvements in efficiency and privacy

Concerning privacy issues, we have introduced several private contract schemes that utilize ZKP in Section 6.1.2. Schemes discussed here also use cryptographic techniques such as ZKP. Different from the prior solutions, however, the core functionalities of the original execution mechanisms are modified and extended to support efficient and privacy-preserving executions of contracts, as discussed in the following.

To improve efficiency and privacy during smart contract executions, Arbitrum[191] emerges with a redesigned virtual machine. In Arbitrum, users delegate the off-chain executions of smart contracts to trusted nodes. With the one-step proof delivered by the Arbitrum virtual machine, the correctness is guaranteed. Such a proof only leaks a small part of privacy, and since the computation is off-chain, no extra information of the contract will be revealed. Moreover, the authors claim that with techniques such as Bulletproofs[239] and zk-SNARKs,[223] the leakage of privacy could be further reduced. Arbitrum requires a reasonable incentive and penalty mechanism to ensure the correct execution offered by rational participants. Since not all nodes execute the same smart contracts, efficiency is also improved.

As mentioned at the beginning of this section, smart contracts' complexity is limited due to the execution mechanisms (e.g., the gas limit). For traditional smart contracts, the execution result is easy to verify. However, for complex contracts the verification procedure is non-trivial and consumes a large number of resources, making it impossible to be performed on-chain. YODA[192] is proposed to help reach an agreement on the execution results of such complex contracts. It introduces a non-deterministic off-chain execution mechanism, with randomly selected nodes and a probability model to determine the execution result. The most prominent feature of YODA is that it eliminates the step of verifying the results on-chain, thereby avoiding the time delay caused by the on-chain settlement.

However, Wüst et al.[194] argue that YODA and Arbitrum are not suitable for concurrent execution of interactive complex smart contracts designated to different groups of miners. For instance, if a contract $C_A$ is executed by a group $\mathcal{G}_A$, and $C_B$ is assigned to $\mathcal{G}_B$, it is infeasible for $C_A$ to call $C_B$ in these systems. To address this problem the authors propose the ACE scheme (asynchronous and concurrent execution), whereby the contract execution procedure is extracted from the traditional miners. That is, miners are only responsible for reaching consensus on the ordering of valid transactions, and the appointed executors are informed by the on-chain information and interact with each other (including the communication across groups) to execute the contracts concurrently. ACE is proven to enable safe cross-group calls without assuming all groups to be reliable, and contracts in ACE are proven to stay responsible when an honest quorum of executors exists. However, such a mechanism makes the contract information publicly available, and it cannot protect the privacy of contracts do as YODA and Arbitrum.

To fully protect users' privacy, Zerocash[232] introduces the zk-SNARK scheme into its underlying execution mechanism but gives up the support of smart contracts. Bowe et al.[193] extend Zerocash and propose Zexe, which utilizes the technique of two-layer recursive proof with properly selected cryptographic parameters to achieve the succinct zero-knowledge proof of arbitrary predicates defined by users. With such recursive proofs, the contents and results of smart contracts can be entirely hidden. The overhead of Zexe is comparable with that of Zerocash[232] and Hawk.[140] Besides, the authors further provide a security proof within the UC model.

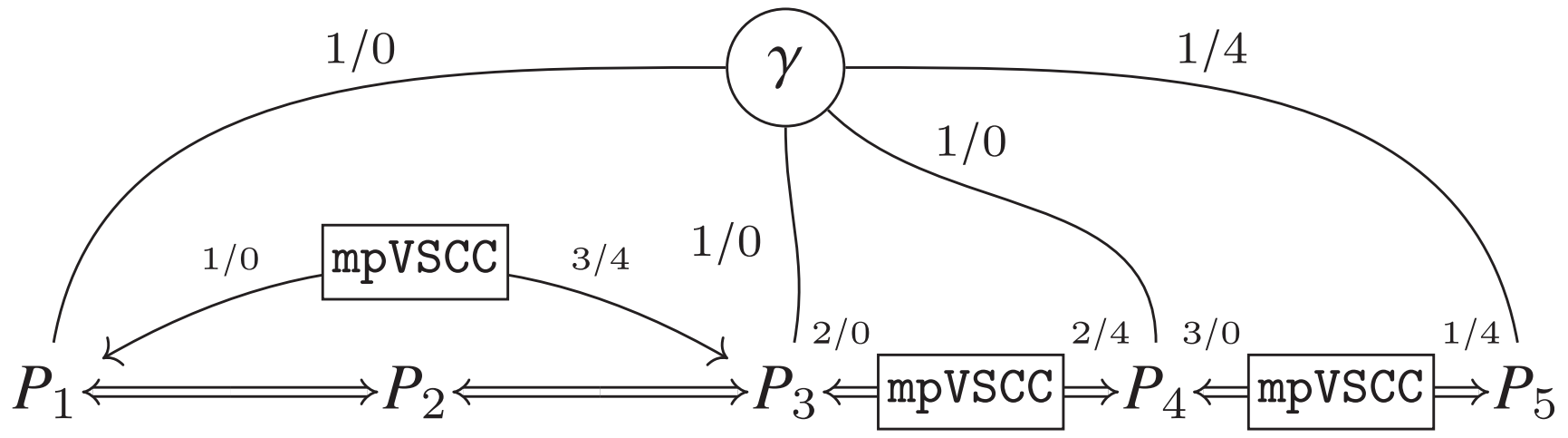

**Figure 19. An example of a multi-party virtual state channel**
The mpVSCC between them refers to the instance of the multi-party virtual state channel contract. The $x/y$ at the end of each channel indicates the initial/final amount of participants in a contract. Figure reprinted with permission from Dziembowski et al.[181]

To improve the efficiency and throughput of blockchains, Poon and Buterin[254] constructed a child-chain scheme Plasma, with a series of contracts anchored on Ethereum. Plasma uses the bitmap to map the funds spent on a single bit, reducing the transaction size. It further alleviates the problems of transaction congestion and limited throughput on Ethereum. Participants can submit fraud proofs to the main chain (Ethereum) to enforce the correct execution in the child-chain. However, such systems focus more on the security of consensus and less on smart contracts' execution mechanism. Therefore, such schemes are beyond the scope of this review. Similar child-chain schemes such as Cosmos,[255] Polkadot,[256] and side-chain schemes[235,236] are also beyond our scope.

As we have observed, schemes from various perspectives have been proposed to improve efficiency and privacy. We remark that since Hawk,[140] Arbitrum,[191] and YODA,[192] there is a trend toward moving the executions of smart contracts off-chain, among several user-designated executors, to keep the contract contents private and allow parallel executions of independent contracts by multiple sets of executors. However, as pointed out by Wüst et al.,[194] the aforementioned proposals do not consider the invocation among contracts, which may involve different sets of executors. We remark that this problem could be further considered in the future, since Wüst et al.[194] sacrificed the privacy improvements introduced by the prior works. Moreover, cryptographic schemes such as ZKP and SMPC also have prominent influences in developing alternative contract execution systems.

## 7. DISCUSSION

In this review, we have discussed and categorized 159 papers (or online resources) on smart contract construction and executions up to August 2020. In the following, we present the comparison with related work and our limitations. We then present the challenges and future research directions on the construction and executions of smart contracts.

### 7.1. Comparison with related work

Compared with the related surveys on the features of contract platforms,[26,28,33] properties of the contracts,[12,28,33–38] and related analysis tools[33,37–42] (see Section 2.3), this paper covers more aspects of smart contracts, elaborates the existing schemes in detail, and points out several research directions for each category. We take efforts to form a systematization of knowledge on contract construction and executions. We fill the gap between research and development caused by the quick evolution of smart contract technology.

### 7.2. Our limitations

First, most of our discussion is related to the studies on Bitcoin and Ethereum. Although the research on these two platforms contributes to the majority of the literature, other platforms such as Hyperledger Fabric are also attractive for business applications. We do not cover the smart contract schemes on other platforms owing to time and space, and leave them for future work.

Second, as mentioned in Section 1, we do not cover the high-level descriptions of smart contracts with other technologies, such as artificial intelligence, cloud computing, and the Internet of Things. These are promising application scenarios for smart contracts, but relevant papers usually fail to address the construction and execution of smart contracts. We leave the classification of these schemes for future work.

### 7.3. Challenges

We have informally mentioned several challenges that hinder the development and adoption of smart contracts in Section 4, Section 5, and Section 6. The security requirements and criteria on smart contracts are quite distinct from those on general computer programs, making the construction of smart contracts skill oriented. There are also many open problems such as privacy leakage, execution efficiency, and contract complexity, which have attracted widespread attention. To address the challenges more clearly, we summarize the major limitations in the following.

(1) *Frequently occurring vulnerabilities*. Smart contracts are believed to have great potential influence in the financial area. However, users have to face the risks brought about by poor development practices. Various attacks have been claimed (see Section 5.2.1 and previous works[12,89–94]), which may influence the adoption of smart contracts in business.
(2) *Incomplete design paradigms*. Several groups[22,85,86] have seen the significance of design paradigms for developers to understand potential risks. However, since the smart contract technology is still under rapid evolution, paradigms may also change quickly and are far from adequate. Besides, several concrete paradigms in the literature (see Section 5.1.1) might be found to be inefficient or vulnerable in the future.

(3) *Inefficient analysis tools*. There have been tens of analysis tools aimed at detecting vulnerabilities in smart contracts (see Section 4.2.1 and Section 5.2.1). However, most of them are inefficient as they require extra effort in describing desired properties in a specific language. Besides, there is a trade-off between high accuracy and full coverage, and users have to choose from these tools.

(4) *Low processing rate and limited complexity*. The current transaction delay is relatively high, making smart contracts infeasible for the applications subject to time delay. Moreover, contracts' complexity is limited (see Section 6). Many schemes have been introduced, such as off-chain channels and alternative systems (see Section 6.2 and[191,192,194]). However, these schemes may introduce new problems.[59,246,247] Besides, the alternatives may require additional techniques such as randomness, requiring massive overheads.

(5) *Lack of privacy*. Privacy is another issue commonly discussed (see Section 6.1 and Section 6.3.3). The off-chain networks and other alternative systems may help protect privacy but again, most of them rely on additional cryptographic schemes[223,224,239] that are still under development.

### 7.4. Future research directions

We enumerate several future research directions summarized from our analysis at the end of each taxonomy, from the perspective of the construction and execution of smart contracts.

(1) *Design paradigms for script-based blockchains*. With the emergence and implementation of off-chain channels,[25,59,169–173] designing a fair and economic off-chain network on script-based blockchains becomes a promising research direction. Scriptless schemes[57–59] in script-based blockchains also deserve further investigation.

(2) *Design tools for script-based blockchains*. There is not much research on the design tools for script-based blockchains. The analysis techniques for Turing-complete blockchains (see Section 5.2.1) may be migrated into script-based blockchains. Moreover, high-level languages with formal proofs[66,67] is a promising research direction.

(3) *Design paradigms for Turing-complete blockchains*. The investigation of smart contracts' best practices[22,85,86] and design paradigms[28,87,88] is a new research direction, requiring massive and continuous efforts. Additionally, mitigation of practical cryptographic protocols[257] to Turing-complete blockchains might be a research direction in the future. Moreover, contracts that are provably secure or private within the UC model[197] is a trend for theoretic research.[173,177–181]

(4) *Design tools for Turing-complete blockchains*. Apart from commonly used techniques such as symbolic execution,[89,115–123] the fuzzing test[137–139] is another prevalent technique. Lightweight and scalable security analysis could be further considered to meet the potential need for the growth of contract size. Moreover, analysis tools providing useful suggestions or counter-examples for correction are attractive, as reported previously.[108,112,121,136]. Programming languages[149,152,155,156] that support formal proofs are also promising. Finally, auxiliary frameworks with high accessibility and expressivity (e.g., Eberhardt and Tai[141]) could be further considered in the future.

(5) *Private contract systems*. In recent years, some researchers have combined various cryptographic schemes such as secure multi-party computation[160–162] and ZKP[6,140,141] to execute private contracts. Exploring efficient and economic solutions for privacy-preserving smart contracts is a promising research direction. Cryptographic schemes such as blind signatures[171] and ring signatures[258] might be adopted to obtain better anonymity. Additionally, inspired by Kerber et al.,[259] we remark that private contracts with formal proofs (especially within the UC model[197]) are also promising for theoretical research. Finally, compilers that convert general functions into protocols in private contract systems also deserve consideration.

## 8. CONCLUSION

In this paper, we have made a categorization of the 159 studies collected. For the schemes on smart contracts construction, we first group them into script-based and Turing-complete blockchains. Schemes in these two kinds of platforms are then divided into design paradigms and tools. The design paradigms refer to practical patterns for common smart contract applications, and the design tools refer to analysis or auxiliary tools that detect vulnerabilities or assist developers during contract construction. We categorize the related extensions and alternative systems for contract executions into private contracts with extra tools, off-chain channels, and extensions on core functionalities. Each category is further divided into smaller aspects according to the techniques they use (e.g., SMPC protocols, ZKP, and TEE) or their targets (e.g., privacy and efficiency). This work is aimed at providing insights for new researchers and developers in this field.

According to our survey, problems such as privacy leakage, execution efficiency, and contract complexity have restricted the application scenarios of smart contracts. We conclude the challenges for smart contracts should be universally adopted into five aspects: (1) frequently occurred vulnerabilities; (2) incomplete design paradigms; (3) inefficient analysis tools; (4) low processing rate and limited contract complexity; (5) lack of privacy. These challenges have attracted widespread attention in the literature and are still open problems that deserve future research.

Future work should pay attention to fair and economic off-chain network schemes, high-level languages with formal proofs, best practices, practical implementation of cryptographic protocols, provably secure or provably private contracts, scalable or automatic analysis tools, private contracts with cryptographic techniques, and practical compilers for private contracts. These directions could alleviate the concern of users and promote the development of smart contracts.

Table 8. Comparison of different types of blockchains

| Type | Execute transactions | Access data | Generate transactions | Examples |
|---|---|---|---|---|
| Public | anyone | anyone | anyone | Bitcoin, Ethereum |
| Open consortium | permitted | registered/anyone | registered/anyone | Ripple, Libra |
| Closed consortium | permitted | permitted | permitted | Corda, Quorum, Hyperledger Fabric |
| Private | owner | owner | owner | locally running blockchain |

The word "registered" here refers to a less restricted condition than "permitted," and involves more participants.

## DEFINITIONS

In this section, we present and rephrase 17 essential definitions and concepts frequently used in our work. Definitions that are only used once are introduced when they first appear in context.

With the development of blockchain technology, there have been several kinds of blockchains with distinct properties. In this paper, we categorize blockchains into three categories—public, consortium, and private blockchains—according to the works of Bano et al., Yagas et al., and Garay et al.[260–262]

### Definition 1 (public blockchain)

In a public blockchain, any node is permitted to join the maintenance of data on the blockchain, and the data are publicly accessible and verifiable. Anyone is allowed to deploy, call, and execute smart contracts through generating and executing transactions.

### Definition 2 (consortium blockchain)

In a consortium blockchain, the nodes that are responsible for executing transactions (i.e., maintaining the data) are determined in advance. How the blockchain data (including transactions and smart contracts) can be proposed and accessed depends on the openness of the consortium:

- Closed consortium: only predetermined nodes can propose records and can access the data
- Open consortium: anyone can register (through a centralized party) to obtain the right of proposing or accessing records on the blockchain

### Definition 3 (private blockchain)

A private blockchain refers to one that is completely controlled by an individual party. It is only used to record private information, and only the owner has the right to access and maintain the data. In practice, common public blockchains include Bitcoin,[1] Ethereum,[2,3] and so forth, and they put no limits on the entering of the P2P network. Common closed consortium blockchains include Corda,[5] Quorum,[6] and Hyperledger Fabric.[7] Such closed consortium blockchains are mainly designed for business partners to cooperate better. Common open consortium blockchains include Ripple[263] and Libra,[264] and these open consortium blockchains offer more transparency by allowing anyone to register to join the network while keeping the rights of maintaining the membership of all the participants. Both public and consortium blockchains mentioned here can be considered as private blockchains when running locally by an individual party.

A brief comparison of the blockchains discussed above is shown in Table 8. Note that the word "registered" means a party should register to a centralized organization before they have the right to access data or generate transactions, while "permitted" refers to a more restricted authorization to determine the membership and occurs less often in practice. In other words, the restriction on permission is stronger than the registration procedure in the context.

Next, to better understand the execution mechanism of the blockchain, we give the definition of consensus that is one of the key techniques within the blockchain, according to the works of Garay et al.[265] and Pass et al.[266]:

### Definition 4 (consensus)

A consensus mechanism enables all participating nodes, whether honest or malicious, to agree on the contents of a blockchain. In a consensus mechanism, the following properties must be satisfied:

- Liveness: any transaction should be finally processed
- Persistence: if an honest party validates a transaction (accept or reject), all other honest nodes will eventually have the same operation

In many related studies, the notion of miners is used to refer to the participants in a blockchain system. We give a simple definition of it as follows.

### Definition 5 (miner)

A miner refers to a node providing its non-trivial work in a consensus mechanism for the rewards in a blockchain. During the maintenance of a blockchain within the consensus mechanism, there are cases when nodes disagree on the final results. This is called a fork in the context of blockchain:

### Definition 6 (fork)

A fork refers to a disagreement on blockchain records among participating nodes.

Forks are usually temporal and will finally be eliminated by the consensus rule. However, under other circumstances a fork may be deliberately triggered to launch an update of the blockchain system. The concept of the fork can be further divided into soft and hard forks, as defined below.

#### *Definition 7 (soft fork)*

A soft fork refers to a fork caused by the update of backward compatible consensus.

#### *Definition 8 (hard fork)*

A hard fork refers to the fork caused by the update of non-backward compatible consensus.

Note that after a soft fork,[267] some transactions or blocks that are valid under the old rules may become invalid while after a hard fork,[267] the transactions or blocks under the new rules become invalid under the old rules.

A soft fork is mainly used to introduce new types of transactions or to fix some bugs in the consensus protocol. It does not require all nodes to switch to the new consensus. Nodes running the old consensus can still recognize the transactions and blocks under the new rules while in comparison, a hard fork usually arises when big events (e.g., the DAO attack) or major disputes in the community occur, and all nodes have to choose one of the forks and end up with two distinct blockchains that are not compatible with each other.

In blockchain systems, users and smart contracts rely on transactions to contact with each other. Therefore, we give a universal definition of a transaction as the following.

### Definition 9 (transaction)

A transaction $Tx$ is a tuple of five elements, i.e., $Tx = (t, in, out, s, pld)$ where $Tx.t$ is the timestamp that a miner receives $Tx$. We assume that at most one $Tx$ could be received at time $t$, namely, $\forall i \neq j, Tx_i.t \neq Tx_j.t$ always holds. With this assumption, transactions will be executed in chronological order (this order may vary among miners). $Tx.in$ (resp. $Tx.out$) is the input (resp. output) of the transaction. $Tx.s$ is the signature of the transaction, which is used to show the ownership of the fund to be transferred in the transaction. $Tx.pld$ refers to arbitrary messages appended to the transaction, and is called payload data in this paper. In fact, the specific contents of a transaction (i.e., the format and structure of each element) vary among blockchains according to the underlying user model. Taking Bitcoin and Ethereum as examples, Bitcoin adopts the unspent transaction output (UTXO) model, while Ethereum uses the account model. This is one of the key differences between these two platforms, and most existing blockchains also adopt either one of these two models. Therefore, we give the definitions of the UTXO and the account model in the following.

### Definition 10 (UTXO model)

In the UTXO model, unspent money is stored in UTXOs. Each transaction consumes existing UTXOs and generates new UTXOs, except the coinbase transaction that assigns the miner a UTXO without inputs as a reward. For a UTXO $U$, it contains information such as the source addresses and the values. For a transaction $Tx$ within the UTXO model, the sum of values in the output UTXOs must be less than or equal to that in the input UTXOs, i.e.: $\sum_{U \in Tx.out} U.v \leq \sum_{U \in Tx.in} U.v$ where $U.v$ refers to the value of $U$ and the extra value in the input is collected by the miners as the execution fee.

### Definition 11 (account model)

In the account model, each user or contract has a fixed account and address. The account records the balance, the contract codes, and the state data specified in the creating transaction. The balance $F_{balance}(a)$ in the account corresponding to address a must be non-negative. In addition, for a transaction to be valid, the input amount $Tx.in$ to be spent should be less than or equal to the balance in the account, i.e., $F_{balance}(a) \geq 0, F_{value}(Tx.in) \leq F_{balance}(a)$, where $F_{value}(Tx.in)$ indicates the value contained in $Tx.in$.

As mentioned previously, the transaction data structure is different among blockchains within these two models. Specifically, in the UTXO model, $Tx.in$ includes a set of UTXOs to be spent, while in the account model it directly refers to the value to be transferred. Similarly, $Tx.out$ includes a new UTXO set in the UTXO model, while it includes responses from the target address (e.g., returned messages from a smart contract) in the account model.

With the above definitions, we are able to give a formal definition of smart contracts. We remark that our definition is inspired by the description of the world state and transactions in Ethereum Yellow Paper,[3] and the ideal smart contract functionality $\mathcal{F}^*_{StCon}$ in Bentov et al.[178]

### Definition 12 (smart contract)

A smart contract refers to a computer program $C$ deployed on a blockchain with public interfaces and state variables, satisfying

$$C(S_i, Tx_i) = (S_j, R_i),$$

where $\mathcal{S} = \{S_{i \in \mathbb{N}^*}\}$ is the set of all possible states in $C$, $\mathcal{T} = \{Tx_{i \in \mathbb{N}^*} = (t, in, out, s, pld)_{i \in \mathbb{N}^*}\}$ is the transaction set, and $\mathcal{R} = \{R_{i \in \mathbb{N}^*}\}$ is the set of possible responses from the contract, e.g., the success or failure symbol of execution, or any other predefined values. After $C$ is called by a valid $Tx_i$, the new state $S_j$, and the response $R_i$ are produced accordingly.

Studies on smart contracts mainly focus on two aspects, which are defined as security and correctness as follows.

### Definition 13 (security of smart contracts)

The security of smart contracts refers to the ability to resist unauthorized state change, including fund transfer, state tampering, and accidental self-destruction.

### Definition 14 (correctness of smart contracts)

The correctness of smart contracts refers to the ability to correctly realize the expected functionality.

To ensure the security and correctness of smart contracts and to achieve other desired properties such as privacy and efficiency, cryptographic schemes and hardware equipment may be introduced. Here we briefly give the definitions of secure multi-party computation, zero-knowledge proof, and trusted execution environment in the following.

### Definition 15 (secure multi-party computation[268])

In a secure multi-party computation protocol $\pi$, participants $P_1, P_2, \cdots, P_n$ can jointly evaluate a probabilistic polynomial time function $f(x_1, x_2, \cdots, x_n) = (y_1, y_2, \cdots, y_n)$ where $x_i$ (resp. $y_i$) is the secret input (resp. output) of $P_i$ ($i = 1, 2, \cdots, n$), and the following two properties hold:

- Correctness: each $P_i$ gets the correct result
- Privacy: any $P_i$ cannot get extra information except his own input and output, especially the inputs and outputs of other participants $P_j$ where $j \neq i$

### Definition 16 (zero-knowledge proof[269])

In a proof system $\langle P, V \rangle(x)$, a prover $P$ proves to a verifier $V$ that $x$ belongs to a language $L$, which is an $NP$ problem, i.e., $x \in L$, $L \in NP$. A protocol $\pi$ is said to be a zero-knowledge proof protocol if the following three properties are satisfied:

- Completeness: any true statement can be accepted with an overwhelming probability

- Soundness: any false statement can only be accepted with a negligible probability
- Zero-knowledge: any probabilistic polynomial time verifier cannot get extra information other than $x \in L$, and its view is indistinguishable from that of a simulator $F_{sim}$

**Definition 17 (trusted execution environment[270])**

Trusted execution environment is a kind of hardware equipment, usually an enclave in the memory, which ensures that the execution environment is not influenced or manipulated. Trusted execution environment guarantees the reliability of the execution results and the privacy of executed contents.

## ACKNOWLEDGMENTS

The authors would like to thank the reviewers and editors for their valuable comments and guidance to make our work more comprehensive. This work is supported in part by the National Key R&D Program of China (2017YFB1400702), in part by the National Natural Science Foundation of China (61972017, 61972018, 61972014, 72031001), in part by the National Cryptography Development Fund (MMJJ20180215), and in part by the Fundamental Research Funds for the Central Universities (YWF-20-BJ-J-1039).

## AUTHOR CONTRIBUTIONS

Conceptualization, B.H. and Z.Z.; methodology, B.H. and Z.Z.; investigation, B.H.; validation, Z.Z.; writing – original draft, B.H.; writing – review & editing, B.H., Z.Z., Y.L., and J.Y.; visualization, B.H., Y.L., and J.Y.; supervision, Z.Z., J.L., R.L., and X.L.; funding acquisition, J.L. and Z.Z.

## REFERENCES

1. Nakamoto, S. (2008). Bitcoin: a peer-to-peer electronic cash system. https://bitcoin.org/bitcoin.pdf.
2. Buterin, V. (2014). A next-generation smart contract and decentralized application platform. https://whitepaperdatabase.com/wp-content/uploads/2017/09/Ethereum-ETH-whitepaper.pdf.
3. Wood, G. (2014). Ethereum: a secure decentralised generalised transaction ledger. https://files.gitter.im/ethereum/yellowpaper/VIyt/Paper.pdf.
4. N. Szabo, Formalizing and securing relationships on public networks, First Monday 2 .
5. Brown, R.G., Carlyle, J., Grigg, I., and Hearn, M. (2016). Corda: an introduction. https://docs.corda.net/_static/corda-introductory-whitepaper.pdf.
6. Harris, O. (2016). Quorum. https://github.com/jpmorganchase/quorum/wiki.
7. Androulaki, E., Barger, A., Bortnikov, V., Cachin, C., Christidis, K., Caro, A.D., Enyeart, D., Ferris, C., Laventman, G., Manevich, Y., et al. (2018). Hyperledger Fabric: a distributed operating system for permissioned blockchains. In Proceedings of the Thirteenth EuroSys Conference, EuroSys 2018, Porto, Portugal, April 23–26, 2018, pp. 30:1–30:15.
8. (2020). Hyperledger fabric case studies. https://www.hyperledger.org/learn/case-studies.
9. Ron, D., and Shamir, A. (2013). Quantitative analysis of the full bitcoin transaction graph. In Financial Cryptography and Data Security - 17th International Conference, FC 2013, Okinawa, Japan, April 1–5, 2013, Revised Selected Papers, Vol. 7859 of Lecture Notes in Computer Science (Springer), pp. 6–24.
10. Meiklejohn, S., Pomarole, M., Jordan, G., Levchenko, K., McCoy, D., Voelker, G.M., and Savage, S. (2016). A fistful of bitcoins: characterizing payments among men with no names. Commun. ACM *59*, 86–93.
11. Siegel, D. (2016). Understanding the DAO attack. https://www.coindesk.com/understanding-dao-hack-journalists/.
12. Atzei, N., Bartoletti, M., and Cimoli, T. (2017). A survey of attacks on Ethereum smart contracts (Sok). In Principles of Security and Trust - 6th International Conference, POST 2017, Held as Part of the European Joint Conferences on Theory and Practice of Software, ETAPS 2017, Uppsala, Sweden, April 22–29, 2017, Proceedings, pp. 164–186.
13. Conti, M., E, S.K., Lal, C., and Ruj, S. (2018). A survey on security and privacy issues of bitcoin. IEEE Commun. Surv. Tutorials *20*, 3416–3452.
14. Zheng, Z., Xie, S., Dai, H., Chen, X., and Wang, H. (2018). Blockchain challenges and opportunities: a survey. Int. J. Web Grid Serv. *14*, 352–375.
15. Kitchenham, B. (2004). Procedures for performing systematic reviews, joint Technical Report. http://www.inf.ufsc.br/~aldo.vw/kitchenham.pdf.
16. Nakamoto, S. (2009). Bitcointalk: Bitcoin forum. bitcointalk.org, accessed August 1, 2020.
17. Bitcoin Wiki, 2010. en.bitcoin.it, accessed August 1, 2020.
18. Tschorsch, F., and Scheuermann, B. (2016). Bitcoin and beyond: a technical survey on decentralized digital currencies. IEEE Commun. Surv. Tutorials *18*, 2084–2123.
19. Maxwell, G. (2011). Bitcoin wiki: zero knowledge contingent payment. https://en.bitcoin.it/wiki/Zero_Knowledge_Contingent_Payment.
20. Hearn, M. (2012). Bitcoin wiki: contracts. https://en.bitcoin.it/wiki/Contract.
21. BitcoinWiki. (2020). Bitcoin wiki: script (MediaWiki). https://en.bitcoin.it/wiki/Script.
22. Holscher, E. (2016). Solidity docs. https://solidity.readthedocs.io/en/latest/solidity-by-example.html.
23. Buterin, V. (2017). Serpent language. https://github.com/ethereum/serpent.
24. Bitcoin Wiki. (2015). Bitcoin wiki: multisignature. https://en.bitcoin.it/wiki/Multisignature.
25. Poon, J., and Dryja, T. (2016). The Bitcoin Lightning Network: scalable off-chain instant payments. https://www.bitcoinlightning.com/wp-content/uploads/2018/03/lightning-network-paper.pdf.
26. Seijas, P.L., Thompson, S.J., and McAdams, D. (2016). Scripting smart contracts for distributed ledger technology. http://eprint.iacr.org/2016/1156.
27. Nxt Community (2014). Nxt whitepaper. https://nxtwiki.org.
28. Bartoletti, M., and Pompianu, L. (2017). An empirical analysis of smart contracts: platforms, applications, and design patterns. In Financial Cryptography and Data Security - FC 2017 International Workshops, WAHC, BITCOIN, VOTING, WTSC, and TA, Sliema, Malta, April 7, 2017, Revised Selected Papers, pp. 494–509.
29. Counterparty.io. (2014). Counterparty: protocol specification. https://counterparty.io/docs/protocol_specification/.
30. Mazieres, D. (2015). The stellar consensus protocol: a federated model for internet-level consensus. https://www.stellar.org/papers/stellar-consensus-protocol.
31. Davis, S. (2014). Monax. https://monax.io/.
32. Kordek, M. (2016). Lisk. https://lisk.io/documentation/lisk-sdk/index.html.
33. Junis, F., Prasetya, F.M.W., Lubay, F.I., and Sari, A.K. (2019). A revisit on blockchain-based smart contract technology. http://arxiv.org/abs/1907.09199.
34. Alharby, M., and van Moorsel, A. (2017). Blockchain-based smart contracts: a systematic mapping study. http://arxiv.org/abs/1710.06372.
35. Dika, A. (2017). Ethereum Smart Contracts: Security Vulnerabilities and Security Tools, Master's thesis (NTNU).

36. Macrinici, D., Cartofeanu, C., and Gao, S. (2018). Smart contract applications within blockchain technology: a systematic mapping study. Telematics Inform. 35, 2337–2354.

37. Ayman, A., Aziz, A., Alipour, A., and Laszka, A. (2019). Smart contract development in practice: trends, issues, and discussions on stack overflow. http://arxiv.org/abs/1905.08833.

38. Harz, D., and Knottenbelt, W.J. (2018). Towards safer smart contracts: a survey of languages and verification methods. http://arxiv.org/abs/1809.09805.

39. Angelo, M.D., and Salzer, G. (2019). A survey of tools for analyzing Ethereum smart contracts. In IEEE International Conference on Decentralized Applications and Infrastructures, DAPPCON 2019, Newark, CA, USA, April 4–9, 2019, pp. 69–78.

40. Liu, J., and Liu, Z. (2019). A survey on security verification of blockchain smart contracts. IEEE Access 7, 77894–77904.

41. Ante, L. (2020). Smart contracts on the blockchain—a bibliometric analysis and review. https://papers.ssrn.com/sol3/papers.cfm?abstract_id=3576393.

42. Almakhour, M., Sliman, L., Samhat, A.E., and Mellouk, A. (2020). Verification of smart contracts: a survey. Pervasive Mobile Comput. 67, 101227.

43. Bitcoin.org. (2014). Bitcoin release 0.9.0. https://bitcoin.org/en/release/v0.9.0.

44. Bartoletti, M., and Pompianu, L. (2017). An analysis of bitcoin op_return metadata. In Financial Cryptography and Data Security - FC 2017 International Workshops, WAHC, BITCOIN, VOTING, WTSC, and TA, Sliema, Malta, April 7, 2017, Revised Selected Papers, pp. 218–230.

45. Faisal, T., Courtois, N., and Serguieva, A. (2018). The evolution of embedding metadata in blockchain transactions. In 2018 International Joint Conference on Neural Networks, IJCNN 2018, Rio de Janeiro, Brazil, July 8–13, 2018, pp. 1–9.

46. Andrychowicz, M., Dziembowski, S., Malinowski, D., and Mazurek, L. (2014). Secure multiparty computations on bitcoin. In 2014 IEEE Symposium on Security and Privacy, SP 2014, Berkeley, CA, USA, May 18–21, 2014, pp. 443–458.

47. Andrychowicz, M., Dziembowski, S., Malinowski, D., and Mazurek, L. (2014). Fair two-party computations via bitcoin deposits. In Financial Cryptography and Data Security - FC 2014 Workshops, BITCOIN and WAHC 2014, Christ Church, Barbados, March 7, 2014, Revised Selected Papers, pp. 105–121.

48. Bartoletti, M., and Zunino, R. (2017). Constant-deposit multiparty lotteries on bitcoin. In Financial Cryptography and Data Security - FC 2017 International Workshops, WAHC, BITCOIN, VOTING, WTSC, and TA, Sliema, Malta, April 7, 2017, Revised Selected Papers, pp. 231–247.

49. Kumaresan, R., Moran, T., and Bentov, I. (2015). How to use bitcoin to play decentralized poker. In Proceedings of the 22nd ACM SIGSAC Conference on Computer and Communications Security, Denver, CO, USA, October 12–16, 2015, pp. 195–206.

50. Bentov, I., and Kumaresan, R. (2014). How to use bitcoin to design fair protocols. In Advances in Cryptology - CRYPTO 2014 - 34th Annual Cryptology Conference, Santa Barbara, CA, USA, August 17–21, 2014, Proceedings, Part II, pp. 421–439.

51. Kumaresan, R., and Bentov, I. (2016). Amortizing secure computation with penalties. In Proceedings of the 2016 ACM SIGSAC Conference on Computer and Communications Security, Vienna, Austria, October 24–28, 2016, pp. 418–429.

52. Kumaresan, R., Vaikuntanathan, V., and Vasudevan, P.N. (2016). Improvements to secure computation with penalties. In Proceedings of the 2016 ACM SIGSAC Conference on Computer and Communications Security, Vienna, Austria, October 24–28, 2016, pp. 406–417.

53. Kiayias, A., Zhou, H., and Zikas, V. (2016). Fair and robust multi-party computation using a global transaction ledger. In Advances in Cryptology - EUROCRYPT 2016 - 35th Annual International Conference on the Theory and Applications of Cryptographic Techniques, Vienna, Austria, May 8–12, 2016, Proceedings, Part II, pp. 705–734.

54. Pass, R., and Shelat, A. (2015). Micropayments for decentralized currencies. In Proceedings of the 22nd ACM SIGSAC Conference on Computer and Communications Security, Denver, CO, USA, October 12–16, 2015, I. Ray, N. Li, and C. Kruegel, eds. (ACM), pp. 207–218.

55. Hu, K., and Zhang, Z. (2018). Fast lottery-based micropayments for decentralized currencies. In Information Security and Privacy - 23rd Australasian Conference, ACISP 2018, Wollongong, NSW, Australia, July 11–13, 2018, Proceedings, pp. 669–686.

56. Chiesa, A., Green, M., Liu, J., Miao, P., Miers, I., and Mishra, P. (2017). Decentralized anonymous micropayments. In Advances in Cryptology - EUROCRYPT 2017 - 36th Annual International Conference on the Theory and Applications of Cryptographic Techniques, Paris, France, April 30 - May 4, 2017, Proceedings, Part II, pp. 609–642.

57. Banasik, W., Dziembowski, S., and Malinowski, D. (2016). Efficient zero-knowledge contingent payments in cryptocurrencies without scripts. In Computer Security - ESORICS 2016 - 21st European Symposium on Research in Computer Security, Heraklion, Greece, September 26–30, 2016, Proceedings, Part II, pp. 261–280.

58. Poelstra, A. (2017). Scriptless scripts. https://download.wpsoftware.net/bitcoin/wizardry/mw-slides/2017-05-milan-meetup/slides.pdf.

59. Malavolta, G., Moreno-Sánchez, P., Schneidewind, C., Kate, A., and Maffei, M. (2019). Anonymous multi-hop locks for blockchain scalability and interoperability. In 26th Annual Network and Distributed System Security Symposium, NDSS 2019, San Diego, California, USA, February 24–27, 2019.

60. Andrychowicz, M., Dziembowski, S., Malinowski, D., and Mazurek, L. (2014). Modeling bitcoin contracts by timed automata. In 2014. Proceedings, pp. 7–22.

61. Bigi, G., Bracciali, A., Meacci, G., and Tuosto, E. (2015). Validation of decentralised smart contracts through game theory and formal methods. In Programming Languages with Applications to Biology and Security: Essays Dedicated to Pierpaolo Degano on the Occasion of His 65th Birthday, pp. 142–161.

62. Atzei, N., Bartoletti, M., Lande, S., and Zunino, R. (2018). A formal model of bitcoin transactions. In Financial Cryptography and Data Security - 22nd International Conference, FC 2018, Nieuwpoort, Curaçao, February 26–March 2, 2018, Revised Selected Papers, pp. 541–560.

63. Atzei, N., Bartoletti, M., Cimoli, T., Lande, S., and Zunino, R. (2018). Sok: unraveling bitcoin smart contracts. In Principles of Security and Trust - 7th International Conference, POST 2018, Held as Part of the European Joint Conferences on Theory and Practice of Software, ETAPS 2018, Thessaloniki, Greece, April 14–20, 2018, Proceedings, pp. 217–242.

64. Atzei, N. (2018). Balzac: bitcoin abstract language, analyzer and compiler. https://blockchain.unica.it/balzac/.

65. ivy lang.org. (2017). Ivy. https://docs.ivylang.org/bitcoin/.

66. O'Connor, R. (2017). Simplicity: a new language for blockchains. In Proceedings of the 2017 Workshop on Programming Languages and Analysis for Security, PLAS@CCS 2017, Dallas, TX, USA, October 30, 2017, pp. 107–120.

67. Bartoletti, M., and Zunino, R. (2018). Bitml: a calculus for bitcoin smart contracts. In Proceedings of the 2018 ACM SIGSAC Conference on Computer and Communications Security, CCS 2018, Toronto, ON, Canada, October 15–19, 2018, pp. 83–100.

68. Bartoletti, M., Cimoli, T., and Zunino, R. (2018). Fun with bitcoin smart contracts. In Leveraging Applications of Formal Methods, Verification and Validation. Industrial Practice - 8th International Symposium, ISoLA 2018, Limassol, Cyprus, November 5–9, 2018, Proceedings, Part IV, pp. 432–449.

69. Atzei, N., Bartoletti, M., Lande, S., Yoshida, N., and Zunino, R. (2019). Developing secure bitcoin contracts with bitml. In Proceedings of the ACM Joint Meeting on European Software Engineering Conference and Symposium on the Foundations of Software Engineering, ESEC/SIGSOFT FSE 2019, Tallinn, Estonia, August 26–30, 2019, pp. 1124–1128.

70. Miller, A., and Bentov, I. (2017). Zero-collateral lotteries in bitcoin and Ethereum. In 2017 IEEE European Symposium on Security and Privacy

Workshops, EuroS&P Workshops 2017, Paris, France, April 26–28, 2017 (IEEE), pp. 4–13.

71. Okoye, M.C., and Clark, J. (2018). Toward cryptocurrency lending. In Financial Cryptography and Data Security - FC 2018 International Workshops, BITCOIN, VOTING, and WTSC, Nieuwpoort, Curaçao, March 2, 2018, Revised Selected Papers, pp. 367–380.

72. Norta, A., Leiding, B., and Lane, A. (2019). Lowering financial inclusion barriers with a blockchain-based capital transfer system. In IEEE INFOCOM 2019 - IEEE Conference on Computer Communications Workshops, INFOCOM Workshops 2019, Paris, France, April 29 - May 2, 2019 (IEEE), pp. 319–324.

73. Ølnes, S. (2016). Beyond bitcoin enabling smart government using blockchain technology. In Electronic Government - 15th IFIP WG 8.5 International Conference, EGOV 2016, Guimarães, Portugal, September 5–8, 2016, Proceedings, Vol. 9820 of Lecture Notes in Computer Science (Springer), pp. 253–264.

74. Ølnes, S., Ubacht, J., and Janssen, M. (2017). Blockchain in government: benefits and implications of distributed ledger technology for information sharing. Government Inf. Q. *34*, 355–364.

75. Hou, H. (2017). The application of blockchain technology in e-government in China. In 26th International Conference on Computer Communication and Networks, ICCCN 2017, Vancouver, BC, Canada, July 31 - Aug. 3, 2017 (IEEE), pp. 1–4.

76. Abodei, E., Norta, A., Azogu, I., Udokwu, C., and Draheim, D. (2019). Blockchain technology for enabling transparent and traceable government collaboration in public project processes of developing economies. In Digital Transformation for a Sustainable Society in the 21st Century - 18th IFIP WG 6.11 Conference on E-Business, E-Services, and E-Society, I3E 2019, Trondheim, Norway, September 18–20, 2019, Proceedings, Vol. 11701 of Lecture Notes in Computer Science (Springer), pp. 464–475.

77. Krogsbøll, M., Borre, L.H., Slaats, T., and Debois, S. (2020). Smart contracts for government processes: case study and prototype implementation (short paper). In Financial Cryptography and Data Security - 24th International Conference, FC 2020, Kota Kinabalu, Malaysia, February 10–14, 2020 Revised Selected Papers, Vol. 12059 of Lecture Notes in Computer Science (Springer), pp. 676–684.

78. Blass, E., and Kerschbaum, F. (2018). Strain: a secure auction for blockchains. In Computer Security - 23rd European Symposium on Research in Computer Security, ESORICS 2018, Barcelona, Spain, September 3–7, 2018, Proceedings, Part I, pp. 87–110.

79. Galal, H.S., and Youssef, A.M. (2018). Verifiable sealed-bid auction on the Ethereum blockchain. In Financial Cryptography and Data Security - FC 2018 International Workshops, BITCOIN, VOTING, and WTSC, Nieuwpoort, Curaçao, March 2, 2018, Revised Selected Papers, pp. 265–278.

80. Eberhardt, J., and Tai, S. (2017). On or off the blockchain? insights on off-chaining computation and data. In Service-Oriented and Cloud Computing - 6th IFIP WG 2.14 European Conference, ESOCC 2017, Oslo, Norway, September 27–29, 2017, Proceedings, pp. 3–15.

81. Molina-Jiménez, C., Solaiman, E., Sfyrakis, I., Ng, I., and Crowcroft, J. (2018). On and off-blockchain enforcement of smart contracts. In Euro-Par 2018: Parallel Processing Workshops - Euro-Par 2018 International Workshops, Turin, Italy, August 27–28, 2018, Revised Selected Papers, Vol. 11339 of Lecture Notes in Computer Science (Springer), pp. 342–354.

82. Molina-Jiménez, C., Sfyrakis, I., Solaiman, E., Ng, I., Wong, M.W., Chun, A., and Crowcroft, J. (2018). Implementation of smart contracts using hybrid architectures with on and off-blockchain components. In 8th IEEE International Symposium on Cloud and Service Computing, SC2 2018, Paris, France, November 18–21, 2018 (IEEE), pp. 83–90.

83. Li, C., Palanisamy, B., and Xu, R. (2019). Scalable and privacy-preserving design of on/off-chain smart contracts. In 35th IEEE International Conference on Data Engineering Workshops, ICDE Workshops 2019, Macao, China, April 8–12, 2019 (IEEE), pp. 7–12.

84. Norta, A., Hawthorne, D., and Engel, S.L. (2018). A privacy-protecting data-exchange wallet with ownership- and monetization capabilities. In 2018 International Joint Conference on Neural Networks, IJCNN 2018, Rio de Janeiro, Brazil, July 8–13, 2018 (IEEE), pp. 1–8.

85. Diligence, C. (2020). Ethereum smart contract security best practices. https://consensys.github.io/smart-contract-best-practices/.

86. OpenZeppelin. (2020). OpenZeppelin: contracts. https://github.com/OpenZeppelin/openzeppelin-contracts.

87. Wöhrer, M., and Zdun, U. (2018). Design patterns for smart contracts in the Ethereum ecosystem. In IEEE International Conference on Internet of Things (iThings) and IEEE Green Computing and Communications (GreenCom) and IEEE Cyber, Physical and Social Computing (CPSCom) and IEEE Smart Data (SmartData), iThings/GreenCom/CPSCom/SmartData 2018, Halifax, NS, Canada, July 30 - August 3, 2018, pp. 1513–1520.

88. Wöhrer, M., and Zdun, U. (2018). Smart contracts: security patterns in the Ethereum ecosystem and solidity. In 2018 International Workshop on Blockchain Oriented Software Engineering, IWBOSE@SANER 2018, Campobasso, Italy, March 20, 2018, pp. 2–8.

89. Luu, L., Chu, D., Olickel, H., Saxena, P., and Hobor, A. (2016). Making smart contracts smarter. In Proceedings of the 2016 ACM SIGSAC Conference on Computer and Communications Security, Vienna, Austria, October 24–28, 2016, pp. 254–269.

90. Grishchenko, I., Maffei, M., and Schneidewind, C. (2018). A semantic framework for the security analysis of Ethereum smart contracts. In Principles of Security and Trust - 7th International Conference, POST 2018, Held as Part of the European Joint Conferences on Theory and Practice of Software, ETAPS 2018, Thessaloniki, Greece, April 14–20, 2018, Proceedings, pp. 243–269.

91. Mense, A., and Flatscher, M. (2018). Security vulnerabilities in Ethereum smart contracts. In Proceedings of the 20th International Conference on Information Integration and Web-Based Applications & Services, iiWAS 2018, Yogyakarta, Indonesia, November 19–21, 2018 (ACM), pp. 375–380.

92. Dika, A., and Nowostawski, M. (2018). Security vulnerabilities in Ethereum smart contracts. In IEEE International Conference on Internet of Things (iThings) and IEEE Green Computing and Communications (GreenCom) and IEEE Cyber, Physical and Social Computing (CPSCom) and IEEE Smart Data (SmartData), iThings/GreenCom/CPSCom/SmartData 2018, Halifax, NS, Canada, July 30 - August 3, 2018 (IEEE), pp. 955–962.

93. Pérez, D., and Livshits, B. (2021). Smart contract vulnerabilities: vulnerable does not imply exploited. In 30th USENIX Security Symposium, USENIX Security 2021, Vancouver, B.C, Canada, August 11–13, 2021.

94. Groce, A., Feist, J., Grieco, G., and Colburn, M. (2020). What are the actual flaws in important smart contracts (and how can we find them)? In Financial Cryptography and Data Security - 24th International Conference, FC 2020, Kota Kinabalu, Malaysia, February 10–14, 2020 Revised Selected Papers, Vol. 12059 of Lecture Notes in Computer Science (Springer), pp. 634–653.

95. Delmolino, K., Arnett, M., Kosba, A.E., Miller, A., and Shi, E. (2016). Step by step towards creating a safe smart contract: lessons and insights from a cryptocurrency lab. In Financial Cryptography and Data Security - FC 2016 International Workshops, BITCOIN, VOTING, and WAHC, Christ Church, Barbados, February 26, 2016, Revised Selected Papers, pp. 79–94.

96. Angelo, M.D., Sack, C., and Salzer, G. (2019). Sok: development of secure smart contracts - lessons from a graduate course. In Financial Cryptography and Data Security - FC 2019 International Workshops, VOTING and WTSC, St. Kitts, St. Kitts and Nevis, February 18–22, 2019, Revised Selected Papers, pp. 91–105.

97. Clack, C.D., Bakshi, V.A., and Braine, L. (2016). Smart contract templates: foundations, design landscape and research directions. http://arxiv.org/abs/1608.00771.

98. Clack, C.D., Bakshi, V.A., and Braine, L. (2016). Smart contract templates: essential requirements and design options. http://arxiv.org/abs/1612.04496.

99. Marino, B., and Juels, A. (2016). Setting standards for altering and undoing smart contracts. In Rule Technologies. Research, Tools, and Applications - 10th International Symposium, RuleML 2016, Stony Brook, NY, USA, July 6–9, 2016. Proceedings, pp. 151–166.

100. Grossman, S., Abraham, I., Golan-Gueta, G., Michalevsky, Y., Rinetzky, N., Sagiv, M., and Zohar, Y. (2018). Online detection of effectively callback free objects with applications to smart contracts. In Proceedings of the ACM on Programming Languages, *2* (POPL), pp. 48:1–48:28.

101. Liu, C., Liu, H., Cao, Z., Chen, Z., Chen, B., and Roscoe, B. (2018). Reguard: finding reentrancy bugs in smart contracts. In Proceedings of the 40th International Conference on Software Engineering: Companion Proceeedings, ICSE 2018, Gothenburg, Sweden, May 27 - June 03, 2018, pp. 65–68.

102. Rodler, M., Li, W., Karame, G.O., and Davi, L. (2019). Sereum: protecting existing smart contracts against re-entrancy attacks. In 26th Annual Network and Distributed System Security Symposium, NDSS 2019, San Diego, California, USA, February 24–27, 2019.

103. Chen, T., Li, X., Luo, X., and Zhang, X. (2017). Under-optimized smart contracts devour your money. In IEEE 24th International Conference on Software Analysis, Evolution and Reengineering, SANER 2017, Klagenfurt, Austria, February 20–24, 2017, pp. 442–446.

104. Chen, T., Li, Z., Zhou, H., Chen, J., Luo, X., Li, X., and Zhang, X. (2018). Towards saving money in using smart contracts. In Proceedings of the 40th International Conference on Software Engineering: New Ideas and Emerging Results, ICSE (NIER) 2018, Gothenburg, Sweden, May 27 - June 03, 2018, pp. 81–84.

105. Marescotti, M., Blicha, M., Hyvärinen, A.E.J., Asadi, S., and Sharygina, N. (2018). Computing exact worst-case gas consumption for smart contracts. In Leveraging Applications of Formal Methods, Verification and Validation. Industrial Practice - 8th International Symposium, ISoLA 2018, Limassol, Cyprus, November 5–9, 2018, Proceedings, Part IV, pp. 450–465.

106. Grech, N., Kong, M., Jurisevic, A., Brent, L., Scholz, B., and Smaragdakis, Y. (2018). Madmax: surviving out-of-gas conditions in Ethereum smart contracts. In Proceedings of the ACM on Programming Languages, *2* (OOPSLA), pp. 116:1–116:27.

107. Albert, E., Gordillo, P., Rubio, A., and Sergey, I. (2019). Running on fumes—preventing out-of-gas vulnerabilities in Ethereum smart contracts using static resource analysis. In Verification and Evaluation of Computer and Communication Systems - 13th International Conference, VECoS 2019, Porto, Portugal, October 9, 2019, Proceedings, pp. 63–78.

108. Albert, E., Correas, J., Gordillo, P., Román-Díez, G., and Rubio, A. (2020). GASOL: gas analysis and optimization for Ethereum smart contracts. In Tools and Algorithms for the Construction and Analysis of Systems - 26th International Conference, TACAS 2020, Held as Part of the European Joint Conferences on Theory and Practice of Software, ETAPS 2020, Dublin, Ireland, April 25–30, 2020, Proceedings, Part II, pp. 118–125.

109. Albert, E., Gordillo, P., Rubio, A., and Schett, M.A. (2020). Synthesis of super-optimized smart contracts using max-smt. In Computer Aided Verification - 32nd International Conference, CAV 2020, Los Angeles, CA, USA, July 21–24, 2020, Proceedings, Part I, Vol. 12224 of Lecture Notes in Computer Science (Springer), pp. 177–200.

110. Chen, T., Feng, Y., Li, Z., Zhou, H., Luo, X., Li, X., Xiao, X., Chen, J., and Zhang, X. (2020). Gaschecker: scalable analysis for discovering gas-inefficient smart contracts. IEEE Trans. Emerging Top. Comput. 1–14.

111. Nikolić, I., Kolluri, A., Sergey, I., Saxena, P., and Hobor, A. (2018). Finding the greedy, prodigal, and suicidal contracts at scale. In Proceedings of the 34th Annual Computer Security Applications Conference, ACSAC 2018, San Juan, PR, USA, December 03–07, 2018, pp. 653–663.

112. Kolluri, A., Nikolić, I., Sergey, I., Hobor, A., and Saxena, P. (2019). Exploiting the laws of order in smart contracts. In Proceedings of the 28th ACM SIGSOFT International Symposium on Software Testing and Analysis, ISSTA 2019, Beijing, China, July 15–19, 2019, pp. 363–373.

113. Torres, C.F., Schütte, J., and State, R. (2018). Osiris: hunting for integer bugs in Ethereum smart contracts. In Proceedings of the 34th Annual Computer Security Applications Conference, ACSAC 2018, San Juan, PR, USA, December 03–07, 2018, pp. 664–676.

114. So, S., Lee, M., Park, J., Lee, H., and Oh, H. (2020). VERISMART: a highly precise safety verifier for Ethereum smart contracts. In 2020 IEEE Symposium on Security and Privacy, SP 2020, San Francisco, CA, USA, May 18–21, 2020 (IEEE), pp. 1678–1694.

115. Albert, E., Gordillo, P., Livshits, B., Rubio, A., Sergey, I., and Ethir. (2018). A framework for high-level analysis of Ethereum bytecode. In Automated Technology for Verification and Analysis - 16th International Symposium, ATVA 2018, Los Angeles, CA, USA, October 7–10, 2018, Proceedings, pp. 513–520.

116. Albert, E., Correas, J., Gordillo, P., Román-Díez, G., and Rubio, A. (2019). SAFEVM: a safety verifier for Ethereum smart contracts. In Proceedings of the 28th ACM SIGSOFT International Symposium on Software Testing and Analysis, ISSTA 2019, Beijing, China, July 15–19, 2019, pp. 386–389.

117. Mueller, B. (2018). Smashing Ethereum Smart Contracts for Fun and ACTUAL Profit (HITB SECCONF Amsterdam). https://conference.hitb.org/hitbsecconf2018ams/sessions/smashing-ethereum-smart-contracts-for-fun-and-actual-profit/.

118. Mossberg, M., Manzano, F., Hennenfent, E., Groce, A., Grieco, G., Feist, J., Brunson, T., and Dinaburg, A. (2019). Manticore: a user-friendly symbolic execution framework for binaries and smart contracts. In 34th IEEE/ACM International Conference on Automated Software Engineering, ASE 2019, San Diego, CA, USA, November 11–15, 2019 (IEEE), pp. 1186–1189.

119. Krupp, J., and Rossow, C. (2018). teether: gnawing at Ethereum to automatically exploit smart contracts. In 27th USENIX Security Symposium, USENIX Security 2018, Baltimore, MD, USA, August 15–17, 2018, pp. 1317–1333.

120. Chang, J., Gao, B., Xiao, H., Sun, J., Cai, Y., and Yang, Z. (2019). scompile: critical path identification and analysis for smart contracts. In Formal Methods and Software Engineering - 21st International Conference on Formal Engineering Methods, ICFEM 2019, Shenzhen, China, November 5–9, 2019, Proceedings, pp. 286–304.

121. Feng, Y., Torlak, E., and Bodík, R. (2019). Precise attack synthesis for smart contracts. http://arxiv.org/abs/1902.06067.

122. Tsankov, P., Dan, A.M., Drachsler-Cohen, D., Gervais, A., Bünzli, F., and Vechev, M.T. (2018). Securify: practical security analysis of smart contracts. In Proceedings of the 2018 ACM SIGSAC Conference on Computer and Communications Security, CCS 2018, Toronto, ON, Canada, October 15–19, 2018, pp. 67–82.

123. Permenev, A., Dimitrov, D., Tsankov, P., Drachsler-Cohen, D., and Vechev, M.T. (2020). Verx: safety verification of smart contracts. In 2020 IEEE Symposium on Security and Privacy, SP 2020, San Francisco, CA, USA, May 18–21, 2020 (IEEE), pp. 1661–1677.

124. Tikhomirov, S., Voskresenskaya, E., Ivanitskiy, I., Takhaviev, R., Marchenko, E., and Alexandrov, Y. (2018). Smartcheck: static analysis of Ethereum smart contracts. In 1st IEEE/ACM International Workshop on Emerging Trends in Software Engineering for Blockchain, WETSEB@ICSE 2018, Gothenburg, Sweden, May 27 - June 3, 2018, pp. 9–16.

125. Lu, N., Wang, B., Zhang, Y., Shi, W., and Esposito, C. (2019). NeuCheck: a more practical Ethereum smart contract security analysis tool. Softw. Pract. Experience. https://doi.org/10.1002/spe.2745.

126. Grishchenko, I., Maffei, M., and Schneidewind, C. (2018). Ethertrust: sound static analysis of Ethereum bytecode. https://secpriv.tuwien.ac.at/fileadmin/t/secpriv/Papers/post2018-tr.pdf.

127. Grishchenko, I., Maffei, M., and Schneidewind, C. (2018). Foundations and tools for the static analysis of Ethereum smart contracts. In Computer Aided Verification - 30th International Conference, CAV 2018, Held as Part of the Federated Logic Conference, FloC 2018, Oxford, UK, July 14–17, 2018, Proceedings, Part I, pp. 51–78.

128. Brent, L., Jurisevic, A., Kong, M., Liu, E., Gauthier, F., Gramoli, V., Holz, R., and Scholz, B. (2018). Vandal: a scalable security analysis framework for smart contracts. http://arxiv.org/abs/1809.03981.

129. Grech, N., Brent, L., Scholz, B., and Smaragdakis, Y. (2019). Gigahorse: thorough, declarative decompilation of smart contracts. In Proceedings of the 41st International Conference on Software Engineering, ICSE 2019, Montreal, QC, Canada, May 25–31, 2019 (IEEE/ACM), pp. 1176–1186.

130. Schneidewind, C., Grishchenko, I., Scherer, M., Maffei, M., and ethor. (2020). Practical and provably sound static analysis of Ethereum smart contracts. In CCS '20: 2020 ACM SIGSAC Conference on Computer and Communications Security, Virtual Event, USA, November 9–13, 2020 (ACM), pp. 621–640.

131. Feist, J., Grieco, G., and Groce, A. (2019). Slither: a static analysis framework for smart contracts. In Proceedings of the 2nd International Workshop on Emerging Trends in Software Engineering for Blockchain, WETSEB@ICSE 2019, Montreal, QC, Canada, May 27, 2019, pp. 8–15.

132. Zhou, E., Hua, S., Pi, B., Sun, J., Nomura, Y., Yamashita, K., and Kurihara, H. (2018). Security assurance for smart contract. In 9th IFIP International Conference on New Technologies, Mobility and Security, NTMS 2018, Paris, France, February 26–28, 2018, pp. 1–5.

133. Nehai, Z., Piriou, P., and Daumas, F.F. (2018). Model-checking of smart contracts. In IEEE International Conference on Internet of Things (iThings) and IEEE Green Computing and Communications (GreenCom) and IEEE Cyber, Physical and Social Computing (CPSCom) and IEEE Smart Data (SmartData), iThings/GreenCom/CPSCom/SmartData 2018, Halifax, NS, Canada, July 30 - August 3, 2018, pp. 980–987.

134. Kalra, S., Goel, S., Dhawan, M., and Sharma, S. (2018). ZEUS: analyzing safety of smart contracts. In 25th Annual Network and Distributed System Security Symposium, NDSS 2018, San Diego, California, USA, February 18–21, 2018.

135. Nehai, Z., and Bobot, F. (2019). Deductive proof of Ethereum smart contracts using why3. http://arxiv.org/abs/1904.11281.

136. Alt, L., and Reitwießner, C. (2018). Smt-based verification of solidity smart contracts. In Leveraging Applications of Formal Methods, Verification and Validation. Industrial Practice - 8th International Symposium, ISoLA 2018, Limassol, Cyprus, November 5–9, 2018, Proceedings, Part IV, pp. 376–388.

137. Jiang, B., Liu, Y., and Chan, W.K. (2018). Contractfuzzer: fuzzing smart contracts for vulnerability detection. In Proceedings of the 33rd ACM/IEEE International Conference on Automated Software Engineering, ASE 2018, Montpellier, France, September 3–7, 2018, pp. 259–269.

138. He, J., Balunovic, M., Ambroladze, N., Tsankov, P., and Vechev, M.T. (2019). Learning to fuzz from symbolic execution with application to smart contracts. In Proceedings of the 2019 ACM SIGSAC Conference on Computer and Communications Security, CCS 2019, London, UK, November 11–15, 2019, pp. 531–548.

139. Grieco, G., Song, W., Cygan, A., Feist, J., and Groce, A. (2020). Echidna: effective, usable, and fast fuzzing for smart contracts. In ISSTA '20: 29th ACM SIGSOFT International Symposium on Software Testing and Analysis, Virtual Event, USA, July 18–22, 2020 (ACM), pp. 557–560.

140. Kosba, A.E., Miller, A., Shi, E., Wen, Z., and Papamanthou, C. (2016). Hawk: the blockchain model of cryptography and privacy-preserving smart contracts. In IEEE Symposium on Security and Privacy, SP 2016, San Jose, CA, USA, May 22–26, 2016, pp. 839–858.

141. Eberhardt, J., and Tai, S. (2018). Zokrates: scalable privacy-preserving off-chain computations. In IEEE International Conference on Internet of Things (iThings) and IEEE Green Computing and Communications (GreenCom) and IEEE Cyber, Physical and Social Computing (CPSCom) and IEEE Smart Data (SmartData), iThings/GreenCom/CPSCom/SmartData 2018, Halifax, NS, Canada, July 30 - August 3, 2018, pp. 1084–1091.

142. Bhargavan, K., Delignat-Lavaud, A., Fournet, C., Gollamudi, A., Gonthier, G., Kobeissi, N., Kulatova, N., Rastogi, A., Sibut-Pinote, T., Swamy, N., and Béguelin, S.Z. (2016). Formal verification of smart contracts: short paper. In Proceedings of the 2016 ACM Workshop on Programming Languages and Analysis for Security, PLAS@CCS 2016, Vienna, Austria, October 24, 2016, pp. 91–96.

143. Chatterjee, K., Goharshady, A.K., and Velner, Y. (2018). Quantitative analysis of smart contracts. In Programming Languages and Systems - 27th European Symposium on Programming, ESOP 2018, Held as Part of the European Joint Conferences on Theory and Practice of Software, ETAPS 2018, Thessaloniki, Greece, April 14–20, 2018, Proceedings, pp. 739–767.

144. Mavridou, A., and Laszka, A. (2018). Designing secure Ethereum smart contracts: a finite state machine based approach. In Financial Cryptography and Data Security - 22nd International Conference, FC 2018, Nieuwpoort, Curaçao, February 26 - March 2, 2018, Revised Selected Papers, pp. 523–540.

145. Mavridou, A., Laszka, A., Stachtiari, E., and Dubey, A. (2019). Verisolid: correct-by-design smart contracts for Ethereum. In Financial Cryptography and Data Security - 23rd International Conference, FC 2019, Frigate Bay, St. Kitts and Nevis, February 18–22, 2019, Revised Selected Papers, pp. 446–465.

146. Xu, W., and Fink, G.A. (2019). Building executable secure design models for smart contracts with formal methods. In Financial Cryptography and Data Security - FC 2019 International Workshops, VOTING and WTSC, St. Kitts, St. Kitts and Nevis, February 18–22, 2019, Revised Selected Papers, pp. 154–169.

147. Banach, R. (2019). Verification-led smart contracts. In Financial Cryptography and Data Security - FC 2019 International Workshops, VOTING and WTSC, St. Kitts, St. Kitts and Nevis, February 18–22, 2019, Revised Selected Papers, pp. 106–121.

148. Spoto, F. (2019). A java framework for smart contracts. In Financial Cryptography and Data Security - FC 2019 International Workshops, VOTING and WTSC, St. Kitts, St. Kitts and Nevis, February 18–22, 2019, Revised Selected Papers, pp. 122–137.

149. Yang, Z., and Lei, H. (2019). Fether: an extensible definitional interpreter for smart-contract verifications in coq. IEEE Access *7*, 37770–37791.

150. Pettersson, J., and Edström, R. (2016). Safer Smart Contracts through Type-Driven Development, Master's Thesis, Master's thesis (Chalmers University of Technology & University of Gothenburg).

151. Biryukov, A., Khovratovich, D., and Tikhomirov, S. (2017). Findel: secure derivative contracts for Ethereum. In Financial Cryptography and Data Security - FC 2017 International Workshops, WAHC, BITCOIN, VOTING, WTSC, and TA, Sliema, Malta, April 7, 2017, Revised Selected Papers, pp. 453–467.

152. Yang, Z., and Lei, H. (2018). Lolisa: formal syntax and semantics for a subset of the solidity programming language. http://arxiv.org/abs/1803.09885.

153. Schrans, F., Eisenbach, S., and Drossopoulou, S. (2018). Writing safe smart contracts in flint. In Conference Companion of the 2nd International Conference on Art, Science, and Engineering of Programming, Nice, France, April 09–12, 2018, pp. 218–219.

154. Crafa, S., Pirro, M.D., and Zucca, E. (2019). Is solidity solid enough? In Financial Cryptography and Data Security - FC 2019 International Workshops, VOTING and WTSC, St. Kitts, St. Kitts and Nevis, February 18–22, 2019, Revised Selected Papers, pp. 138–153.

155. Sergey, I., Kumar, A., and Hobor, A. (2018). Scilla: a smart contract intermediate-level language. http://arxiv.org/abs/1801.00687.

156. Sergey, I., Nagaraj, V., Johannsen, J., Kumar, A., Trunov, A., and Hao, K.C.G. (2019). Safer smart contract programming with scilla. In Proceedings of the ACM on Programming Languages, *3* (OOPSLA), pp. 185:1–185:30.

157. Hirai, Y. (2017). Defining the Ethereum virtual machine for interactive theorem provers. In Financial Cryptography and Data Security - FC 2017 International Workshops, WAHC, BITCOIN, VOTING, WTSC, and TA, Sliema, Malta, April 7, 2017, Revised Selected Papers, pp. 520–535.

158. Amani, S., Bégel, M., Bortin, M., and Staples, M. (2018). Towards verifying Ethereum smart contract bytecode in isabelle/hol. In Proceedings of the 7th ACM SIGPLAN International Conference on Certified Programs and Proofs, CPP 2018, Los Angeles, CA, USA, January 8–9, 2018, pp. 66–77.

159. Hildenbrandt, E., Saxena, M., Rodrigues, N., Zhu, X., Daian, P., Guth, D., Moore, B.M., Park, D., Zhang, Y., Stefanescu, A., and Rosu, G. (2018). KEVM: a complete formal semantics of the Ethereum virtual machine. In 31st IEEE Computer Security Foundations Symposium, CSF 2018, Oxford, United Kingdom, July 9–12, 2018, pp. 204–217.

160. Zyskind, G., and Pentland, A. (2018). Enigma: decentralized computation platform with guaranteed privacy. In New Solutions for Cybersecurity, H. Shrobe, D.L. Shrier, and A. Pentland, eds. (MIT Press), pp. 425–454.

161. Choudhuri, A.R., Green, M., Jain, A., Kaptchuk, G., and Miers, I. (2017). Fairness in an unfair world: fair multiparty computation from public bulletin boards. In Proceedings of the 2017 ACM SIGSAC Conference on Computer and Communications Security, CCS 2017, Dallas, TX, USA, October 30–November 3, 2017, pp. 719–728.

162. Sánchez, D.C. (2018). Raziel: private and verifiable smart contracts on blockchains. http://arxiv.org/abs/1807.09484.

163. Brandenburger, M., Cachin, C., Kapitza, R., and Sorniotti, A. (2018). Blockchain and trusted computing: problems, pitfalls, and a solution for hyperledger fabric. http://arxiv.org/abs/1805.08541.

164. Bowman, M., Miele, A., Steiner, M., and Vavala, B. (2018). Private data objects: an overview. http://arxiv.org/abs/1807.05686.

165. Cheng, R., Zhang, F., Kos, J., He, W., Hynes, N., Johnson, N.M., Juels, A., Miller, A., and Song, D. (2019). Ekiden: a platform for confidentiality-preserving, trustworthy, and performant smart contracts. In IEEE European Symposium on Security and Privacy, EuroS&P 2019, Stockholm, Sweden, June 17–19, 2019, pp. 185–200.

166. Das, P., Eckey, L., Frassetto, T., Gens, D., Hostáková, K., Jauernig, P., Faust, S., and Sadeghi, A. (2019). Fastkitten: practical smart contracts on bitcoin. In 28th USENIX Security Symposium, USENIX Security 2019, Santa Clara, CA, USA, August 14–16, 2019, pp. 801–818.

167. Kaptchuk, G., Green, M., and Miers, I. (2019). Giving state to the stateless: augmenting trustworthy computation with ledgers. In 26th Annual Network and Distributed System Security Symposium, NDSS 2019, San Diego, California, USA, February 24–27, 2019.

168. Lind, J., Naor, O., Eyal, I., Kelbert, F., Sirer, E.G., and Pietzuch, P.R. (2019). Teechain: a secure payment network with asynchronous blockchain access. In Proceedings of the 27th ACM Symposium on Operating Systems Principles, SOSP 2019, Huntsville, ON, Canada, October 27–30, 2019, pp. 63–79.

169. Decker, C., and Wattenhofer, R. (2015). A fast and scalable payment network with bitcoin duplex micropayment channels. In Stabilization, Safety, and Security of Distributed Systems - 17th International Symposium, SSS 2015, Edmonton, AB, Canada, August 18–21, 2015, Proceedings, pp. 3–18.

170. McCorry, P., Möser, M., Shahandashti, S.F., and Hao, F. (2016). Towards bitcoin payment networks. In Information Security and Privacy - 21st Australasian Conference, ACISP 2016, Melbourne, VIC, Australia, July 4–6, 2016, Proceedings, Part I, pp. 57–76.

171. Heilman, E., Baldimtsi, F., and Goldberg, S. (2016). Blindly signed contracts: anonymous on-blockchain and off-blockchain bitcoin transactions. In Financial Cryptography and Data Security - FC 2016 International Workshops, BITCOIN, VOTING, and WAHC, Christ Church, Barbados, February 26, 2016, Revised Selected Papers, pp. 43–60.

172. Green, M., and Miers, I. (2017). Bolt: anonymous payment channels for decentralized currencies. In Proceedings of the 2017 ACM SIGSAC Conference on Computer and Communications Security, CCS 2017, Dallas, TX, USA, October 30 - November 03, 2017, pp. 473–489.

173. Malavolta, G., Moreno-Sánchez, P., Kate, A., Maffei, M., and Ravi, S. (2017). Concurrency and privacy with payment-channel networks. In Proceedings of the 2017 ACM SIGSAC Conference on Computer and Communications Security, CCS 2017, Dallas, TX, USA, October 30 - November 03, 2017, pp. 455–471.

174. Tremback, J., and Hess, Z. (2015). Universal payment channels. http://jtremback.github.io/universal-payment-channels/universal-payment-channels.pdf.

175. Peterson, D. (2016). Sparky: a lightning network in two pages of solidity. https://www.blunderingcode.com/a-lightning-network-in-two-pages-of-solidity/.

176. Raiden Network (2017). What is the Raiden network?. https://raiden.network/101.html.

177. Dziembowski, S., Eckey, L., Faust, S., and Malinowski, D. (2019). Perun: virtual payment hubs over cryptocurrencies. In 2019 IEEE Symposium on Security and Privacy, SP 2019, San Francisco, CA, USA, May 19–23, 2019 (IEEE), pp. 106–123.

178. Bentov, I., Kumaresan, R., and Miller, A. (2017). Instantaneous decentralized poker. In Advances in Cryptology - ASIACRYPT 2017 - 23rd International Conference on the Theory and Applications of Cryptology and Information Security, Hong Kong, China, December 3–7, 2017, Proceedings, Part II, pp. 410–440.

179. Dziembowski, S., Faust, S., and Hostáková, K. (2018). General state channel networks. In Proceedings of the 2018 ACM SIGSAC Conference on Computer and Communications Security, CCS 2018, Toronto, ON, Canada, October 15–19, 2018 (ACM), pp. 949–966.

180. Miller, A., Bentov, I., Bakshi, S., Kumaresan, R., and McCorry, P. (2019). Sprites and state channels: payment networks that go faster than lightning. In Financial Cryptography and Data Security - 23rd International Conference, FC 2019, Frigate Bay, St. Kitts and Nevis, February 18–22, 2019, Revised Selected Papers, pp. 508–526.

181. Dziembowski, S., Eckey, L., Faust, S., Hesse, J., and Hostáková, K. (2019). Multi-party virtual state channels. In Advances in Cryptology - EUROCRYPT 2019 - 38th Annual International Conference on the Theory and Applications of Cryptographic Techniques, Darmstadt, Germany, May 19–23, 2019, Proceedings, Part I, pp. 625–656.

182. Close, T., and Stewart, A. (2018). Forcemove: an n-party state channel protocol. https://magmo.com/force-move-games.pdf.

183. McCorry, P., Buckland, C., Bakshi, S., Wüst, K., and Miller, A. (2019). You sank my battleship! A case study to evaluate state channels as a scaling solution for cryptocurrencies. In Financial Cryptography and Data Security - FC 2019 International Workshops, VOTING and WTSC, St. Kitts, St. Kitts and Nevis, February 18–22, 2019, Revised Selected Papers, Vol. 11599 of Lecture Notes in Computer Science (Springer), pp. 35–49.

184. Buckland, C., and McCorry, P. (2019). Two-party state channels with assertions. In Financial Cryptography and Data Security - FC 2019 International Workshops, VOTING and WTSC, St. Kitts, St. Kitts and Nevis, February 18–22, 2019, Revised Selected Papers, pp. 3–11.

185. McCorry, P., Bakshi, S., Bentov, I., Meiklejohn, S., and Miller, A. (2019). Pisa: arbitration outsourcing for state channels. In Proceedings of the 1st ACM Conference on Advances in Financial Technologies, AFT 2019, Zurich, Switzerland, October 21–23, 2019, pp. 16–30.

186. Möser, M., Eyal, I., and Sirer, E.G. (2016). Bitcoin covenants. In Financial Cryptography and Data Security - FC 2016 International Workshops, BITCOIN, VOTING, and WAHC, Christ Church, Barbados, February 26, 2016, Revised Selected Papers, pp. 126–141.

187. O'Connor, R., and Piekarska, M. (2017). Enhancing bitcoin transactions with covenants. In Financial Cryptography and Data Security - FC 2017 International Workshops, WAHC, BITCOIN, VOTING, WTSC, and TA, Sliema, Malta, April 7, 2017, Revised Selected Papers, pp. 191–198.

188. Fynn, E., Bessani, A., and Pedone, F. (2020). Smart contracts on the move. In 50th Annual IEEE/IFIP International Conference on Dependable Systems and Networks, DSN 2020, Valencia, Spain, June 29 - July 2, 2020 (IEEE), pp. 233–244.

189. Westerkamp, M. (2019). Verifiable smart contract portability. In IEEE International Conference on Blockchain and Cryptocurrency, ICBC 2019, Seoul, Korea (South), May 14–17, 2019, pp. 1–9.

190. Dickerson, T.D., Gazzillo, P., Herlihy, M., Saraph, V., and Koskinen, E. (2018). Proof-carrying smart contracts. In Financial Cryptography and Data Security - FC 2018 International Workshops, BITCOIN, VOTING, and WTSC, Nieuwpoort, Curaçao, March 2, 2018, Revised Selected Papers, pp. 325–338.

191. Kalodner, H.A., Goldfeder, S., Chen, X., Weinberg, S.M., and Felten, E.W. (2018). Arbitrum: scalable, private smart contracts. In 27th USENIX Security Symposium, USENIX Security 2018, Baltimore, MD, USA, August 15–17, 2018, pp. 1353–1370.

192. Das, S., Ribeiro, V.J., and Anand, A. (2019). YODA: enabling computationally intensive contracts on blockchains with byzantine and selfish nodes. In 26th Annual Network and Distributed System Security Symposium, NDSS 2019, San Diego, California, USA, February 24–27, 2019.

193. Bowe, S., Chiesa, A., Green, M., Miers, I., Mishra, P., and Wu, H. (2020). ZEXE: enabling decentralized private computation. In 2020 IEEE Symposium on Security and Privacy, SP 2020, San Francisco, CA, USA, May 18–21, 2020 (IEEE), pp. 947–964.

194. Wüst, K., Matetic, S., Egli, S., Kostiainen, K., and Capkun, S. (2020). ACE: asynchronous and concurrent execution of complex smart contracts. In CCS '20: 2020 ACM SIGSAC Conference on Computer and Communications Security, Virtual Event, USA, November 9–13, 2020 (ACM), pp. 587–600.

195. Gavin, A. (2012). Bip16: pay to script hash. https://en.bitcoin.it/wiki/BIP_0016.

196. Bitcoin Forum. (2014). Ascii art. https://bitcointalk.org/index.php?topic=568304.0.

197. Canetti, R. (2001). Universally composable security: a new paradigm for cryptographic protocols. In 42nd Annual Symposium on Foundations of Computer Science, FOCS 2001, 14–17 October 2001, Las Vegas, Nevada, USA, pp. 136–145.

198. Jourenko, M., Kurazumi, K., Larangeira, M., and Tanaka, K. (2019). Sok: a taxonomy for layer-2 scalability related protocols for cryptocurrencies. https://eprint.iacr.org/2019/352.

199. Schnorr, C. (1989). Efficient identification and signatures for smart cards. In Advances in Cryptology - CRYPTO '89, 9th Annual International Cryptology Conference, Santa Barbara, California, USA, August 20–24, 1989, Proceedings, Vol. 435 of Lecture Notes in Computer Science, G. Brassard, ed. (Springer), pp. 239–252.

200. Behrmann, G., David, A., and Larsen, K.G. (2004). A tutorial on uppaal. In Formal Methods for the Design of Real-Time Systems, International School on Formal Methods for the Design of Computer, Communication and Software Systems, SFM-RT 2004, Bertinoro, Italy, September 13–18, 2004, Revised Lectures, pp. 200–236.

201. Coq Development Team (2020). The coq proof assistant. https://coq.inria.fr/.

202. Edgington, B. (2017). Documentation for the lll compiler. https://lll-docs.readthedocs.io/en/latest/index.html.

203. Qureshi, H. (2020). Flash loans: why flash attacks will be the new normal. https://medium.com/dragonfly-research/flash-loans-why-flash-attacks-will-be-the-new-normal-5144e23ac75a.

204. Fischlin, M. (2001). A cost-effective pay-per-multiplication comparison method for millionaires. In Topics in Cryptology - CT-RSA 2001, the Cryptographer's Track at RSA Conference 2001, San Francisco, CA, USA, April 8–12, 2001, Proceedings, pp. 457–472.

205. Pedersen, T.P. (1991). Non-interactive and information-theoretic secure verifiable secret sharing. In Advances in Cryptology - CRYPTO '91, 11th Annual International Cryptology Conference, Santa Barbara, California, USA, August 11–15, 1991, Proceedings, Vol. 576 of Lecture Notes in Computer Science, J. Feigenbaum, ed. (Springer), pp. 129–140.

206. Vogelsteller, F., and Buterin, V. (2015). Eip-20: Erc-20 token standard. https://eips.ethereum.org/EIPS/eip-20.

207. CoinDesk. (2020). What is defi?. https://www.coindesk.com/what-is-defi.

208. Sillaber, C., and Waltl, B. (2017). Life cycle of smart contracts in blockchain ecosystems. Datenschutz und Datensicherheit *41*, 497–500.

209. Filliâtre, J., and Paskevich, A. (2013). Why3—where programs meet provers. In Programming Languages and Systems - 22nd European Symposium on Programming, ESOP 2013, Held as Part of the European Joint Conferences on Theory and Practice of Software, ETAPS 2013, Rome, Italy, March 16–24, 2013. Proceedings, pp. 125–128.

210. (2017). Vyper documentation. https://vyper.readthedocs.io/en/stable/.

211. Massalin, H. (1987). Superoptimizer—a look at the smallest program. In Proceedings of the Second International Conference on Architectural Support for Programming Languages and Operating Systems (ASPLOS II), Palo Alto, California, USA, October 5–8, 1987, R.H. Katz and M. Freeman, eds. (ACM Press), pp. 122–126.

212. Zhu, R., Ding, C., and Huang, Y. (2019). Efficient publicly verifiable 2pc over a blockchain with applications to financially-secure computations. In Proceedings of the 2019 ACM SIGSAC Conference on Computer and Communications Security, CCS 2019, London, UK, November 11–15, 2019 (ACM), pp. 633–650.

213. Solar-Lezama, A., Tancau, L., Bodík, R., Seshia, S.A., and Saraswat, V.A. (2006). Combinatorial sketching for finite programs. In Proceedings of the 12th International Conference on Architectural Support for Programming Languages and Operating Systems, ASPLOS 2006, San Jose, CA, USA, October 21–25, 2006, pp. 404–415.

214. ETAPS (2019). Competition on software verification (sv-comp). https://sv-comp.sosy-lab.org/2019/index.php.

215. Mueller, B. (2018). Mythril. https://github.com/ConsenSys/mythril.

216. Mueller, B. (2018). Laser-ethereum: symbolic virtual machine for Ethereum. https://github.com/b-mueller/laser-ethereum.

217. Ethereum Foundation (2019). Ethereum contract ABI. https://github.com/ethereum/wiki/wiki/Ethereum-Contract-ABI.

218. Aho, A.V., Sethi, R., and Ullman, J.D. (1986). Compilers: Principles, Techniques, and Tools, Addison-Wesley Series in Computer Science/World Student Series Edition (Addison-Wesley).

219. Clark, J., and DeRose, S. (1999). Xml path language (xpath). http://new-design.renderx.com/files/demos/xmlspec/xpath/REC-xpath-19991116.pdf.

220. ConsenSys. (2017). Solidity parser. https://github.com/ConsenSys/solidity-parser.

221. Cimatti, A., Clarke, E.M., Giunchiglia, F., and Roveri, M. (1999). NUSMV: a new symbolic model verifier. In Computer Aided Verification, 11th International Conference, CAV '99, Trento, Italy, July 6–10, 1999, Proceedings, pp. 495–499.

222. Lamport, L. (2002). Specifying Systems, the TLA+ Language and Tools for Hardware and Software Engineers (Addison-Wesley).

223. Ben-Sasson, E., Chiesa, A., Genkin, D., Tromer, E., and Virza, M. (2013). Snarks for C: verifying program executions succinctly and in zero knowledge. In Advances in Cryptology - CRYPTO 2013 - 33rd Annual Cryptology Conference, Santa Barbara, CA, USA, August 18–22, 2013. Proceedings, Part II, pp. 90–108.

224. Parno, B., Howell, J., Gentry, C., and Raykova, M. (2013). Pinocchio: nearly practical verifiable computation. In 2013 IEEE Symposium on Security and Privacy, SP 2013, Berkeley, CA, USA, May 19–22, 2013, pp. 238–252.

225. Ahman, D. (2016). F*: verification system for effectful programs. https://fstar-lang.org.

226. Abrial, J. (2010). Modeling in Event-B: System and Software Engineering (Cambridge University Press).

227. Brady, E. (2013). Idris, a general-purpose dependently typed programming language: design and implementation. J. Funct. Program *23*, 552–593.

228. Nipkow, T., Paulson, L.C., and Wenzel, M. (2002). Isabelle/HOL: A Proof Assistant for Higher-Order Logic, Vol. 2283 of Lecture Notes in Computer Science (Springer).

229. Rosu, G., and Serbanuta, T. (2010). An overview of the K semantic framework. J. Log. Algebr. Program *79*, 397–434.

230. Ethereum Foundation (2020). Ethereum tests. https://github.com/ethereum/tests.

231. Buterin, V. (2017). Blockchain and smart contract mechanism design challenges. https://fc17.ifca.ai/wtsc/Vitalik%20Malta.pdf.

232. Ben-Sasson, E., Chiesa, A., Garman, C., Green, M., Miers, I., Tromer, E., and Virza, M. (2014). Zerocash: decentralized anonymous payments from bitcoin. In 2014 IEEE Symposium on Security and Privacy, SP 2014, Berkeley, CA, USA, May 18–21, 2014, pp. 459–474.

233. Yu, H., Zhang, Z., and Liu, J. (2017). Research on scaling technology of bitcoin blockchain. J. Computer Res. Development *54*, 2390–2403.

234. Lombrozo, E., Lau, J., and Wuille, P. (2015). Bip 141: segregated witness (consensus layer). https://github.com/bitcoin/bips/blob/master/bip-0141.mediawiki.

235. Back, A., Corallo, M., Dashjr, L., Friedenbach, M., Maxwell, G., Miller, A., Poelstra, A., Timón, J., and Wuille, P. (2014). Enabling blockchain innovations with pegged sidechains. http://www.opensciencereview.com/papers/123/enablingblockchain-innovations-with-pegged-sidechains.

236. Teutsch, J., Straka, M., and Boneh, D. (2019). Retrofitting a two-way peg between blockchains. http://arxiv.org/abs/1908.03999.

237. Luu, L., Narayanan, V., Zheng, C., Baweja, K., Gilbert, S., and Saxena, P. (2016). A secure sharding protocol for open blockchains. In Proceedings of the 2016 ACM SIGSAC Conference on Computer and Communications Security, Vienna, Austria, October 24–28, 2016 (ACM), pp. 17–30.

238. Necula, G.C. (1997). Proof-carrying code. In Conference Record of POPL'97: The 24th ACM SIGPLAN-SIGACT Symposium on Principles of Programming Languages, Papers Presented at the Symposium, Paris, France, 15–17 January 1997, pp. 106–119.

239. Bünz, B., Bootle, J., Boneh, D., Poelstra, A., Wuille, P., and Maxwell, G. (2017). Bulletproofs: efficient range proofs for confidential transactions. https://cryptopapers.info/assets/pdf/bulletproofs.pdf.

240. Costan, V., and Devadas, S. (2016). Intel sgx explained. https://eprint.iacr.org/2016/086.pdf.

241. ARM Limited (2009). Security technology: building a secure system using trustzone technology. http://infocenter.arm.com/help/topic/com.arm.doc.prd29-genc-009492c/PRD29-GENC-009492C_trustzone_security_whitepaper.pdf.

242. Bulck, J.V., Minkin, M., Weisse, O., Genkin, D., Kasikci, B., Piessens, F., Silberstein, M., Wenisch, T.F., Yarom, Y., and Strackx, R. (2019). Breaking virtual memory protection and the SGX ecosystem with foreshadow. IEEE Micro *39*, 66–74.

243. ElementsProject (2016). A Lightning Network implementation in C. https://github.com/ElementsProject/lightning.

244. ACINQ (2017). A Scala implementation of the Lightning Network. https://github.com/ACINQ/eclair.

245. Boldyreva, A. (2003). Threshold signatures, multisignatures and blind signatures based on the Gap-Diffie-Hellman-group signature scheme. In Public Key Cryptography - PKC 2003, 6th International Workshop on Theory and Practice in Public Key Cryptography, Miami, FL, USA, January 6–8, 2003, Proceedings, pp. 31–46.

246. Khalil, R., and Gervais, A. (2017). Revive: rebalancing off-blockchain payment networks. In Proceedings of the 2017 ACM SIGSAC Conference on Computer and Communications Security, CCS 2017, Dallas, TX, USA, October 30 - November 03, 2017, pp. 439–453.

247. Subramanian, L.M., Eswaraiah, G., and Vishwanathan, R. (2019). Rebalancing in acyclic payment networks. In 17th International Conference on Privacy, Security and Trust, PST 2019, Fredericton, NB, Canada, August 26–28, 2019 (IEEE), pp. 1–5.

248. Coleman, J. (2015). State channels. https://www.jeffcoleman.ca/state-channels/.

249. Allison, I. (2016). Ethereum's Vitalik Buterin explains how state channels address privacy and scalability. https://www.ibtimes.co.uk/ethereums-vitalik-buterin-explains-how-state-channels-address-privacy-scalability-1566068.

250. Coleman, J., Horne, L., and Xuanji, L. (2018). Counterfactual: generalized state channels. https://l4.ventures/papers/statechannels.pdf.

251. Dryja, T., and Milano, S.B. (2016). Unlinkable outsourced channel monitoring. https://diyhpl.us/wiki/transcripts/scalingbitcoin/milan/unlinkable-outsourced-channel-monitoring.

252. Osuntokun, O. (2018). Hardening lightning. https://cyber.stanford.edu/sites/g/files/sbiybj9936/f/hardening_lightning_updated.pdf.

253. Ateniese, G., Magri, B., Venturi, D., and Andrade, E.R. (2017). Redactable blockchain—or—rewriting history in bitcoin and friends. In 2017 IEEE European Symposium on Security and Privacy, EuroS&P 2017, Paris, France, April 26–28, 2017 (IEEE), pp. 111–126.

254. Poon, J., and Buterin, V. (2017). Plasma: scalable autonomous smart contracts. https://www.plasma.io/plasma-deprecated.pdf.

255. Kwon, J., and Buchman, E. (2016). Cosmos: a network of distributed ledgers. https://cosmos.network/whitepaper.

256. Wood, G. (2016). Polkadot: vision for a heterogeneous multi-chain framework. https://polkadot.network/PolkaDotPaper.pdf.

257. Schindler, P., Judmayer, A., Stifter, N., and Weippl, E.R. (2019). ETHDKG: distributed key generation with Ethereum smart contracts. https://eprint.iacr.org/2019/985.

258. Sun, S., Au, M.H., Liu, J.K., and Yuen, T.H. (2017). Ringct 2.0: a compact accumulator-based (linkable ring signature) protocol for blockchain cryptocurrency monero. In Computer Security - ESORICS 2017 - 22nd European Symposium on Research in Computer Security, Oslo, Norway, September 11–15, 2017, Proceedings, Part II, Vol. 10493 of Lecture Notes in Computer Science (Springer), pp. 456–474.

259. Kerber, T., Kiayias, A., and Kohlweiss, M. (2020). Kachina—foundations of private smart contracts, to appear at CSF' 21 https://eprint.iacr.org/2020/543.pdf.

260. Bano, S., Sonnino, A., Al-Bassam, M., Azouvi, S., McCorry, P., Meiklejohn, S., and Danezis, G. (2019). Sok: consensus in the age of blockchains. In Proceedings of the 1st ACM Conference on Advances in Financial Technologies, AFT 2019, Zurich, Switzerland, October 21–23, 2019, pp. 183–198.

261. Yaga, D., Mell, P., Roby, N., and Scarfone, K. (2018). Blockchain technology overview, nISTIR 8202. 10.6028/NIST.IR.8202.

262. Garay, J.A., Kiayias, A., and Sok. (2020). A consensus taxonomy in the blockchain era. In Topics in Cryptology - CT-RSA 2020 - the Cryptographers' Track at the RSA Conference 2020, San Francisco, CA, USA, February 24–28, 2020, Proceedings, pp. 284–318.

263. Mauri, L., Cimato, S., and Damiani, E. (2020). A formal approach for the analysis of the XRP ledger consensus protocol. In Proceedings of the 6th International Conference on Information Systems Security and Privacy, ICISSP 2020, Valletta, Malta, February 25–27, 2020 (SCITEPRESS), pp. 52–63.

264. Libra Association Members (2020). Libra white paper v2.0. https://libra.org/en-US/white-paper/.

265. Garay, J.A., Kiayias, A., and Leonardos, N. (2015). The bitcoin backbone protocol: analysis and applications. In Advances in Cryptology - EUROCRYPT 2015 - 34th Annual International Conference on the Theory and Applications of Cryptographic Techniques, Sofia, Bulgaria, April 26–30, 2015, Proceedings, Part II, pp. 281–310.

266. Pass, R., Seeman, L., and Shelat, A. (2017). Analysis of the blockchain protocol in asynchronous networks. In Advances in Cryptology - EUROCRYPT 2017 - 36th Annual International Conference on the Theory and Applications of Cryptographic Techniques, Paris, France, April 30 - May 4, 2017, Proceedings, Part II, pp. 643–673.

267. Bitcoin Developer Guide. (2020). Consensus rule changes. https://bitcoin.org/en/blockchain-guide#consensus-rule-changes.

268. Canetti, R. (2000). Security and composition of multiparty cryptographic protocols. J. Cryptology *13*, 143–202.

269. Morais, E., Koens, T., Van Wijk, C., and Koren, A. (2019). A survey on zero knowledge range proofs and applications. SN Appl. Sci. *1*, 946.

270. Sabt, M., Achemlal, M., and Bouabdallah, A. (2015). Trusted execution environment: what it is, and what it is not. In 2015 IEEE TrustCom/BigDataSE/ISPA, Helsinki, Finland, August 20–22, 2015, *1*, pp. 57–64.